\documentclass[aps,prb,twocolumn,nofootinbib,
amsmath,amssymb,amsfonts,eqsecnum,superscriptaddress]{revtex4-2}
               
\AtBeginDocument{\usepackage{booktabs}}               
\makeatletter
\g@addto@macro\bfseries{\boldmath}
\makeatother

\usepackage[varg]{txfonts}

\usepackage[T1]{fontenc}
\usepackage[utf8]{inputenc}

\usepackage{hyperref}
\usepackage{amsmath}
\usepackage{color}
\usepackage{graphicx}
\usepackage[percent]{overpic}
\usepackage{mathrsfs}
\usepackage{bm}
\usepackage{braket}
\usepackage{mathtools}
\usepackage[caption=false]{subfig}

\definecolor{cL}{RGB}{59, 83, 140}
\definecolor{cM}{RGB}{33, 145, 141}
\definecolor{cH}{RGB}{95, 202, 98}
\definecolor{cG}{RGB}{204,204,204}

\definecolor{cL}{RGB}{59, 83, 140}
\definecolor{cM}{RGB}{33, 145, 141}
\definecolor{cH}{RGB}{95, 202, 98}
\definecolor{cG}{RGB}{204,204,204}

\begin{document}

\title{Disorder-Free Localization and Many-Body Quantum Scars from Magnetic Frustration}
\author{Paul A. McClarty}
\affiliation{Max-Planck-Institut f\"ur Physik komplexer Systeme, 01187 Dresden, Germany}
\author{Masudul Haque} 
\affiliation{Max-Planck-Institut f\"ur Physik komplexer Systeme, 01187 Dresden, Germany}
\affiliation{Department of Theoretical Physics, Maynooth University, Co. Kildare, Ireland}
\author{Arnab Sen} 
\affiliation{School of Physical Sciences, Indian Association for the Cultivation of Science, Kolkata 700032, India}
\author{Johannes Richter} 
\affiliation{Max-Planck-Institut f\"ur Physik komplexer Systeme, 01187 Dresden, Germany}
\affiliation{Institut f\"ur Physik, Universit\"at Magdeburg, P.O. Box 4120, D-39016 Magdeburg, Germany}

\begin{abstract}
The concept of geometrical frustration has led to rich insights into condensed matter physics, especially as a mechansim to produce exotic low energy states of matter.  Here we show that  frustration provides a natural vehicle to generate models exhibiting anomalous thermalization of various types within high energy states. We consider three classes of non-integrable frustrated spin models: (I) systems with local conserved quantities where the number of symmetry sectors grows exponentially with the system size but more slowly than the Hilbert space dimension, (II) systems with exact eigenstates that are singlet coverings, and (III) flat band systems hosting magnon crystals. We argue that several 1D and 2D models from class (I) exhibit disorder-free localization in high energy states so that information propagation is dynamically inhibited on length scales greater than a few lattice spacings. We further show that models of class (II) and (III) exhibit quantum many-body scars --- eigenstates of non-integrable Hamiltonians with finite energy density and anomalously low entanglement entropy. Our results demonstrate that magnetic frustration supplies a means to systematically construct classes of non-integrable models exhibiting anomalous thermalization in mid-spectrum states. 
\end{abstract}

\date{\today}

\maketitle

\section{Introduction}
\label{sec:Introduction}

There is strong evidence that most eigenstates of non-integrable many-body Hamiltonians, if sufficiently far from the spectral edges, are ``thermal" in the sense that expectation values of local observables on such eigenstates match well to the predictions of statistical mechanics \cite{DAlessio_review2016}. This observation is formalized in the Eigenstate Thermalization Hypothesis (ETH)  \cite{PhysRevA.43.2046,PhysRevE.50.888, Rigol_Nature2008, Reimann_NJP2015, DAlessio_review2016, Deutsch_RepProgPhys2018}, and is tied to the success of random matrix theory in describing some properties of the many-body spectrum, such as level repulsion.  The complete breakdown of thermalization occurs only in extreme instances. One widely known example of anomalous thermalization is in integrable quantum systems where there is no level repulsion between eigenvalues and the long time averages of local observables approach a distribution that is tethered to the presence of an extensive number of conserved quantities \cite{vidmar2016generalized}. Another well-known example is the many-body localized (MBL) phase in interacting disordered systems in which high energy states have area law entanglement and in which an extensive number of local integrals of the motion are emergent \cite{nandkishore2015many}.
In both the MBL phase and in integrable systems, the majority of eigenstates depart very much from random states compared to those of generic non-integrable models. 

In this paper, we discuss two other types of anomalous thermalization: {\it disorder-free localization} and {\it many-body quantum scars}.  Exploiting insights from the field of frustrated quantum magnetism, we show how to design classes of many-body systems that display physics of one of these types. 

Disorder-free localization is a variant of many-body localization in translationally invariant systems \cite{PhysRevLett.118.266601, Yao_Lauman_Cirac_Lukin_Moore_PRL2016, Silva_Mueller_PRB2015, Mondaini_Cai_PRB2017, PhysRevLett.120.030601, PhysRevLett.121.040603, vanNieuwenburg9269,PhysRevX.10.011047, moudgalya2019thermalization, PhysRevLett.124.207602, PhysRevB.99.180302,Kuno_2020,2020arXiv200304901K, Magnifico_Ercolessi_Quantum2020_Schwinger, Schulz_Hooley_Moessner_Pollmann_PRL2019, Ribeiro_Lazarides_Haque_PRL2020, Kuno_Orito_Ichinose_NJP2020_Creuzladder, danieli2020manybody}.  In this phenomenon, information propagation is inhibited by the emergence of a localization length.  In some cases the localization originates from the single-particle eigenstates being localized, e.g. due to a Stark field \cite{Schulz_Hooley_Moessner_Pollmann_PRL2019, Ribeiro_Lazarides_Haque_PRL2020} or due to a flat band \cite{Kuno_Orito_Ichinose_NJP2020_Creuzladder, danieli2020manybody}.  More intricate mechanisms have also been uncovered, for example, Ref.~\onlinecite{PhysRevLett.118.266601} introduces a spin chain coupled to complex fermions with an extensive number of conserved quantities that maps to free fermions in a disorder potential generated by the different configurations of the symmetry sectors so that each sector is Anderson localized.  (In discussing disorder-free localization, we are interested in situations where typical initial states show signatures of localization, in contrast to the freezing of particular initial states such as single-domain-wall states \cite{LeaSantos_JMP2009_transport, Haque_PRA2010_locking, Choudhury_Kim_Zhou_arXiv2018}, which can result from spectral degeneracies.)

In the case of many-body quantum scars, an otherwise apparently unexceptional spectrum of eigenstates is peppered with highly athermal states. 
Such states were found to occur in the PXP chain, a kinetically constrained model of spins one-half \cite{2018NatPh..14..745T, PhysRevB.98.155134, PhysRevB.69.075106, Lesanovsky_Katsura_PRA2012}.  The PXP model is well-realized experimentally with Rydberg atoms \cite{bernien2017probing}. 
These athermal eigenstates are called many-body quantum scars after their non-ergodic counterparts in single particle semi-classical chaos that trace out periodic trajectories in phase space but are perturbatively connected to chaotic states \cite{PhysRevLett.53.1515}.  
Many-body quantum scars are characterized by their anomalously low entanglement and through local observables that strongly depart from random matrix predictions. The dynamics of states prepared with significant overlap with scar eigenstates is also anomalous, involving large amplitude oscillations in the entanglement entropy and in local correlation functions \cite{2018NatPh..14..745T,PhysRevB.98.155134,PhysRevLett.122.220603}.  The reason for this non-thermalizing dynamics is that, for such initial states, the evolution can be thought of  as taking place predominantly within the subspace spanned by the scar states. 
In addition to the PXP chain \cite{2018NatPh..14..745T,PhysRevB.98.155134,PhysRevLett.122.220603,PhysRevLett.122.040603,Khemani_Lauman_Chandran_PRB2019, Lin_Motrunich_PRL2019_exactscars, Iadecola_Schecter_Xu_PRB2019, shiraishi2019connection, Bernevig_Regnault_arXiv1906_thintorus, Lin_Chandran_Motrunich_PRResearch2020, Bull_Desaules_Papic_PRB2020, Papic_Abanin_PRX2020_slow, turner2020correspondence}, a number of other systems have been found to exhibit quantum many-body scar states, including the AKLT chain \cite{PhysRevB.98.235156,PhysRevB.98.235155, shiraishi2019connection, Mark_Lin_Motrunich_PRB2020, Moudgalya_Bernevig_Fendley_Regnault_2020}, the 1D transverse field Ising model with longitudinal field \cite{PhysRevLett.122.130603,PhysRevB.99.195108}, quantum Hall systems in the thin torus limit \cite{Bernevig_Regnault_arXiv1906_thintorus, Nachtergaele_Warzel_arxiv2020_thintorus}, the fermionic Hubbard model \cite{2017ScPP....3...43V,PhysRevLett.123.036403, Mark_Motrunich_2020_etapairing, Moudgalya_Regnault_Bernevig_2020_etapairing}, the spin-1 $XY$ model \cite{Schecter_Iadecola_PRL2019_xy, Pichler_Lukin_Ho_PRB2020_xy,  Mark_Lin_Motrunich_PRB2020},  periodically driven matter \cite{PhysRevB.101.245107,sugiura2019manybody,Pai_Pretko_PRL2019,  mukherjee2020dynamics,Mintert_Knolle_PRL2020, mizuta2020exact},  topologically ordered systems including fracton models \cite{PhysRevResearch.1.033144,PhysRevB.101.174204, Pai_Pretko_PRL2019, shiraishi2019connection}, 2D Rydberg lattices \cite{Papic_Abanin_Serbyn_PRResearch2020, Lin_Calvera_Hsieh_PRB2020}, among other examples \cite{Hudomal_2020,  Schecter_Iadecola_PRL2019_xy,
PhysRevB.101.241111,PhysRevLett.123.030601,PhysRevLett.119.030601,PhysRevX.10.011047,PhysRevLett.124.180604,hart2020random,dooley2020enhancing, Gambassi_Dalmonte_PRX2020, Schoutens_PRB2020_scars, Moudgalya_Bernevig_Fendley_Regnault_2020}.

Geometrical frustration is well-known to lead to many interesting and exotic phenomena, including flat bands, quantum and classical spin liquids and fractionalization \cite{lacroix2011introduction,Starykh_2015,SavaryBalents,Derzhko_2015}. In this paper, we describe how geometrical frustration supplies a mechanism to construct models with anomalous thermalization including both disorder-free localization and many-body scar states. These spin models, as explained in Section~\ref{sec:models}, have antiferromagnetic couplings on lattices of triangular units, which serve as the basic units underlying frustrated magnetism. We introduce three classes of models: (I) non-integrable models with local conservation laws, (II) models with protected singlet coverings  that can be tuned through the spectrum, and (III) flat band models hosting localized magnon states and magnon crystals. 

Models from class (I) are intermediate between non-integrable models that typically have $O(1)$ conservation laws and integrable models in which the number of conserved quantities equals the number of local degrees of freedom so that all states are specified by a quantum number associated to the conserved quantities.   Class (II) contains, among other examples, the Shastry-Sutherland model \cite{shastry1981exact, PhysRevLett.82.3168,albrecht1996first,PhysRevB.72.104425, PhysRevResearch.1.033038,mcclarty2017topological}, which is a foundational model of frustrated magnetism and is  realized to a good approximation in SrCu$_2($BO$_3)_2$.   Some  other models discussed in this paper are also realized in magnetic materials.

In Section~\ref{sec:LIOM}, we discuss the thermalization properties of typical eigenstates in models from class (I), and demonstrate disorder-free localization emerging in one of these models. Then, in Section~\ref{sec:scars2}, we give various examples of models in 1D and 2D exhibiting many-body scar states from class (II) and in Section~\ref{sec:scars3} an example of a model from class (III).  The mechanism that gives low-entanglement scar states for class (II) also gives the exact ground state of the Shastry-Sutherland model for a range of parameter values.  The scars presented for these models can be tuned parametrically relative to the many-body spectrum.  All the quantum scars we present are ``true'' scars in the sense that they are not distinguished by symmetry compared to the surrounding eigenstates, i.e., they are not the extreme eigenvalues (or isolated eigenvalues) in separate symmetry sectors \cite{footnote_symmetrysector_etapairing}.  

Section~\ref{sec:conclusions} provides a summary and some context.

\section{Models and Mechanism}
\label{sec:models}

We now introduce the mechanism that we exploit to write down models exhibiting anomalous mid-spectrum states.  This mechanism is based on the simplest frustrated unit --- three spins coupled by antiferromagnetic Heisenberg exchange.  We will then discuss separately three separate classes of magnetic systems combining such frustrated units.

Consider the Heisenberg model with antiferromagnetic couplings on a triangle of spins one-half with one distinguished bond. The Hamiltonian is \begin{equation}
H_{\Delta}=J \boldsymbol{S}_1\cdot \boldsymbol{S}_2 + J' \boldsymbol{S}_1\cdot \boldsymbol{S}_3 + J' \boldsymbol{S}_2\cdot \boldsymbol{S}_3
\label{eqn:triang}
\end{equation}
with $J,J'>0$.  We refer to the $(\boldsymbol{S}_1,\boldsymbol{S}_2)$ bond as the distinguished bond, the $J$ bond, or the dimer.  For this geometrically frustrated triangular unit, the total spin $(\boldsymbol{S}_1 + \boldsymbol{S}_2)^2$ is a conserved quantity. It follows that the singlet state on the distinguished bond, $\vert 0 \rangle \equiv \frac{1}{\sqrt{2}}(\lvert\uparrow\downarrow\rangle -\lvert\downarrow\uparrow\rangle)$, is protected --- eigenstates will have well-defined total spin on the distinguished bond.  In a loose sense, this feature arises from destructive interference on the two identical $J'$ bonds and so it is destroyed if those bonds are made inequivalent. 

To analyze further the spectrum of the triangular plaquette, we  introduce projectors $P_{S=0}(\boldsymbol{S}_i, \boldsymbol{S}_j)$ and $P_{S=1}(\boldsymbol{S}_i, \boldsymbol{S}_j)$, the total spin $0$ and $1$ projectors for pairs of spins ($i$ and $j$).   We also introduce
\begin{equation*}
P_{S=3/2}(\boldsymbol{S}_i, \boldsymbol{S}_j,\boldsymbol{S}_k) \equiv \frac{1}{3}\left( \boldsymbol{S}_i + \boldsymbol{S}_j + \boldsymbol{S}_k \right)^2 - \frac{1}{4} ,
\end{equation*}
the projector onto the total spin $3/2$ sector of three spins. Then the Hamiltonian can be rewritten as 
\begin{align}
H_{\Delta} & = \frac{3}{2}J' P_{S=3/2}(\boldsymbol{S}_1, \boldsymbol{S}_2,\boldsymbol{S}_3) - \frac{3J'}{4} \nonumber \\ & + (J-J') \left( -\frac{3}{4} P_{S=0}(\boldsymbol{S}_1, \boldsymbol{S}_2) + \frac{1}{4}P_{S=1}(\boldsymbol{S}_1, \boldsymbol{S}_2) \right) .
\label{eq:proj}
\end{align}
The projectors mutually commute.  So, for typical couplings, the spectrum splits up into total spin $3/2$ and $1/2$ sectors, as well as singlet and triplet sectors on the distinguished ($J$) bond. This means that there is a four-fold degenerate spin $3/2$ level.  These four states each has a triplet on the $J$ bond.  There are also two doublets corresponding to total spin $1/2$.  The $J$ bond is a singlet in one of these degenerate pairs and is a triplet in the other pair.  At the fully frustrated point $J=J'$, the last term in Eq.\ \eqref{eq:proj} vanishes, so the two doublets merge --- there is a level crossing at $J'/J=1$.  The singlet is the ground state for $J'/J < 1$.

While we focus in this work on spin-$1/2$ systems,  the existence of a conserved total spin on the distinguished bond generalizes to any spin, $S$: the singlet state of two spins $S$ on the $J$ bond is an exact eigenstate with energy $-JS(S+1)$.    Diagonalization of the spin $S$ Hamiltonian reveals that new protected states (with $J'$-independent energy) can arise for $S\geq 3/2$. 

The triangular plaquette with Heisenberg exchange and one distinguished $J$ bond provides the basic unit to create lattice models with disorder-free localization and many-body scars. We distinguish three different cases. 

\begin{figure*}[!htbp]
\centering
\includegraphics[width=0.95\textwidth]{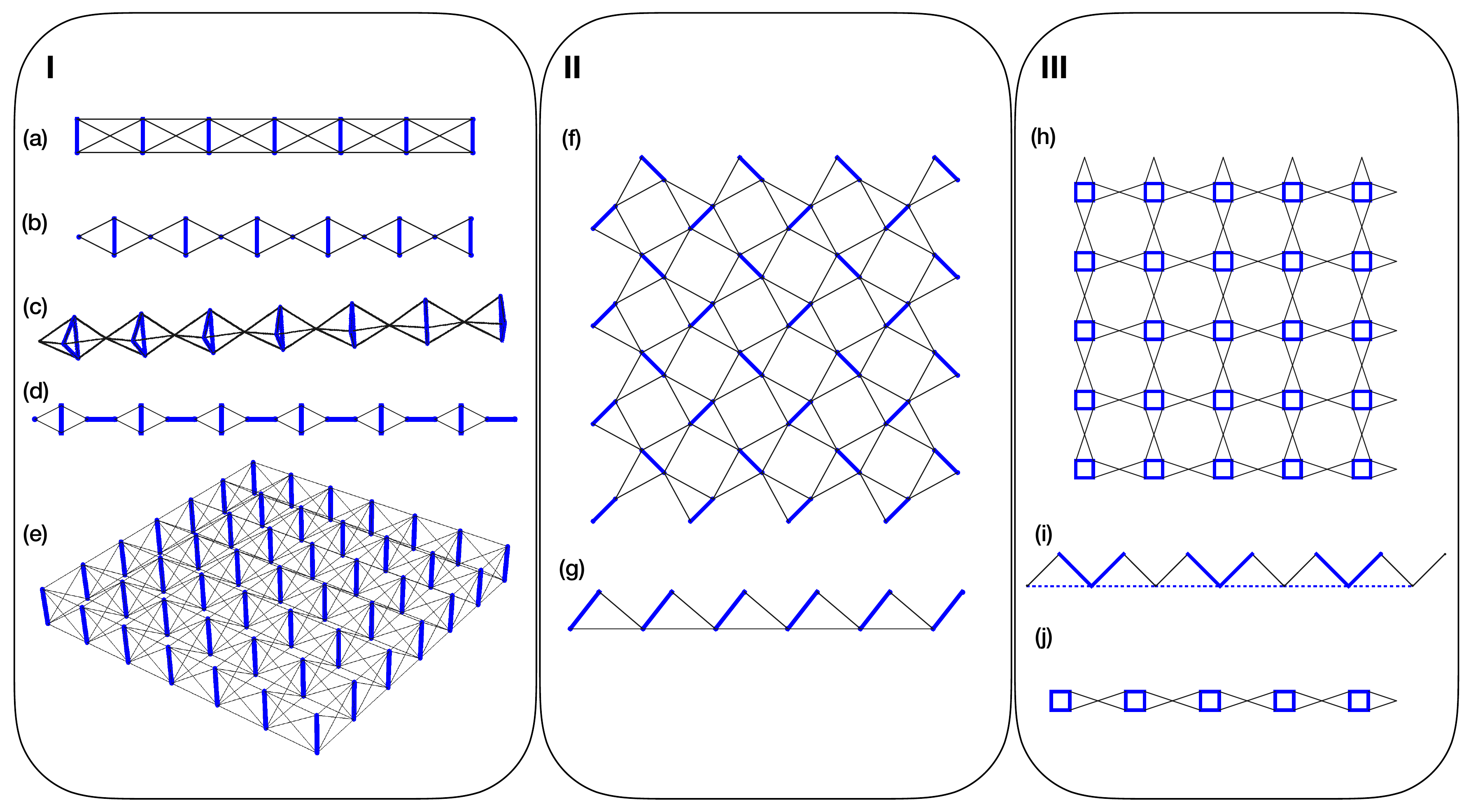}
\caption{Collection of lattices exhibiting athermal states in short-ranged interacting Heisenberg models with $J$ couplings marked in thick blue and $J'$ on the remaining bonds. Class (I): models on lattices (a)-(e) have an extensive number of local integrals of the motion.  (a) Fully-frustrated ladder, (b) diamond chain, (c) pyrochlore chain, (d) orthogonal dimer chain, (e) fully-frustrated bilayer. Class (II) with protected singlet states: (f) Shastry-Sutherland lattice, (g) sawtooth chain with $J$ bond on teeth. Class (III) with localized magnon states: (h) square kagome lattice (i) sawtooth chain with J' bonds on the spine (dotted), J bonds on every second valley which support the localized magnons (thick blue line) and J''=2J' on odd valleys (black solid line) and (j) the bowtie chain.
\label{fig:dimer_lattices}
}
\end{figure*}

{\bf Class (I):} In general, $P_{S=3/2}$ operators on adjacent triangular units do not commute with one another. However, there are various ways to combine the triangles such that the total spin conservation on $J$ bonds is preserved.   For example, this is achieved by connecting triangular units back-to-back and then connecting these four spin structures via the dangling spins.  Examples include the diamond chain in Fig.~\ref{fig:dimer_lattices}(b) and the orthogonal dimer chain in Fig.~\ref{fig:dimer_lattices}(d) \cite{richter1998antiferromagnetic,PhysRevB.62.5558,PhysRevB.65.054420}.  These models have spin conservation on the vertical bonds.  So does the fully frustrated ladder in Figs.~\ref{fig:dimer_lattices}(a)  and the bilayer in Fig.~\ref{fig:dimer_lattices}(e)). Ref.~\onlinecite{Tanaka_2014} argues that the latter model with XXZ couplings is realized in a particular material to a good approximation. In this class of lattices, the frustration mechanism is responsible for an extensive number of conservation laws that is, however, smaller than the number of degrees of freedom. For example, in the orthogonal dimer chain there is one conserved quantity per unit cell of four spins. Such models are intermediate between integrable models --- in which the number of local conserved quantities equals the number of degrees of freedom --- and generic non-integrable systems --- which have $O(1)$ conserved quantities.  We will address the question of whether typical states in class (I) models thermalize.

It is also possible to generalize the frustration mechanism that generates local conservation laws (class (I)) from dimers to trimers, quadrumers and so on. For example, a Heisenberg coupled triangle with all exchange couplings equal to $J$ has a singlet eigenstate when each spin is an integer. If we couple this triangle to one other spin through $J'$ exchange, the singlet remains an exact eigenstate and one can build chains of such units such as the pyrochlore chain Fig.~\ref{fig:dimer_lattices}(c) which belongs to class (I). To generalize this to more spatially extended singlets \cite{PhysRevB.95.224415}, we simply require that the polygonal unit admits a singlet state.  Thus, if the polygon has an odd number of vertices the individual spins have to be integer-valued; there is no such constraint for even numbers of vertices.

{\bf Class (II):} If we relax the constraint that the total spin on each $J$ bond be conserved, we can nevertheless devise lattices that retain the $J$ bond singlet covering as an exact eigenstate. To see how this can be done we take the example of the sawtooth chain \cite{PhysRevB.53.6401,PhysRevB.67.054412}, shown in  Fig.~\ref{fig:dimer_lattices}(g), with Hamiltonian
\begin{equation}
H_{\rm ST}= \sum_i  \left( J \boldsymbol{S}_{i,1}\cdot \boldsymbol{S}_{i,2} + J' \boldsymbol{S}_{i,1}\cdot \boldsymbol{S}_{i+1,1} + J' \boldsymbol{S}_{i,2}\cdot \boldsymbol{S}_{i+1,1} \right).
\label{eqn:sawtooth}
\end{equation}
We write this in terms of projectors as indicated in Eq.~\ref{eq:proj} and note that a state that satisfies $P_{S=3/2}(\boldsymbol{S}_{i,1}, \boldsymbol{S}_{i,2},\boldsymbol{S}_{i+1,1})=0$ and $P_{S=1}(\boldsymbol{S}_{i,1}, \boldsymbol{S}_{i,2})=0$ is an exact eigenstate with energy $-3J' N_{\Delta}/4$,  where $N_{\Delta}$ is the number of triangles. These conditions constrain each triangle to have total spin $S=1/2$ while the $J$ bonds have $S=0$. The product state with singlets on the $J$ bonds has these properties and is the unique state that does. These are many-body quantum scars because dimer coverings are highly atypical states (often with area law entanglement entropy) that can be embedded within the many-body spectrum of a translationally invariant model with no local conservation laws. 

This reasoning is reminiscent of the embedding argument of Shiraishi and Mori \cite{PhysRevLett.119.030601} that gives a systematic way to place athermal states into the spectrum of a many-body Hamiltonian. We briefly review this result. We introduce a set of local projection operators $P_\alpha$ that need not commute. The scar states are those that are annihilated by all the projectors $P_\alpha\vert \Psi\rangle=0$. There is a class of Hamiltonians with such states as eigenstates:
\begin{equation}
H = \sum_{\alpha} P_\alpha \hat{h}_\alpha P_\alpha + H'
\label{eq:Hscar}
\end{equation}
where $[H',P_\alpha] =0$ and $h_{\alpha}$ is an arbitrary local operator. There are many-body scars in the concrete sense described above because
\[
P_\alpha H\vert \Psi\rangle = P_\alpha H' \vert \Psi\rangle = H' P_\alpha  \vert \Psi\rangle = 0.
\]
The example of the sawtooth chain (Fig.~\ref{fig:dimer_lattices}(g)) is a special case of this kind of mechanism where the Hamiltonian is merely a sum of projectors with the remarkable feature that the conditions $P_\alpha\vert \Psi\rangle=0$ are solved by a dimer covering. 

It is evident from the foregoing that the dimer covering eigenstate appears in certain lattices composed of triangular units.  There is a large class of such lattices.  Apart from the sawtooth lattice, we show that the Shastry-Sutherland lattice (Fig.~\ref{fig:dimer_lattices}(f)) and the maple leaf lattice (Fig.~\ref{fig:mllshsu} (left)) exhibit similar physics. 

In addition to $J$-$J'$ Heisenberg models we consider the counterpart XXZ models by including the perturbation
\begin{equation}
H'_{\lambda} = \lambda \sum_i  \left( J S^z_{i,1} S^z_{i,2} + J' S^z_{i,1}S^z_{i+1,1} + J' S^z_{i,2}S^z_{i+1,1} \right).
\label{eqn:sawtoothxxz}
\end{equation}
This perturbation commutes with the projection operators and is therefore equivalent to switching on $H'$ in Eq.~\ref{eq:Hscar}.  The physics we have presented above is thus preserved.  When using an XXZ anisotropy, the total spin is no longer a conserved quantum number.  Hence for nonzero $\lambda$, the spectrum is not separated into sectors corresponding to different values of the total spin.  This is convenient, e.g., when  calculating level statistics.  

{\bf Class (III):} A third class of interesting frustrated models deriving from the $H_{\Delta}$ model on a triangular plaquette is the famous class of models with a flat band of one-magnon states leading to localized multi-magnon states. For a recent review see Ref.~\cite{Derzhko_2015}. An example with localized magnons is the 
Heisenberg J-J'-J'' model on a sawtooth chain (Fig.~\ref{fig:dimer_lattices}(i)) that differs from the case discussed above with J bonds on the left or right jagged edges of the chain \cite{PhysRevLett.88.167207,PhysRevB.70.100403}. Suppose $J=J''=2$ and $J'=1$ starting from the product state $\vert \uparrow\ldots \uparrow\rangle$. Now apply operator $\Sigma^-_i \equiv (S_{i-1}^- - 2S_{i}^- + S_{i+1}^-)$ to state $\vert \uparrow_{i-1}\uparrow_{i}\uparrow_{i+1}\rangle$ where $i$ is a site at the base of one of the ``valleys" on the sawtooth chain. This is a single localized magnon state. It turns out that this model has a pair of exact many-body eigenstates formed by applying $\Sigma^-_i$ to every even, or odd, valley along the chain. This lives within the sector with half of the saturation magnetization. This model has the undesirable feature that the exact localized magnon state is highly fine-tuned $-$ a small change of the $J'/J$ coupling destroys the magnon localization whereas in classes (I) and (II) the protected states are robust to changes in the ratio $J'/J$. Moreover, the localized magnon states live in the ground state of a given symmetry sector. However, there are models where the magnon crystal states are robust to changes in the couplings with three examples given in Fig.~\ref{fig:dimer_lattices}: the square kagome lattice, a variant of the sawtooth model just described, and the bowtie chain.  Later in this paper (Section \ref{sec:scars3}), we study the square kagome example in some detail showing that multiple scar states arise in this model that can be tuned through the spectrum and separated in energy.

\section{Thermalization Dynamics in Frustrated Models with Local Conservation Laws}
\label{sec:LIOM}

In this section, we consider models from class (I) focussing on two examples: the orthogonal dimer chain (Fig.~\ref{fig:dimer_lattices}(d)) and the fully frustrated ladder (Fig.~\ref{fig:dimer_lattices}(a)). We show that the distribution of the entanglement of mid-spectrum eigenstates has a large variance in contrast to usual non-integrable models. Also, both models violate the usual ETH scaling of eigenstate matrix element distributions. These results show that mid-spectrum states of the models are highly unusual although both are non-integrable. We go further and argue that, in fact, these models exhibit a variant of many-body localization albeit in the absence of quenched disorder. We explain that the local conserved quantities fragment the eigenstates in real space leading to a localization length of the order of a few lattice spacings and demonstrate that the dynamics of the fully frustrated ladder is consistent with the picture of disorder-free localization. At the end of the section, we provide concrete examples of two-dimensional translationally invariant spin models that, through a mapping to a percolation picture, can be argued to exhibit similar phenomena. 

\begin{figure}[tbp]
\centering
\includegraphics[width=\columnwidth]{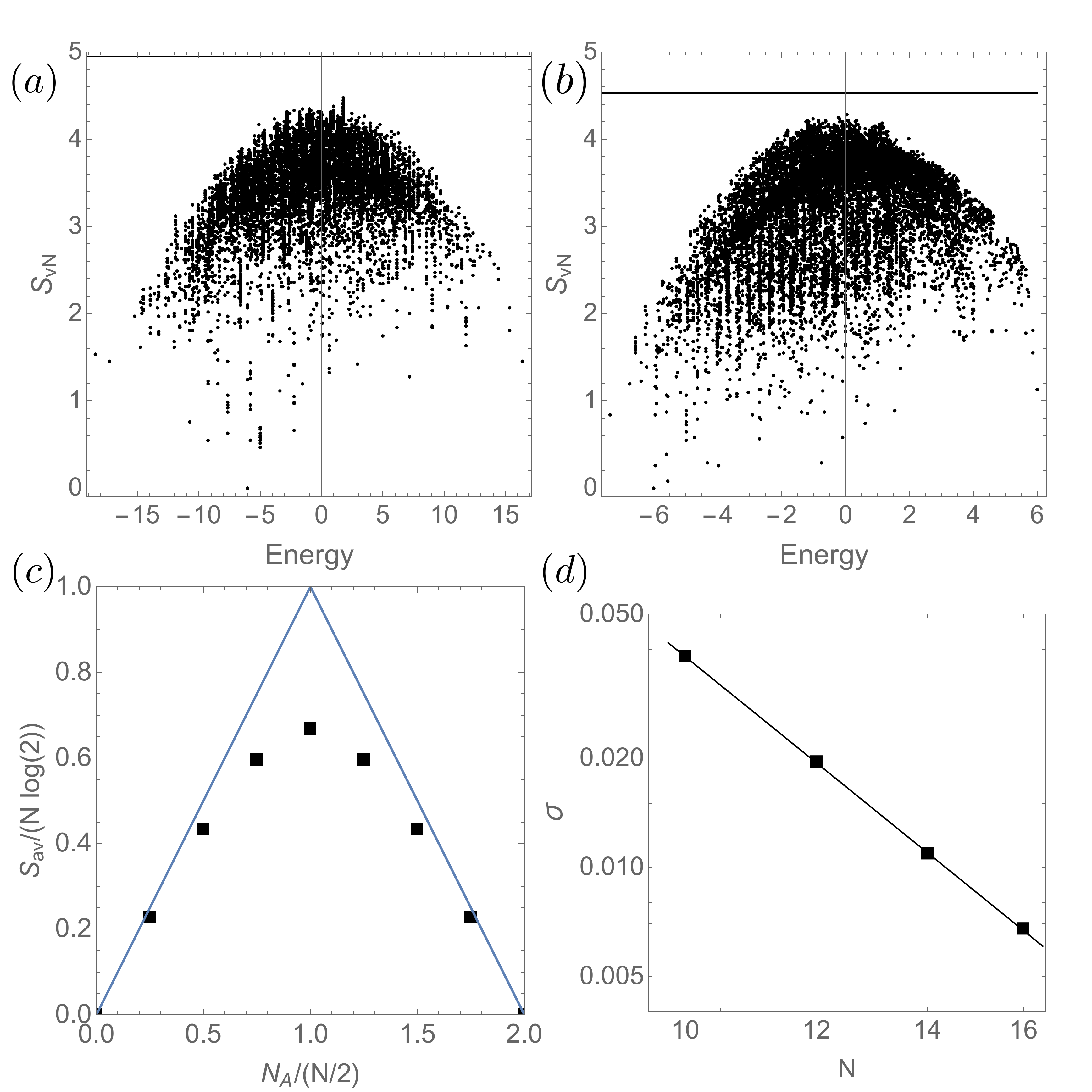}
\caption{Entanglement of eigenstates within total $S^z=0$ sector, for two members of class (I).  (a) Fully frustrated ladder, $J=1$, $J'=2$, $\lambda=0.5$, $N=16$;  an $8,8$ site bipartition. (b) Orthogonal dimer chain, $J=J'=1$, $N=16$; a $9,7$ bipartition.  The average entanglement of a random state with same Hilbert space size and same bipartitions is shown in each case as a horizontal line.  Both plots show an anomalously broad distribution due to a large number of a conserved quantities. 
(c,d) are for the fully frustrated ladder.  
(c) Ratio of the average entanglement in the middle of the spectrum to the maximal entanglement as a function of the subsystem size $N_A$.  Blue lines: envelope of maximal entanglement.  
(d) Scaling of the width $\sigma$ of the distribution of off-diagonal matrix elements of a local operator, consistent with a power law $\sigma \sim N^{-\alpha}$ and $\alpha\approx3.7$. 
\label{fig:liom}
}
\end{figure}

As a concrete example of a model from class (I), we consider the orthogonal dimer chain \cite{richter1998antiferromagnetic,PhysRevB.62.5558,PhysRevB.65.054420} shown in Fig.~\ref{fig:dimer_lattices}(d). In common with other models in this class, this model has total spin conserved on each bond with $J$ exchange. The chain has four sites per unit cell and hence, for spin one-half, $2^{4C}=16^C$ states where $C$ is the number of unit cells. The number of symmetry sectors also grows exponentially in the system size but with a smaller exponent: as $2^C$. This model is distinct from integrable models in which the number of symmetry sectors equals the number of states. For example, in free fermion models each state belongs to a unique quasiparticle number sector while, for the Heisenberg chain, each state has a unique set of Bethe quantum numbers. The orthogonal dimer chain model is also distinct from typical non-integrable models in which the number of conservation law is constant and of order one. A second example of this type of model is the fully frustrated ladder (Fig.~\ref{fig:dimer_lattices}(a)) \cite{PhysRevB.43.8644,honecker2000magnetization,PhysRevB.82.214412} which has two sublattices per unit cell and hence $4^C$ states and $2^C$ symmetry sectors. In all such examples, the size of the subspace within a typical sector grows exponentially with the system size. An obvious question is the extent to which the thermalization properties of this class of models emulates that of well-known integrable and non-integrable models. To start addressing this question, we consider the entanglement of the eigenstates, measured using the von Neumann entropy $S_{\rm vN}=-{\rm Tr}\ \rho_{\rm A}\log \rho_{\rm A}$, where $\rho_{\rm A}$ is the reduced density matrix for subsystem $A$.  The entanglement entropy for a cut through $J'$ bonds on a $16$ site fully frustrated ladder with blocks of $8$ spins on each subsystem is shown in Fig.~\ref{fig:liom}(a). Of the $12870$ states, there is a single one with zero entanglement on this cut. However, the more striking observation is that there is a very large number of low entanglement states similar to analogous results for integrable models \cite{Alba_2009, Beugeling_2015, PhysRevLett.119.020601, LeBlond_Mallaya_Vidmar_Rigol_PRE2019} and the entanglement generally falls significantly below that expected for completely random states (Fig.~\ref{fig:liom}(c)).  The entanglement for the orthogonal dimer chain is similar and shown in Fig.~\ref{fig:liom}(b).

We now investigate whether the eigenstate thermalization hypothesis is obeyed by the eigenstates of models within class (I). A natural expectation would be that thermalization takes place as for non-integrable models within each exponentially large symmetry sector. To add weight to this hypothesis, let us consider the fully frustrated ladder within the sector with all dimer bonds in the total $S=1$ sector. In this sector, the dimer bonds maps to composite spins one and the coupling between them is simply a Heisenberg coupling because $(\boldsymbol{S}_{i,1}+\boldsymbol{S}_{i,2})\cdot(\boldsymbol{S}_{i+1,1}+\boldsymbol{S}_{i+1,2})$ is just the set of $J'$ couplings between rungs of the fully frustrated ladder.  Thus, the ``all triplets'' sector is effectively a Heisenberg-coupled spin one chain (spin one Haldane chain), which is not integrable and hence is expected to obey  ETH.  Thus we have one example of an exponentially large sector that obeys ETH, in a model with an exponentially large number of symmetry sectors.  We can imagine preparing a state in a random state within the sector with all rungs having $S=1$ and with some energy density --- we should find that observables at long times can be described by a statistical ensemble average of eigenstates within this sector at some fixed temperature set by the initial energy density.

The sector with $S=1$ on all rungs has exponentially small weight in the whole Hilbert space.  We must consider all other sectors if we are to understand the gross thermalization properties of the model. The composite spin picture described above sheds light on all the remaining sectors. Each sector has well-defined $S=0$ or $S=1$ on each rung of the ladder. We know that consecutive $S=1$ rungs map to Haldane chains. The presence of $S=0$ rungs has the effect of completely decoupling neighboring Haldane chain fragments. It follows that a state prepared in a given sector cannot completely thermalize because entanglement cannot spread beyond $S=0$ rungs. In other words, there is dynamical localization in each sector. 

Since we are interested in the thermalization of typical states it is necessary to address how the amplitude in such a state is distributed among the configurations with different chain lengths. The distribution of chain lengths must be calculated by weighting the configurations by the dimension of their Hilbert space. This distribution $P(\ell)$ is equivalent to that of the distribution of success run lengths, $\ell$, in Bernoulli trials with a weighted coin producing heads with probability $p=3/4$ and tails with probability $q=1/4$. The distribution of $l$ consecutive $S=1$ rungs is evidently $p^\ell = \exp(-a \ell)$ with $a\approx 0.28$ so short fragments are overwhelmingly important among the set of all symmetry sectors. Indeed, the mean $S=1$ chain length is about $4$. We conclude that typical states $-$ those that can be decomposed into a linear combination of symmetry sectors of roughly equal weight $-$ must be localized apart from the exponentially small tail that lives in the sector with all rungs $S=1$. This is an example of so-called disorder-free localization as the model is translationally invariant. Similarly to the case of MBL high energy states in the spectrum are dynamically localized. However, unlike the MBL phase, the fully frustrated ladder is fine-tuned and the anomalous thermalization properties we have described cannot survive sufficiently large generic perturbations. This localization mechanism bears some resemblance to the Hilbert space fragmentation picture of several recent works \cite{PhysRevX.10.011047,moudgalya2019thermalization,PhysRevLett.124.207602} albeit in a rather different setting.

We now turn to the question of whether signs of this physics can be observed numerically.  We focus now on the fully frustrated ladder, because it has only $2$ sites per unit cell and so we are able to study a wider range of system sizes than in the other models described above.  We first address whether eigenstates of the model obey ETH, meaning that we consider some local operator $\hat{O}$ and compute its eigenstate matrix elements. If ETH is satisfied, as is generally the case in non-integrable models, then \cite{srednicki1996thermal, srednicki1999approach}
\begin{equation}
\langle E_A \vert \hat{O} \vert E_B  \rangle = \delta_{AB} f_{O}^{(1)}(\bar{E}) + e^{-S(\bar{E})/2}f_{O}^{(2)}(\bar{E},\omega) R_{AB}
\end{equation}
where $S\sim \log \mathcal{D}$ is the entropy and $\mathcal{D}$ is the Hilbert space dimension, $\vert E_A  \rangle$ is an energy eigenstate with eigen-energy $E_A$,  $\bar{E}=(E_A + E_B)/2$, and $\omega=E_B - E_A$.  The $f_{O}^{(1/2)}$ are smooth functions, and  $R_{AB}$ is a (pseudo) random variable with zero mean and unit variance.  
A crucial aspect of ETH is the scaling of the width of the distribution of either diagonal or off-diagonal matrix elements: the width  falls off as $e^{-S(\bar{\mathsf{E}})/2}\sim \mathcal{D}^{-1/2}$, i.e., exponentially with system size.   This scaling is
based on the similarity between typical many-body eigenstates and random states
\cite{Marquardt_PRE12, Beugeling_scaling_PRE14, Beugeling_offdiag_PRE2015}.  This  behavior contrasts sharply with integrable systems, which do not obey ETH scaling --- the width of diagonal matrix element distributions generally have power law decay with system size \cite{ziraldo2013relaxation, Beugeling_scaling_PRE14, Alba_PRB15,ArnabSenArnabDas_PRB16, HaqueMcClarty_SYKETH}, and the off-diagonal matrix element generally has a non-gaussian distribution \cite{Beugeling_offdiag_PRE2015, HaqueMcClarty_SYKETH, LeBlond_Mallaya_Vidmar_Rigol_PRE2019}. In the latter studies, one follows the spirit of corresponding investigations of non-integrable models by splitting the spectrum into {\it global} symmetry sectors instead of the local sectors characteristic of integrable systems. In this way, one finds large qualitative departures from standard ETH scaling. The question we address here is whether the same is true also for an instance of a class (I) model.

The local operator we consider is $\frac{1}{2}(S_{i}^{+}S_{j}^{-}+S_{i}^{-}S_{j}^{+})$ where $i$ and $j$ are taken to be sites on neighboring rungs of the ladder. The exchange is taken to be $J=1$, $J'=2$ and we break translational invariance by setting the exchange on bonds between two rungs to have $J=J'$. We have computed the distribution of off-diagonal matrix elements for different system sizes. The distribution is highly peaked at zero with long tails.  The width of the distribution narrows for larger system sizes consistent with power law scaling (Fig.~\ref{fig:liom}(d)) in contrast to the exponential scaling expected for typical non-integrable models.  The violation of ETH scaling by the fully frustrated ladder is consistent with the expectation of localization within mid-spectrum eigenstates. 

\begin{figure*}[tbp]
\centering
\includegraphics[width=0.9\textwidth]{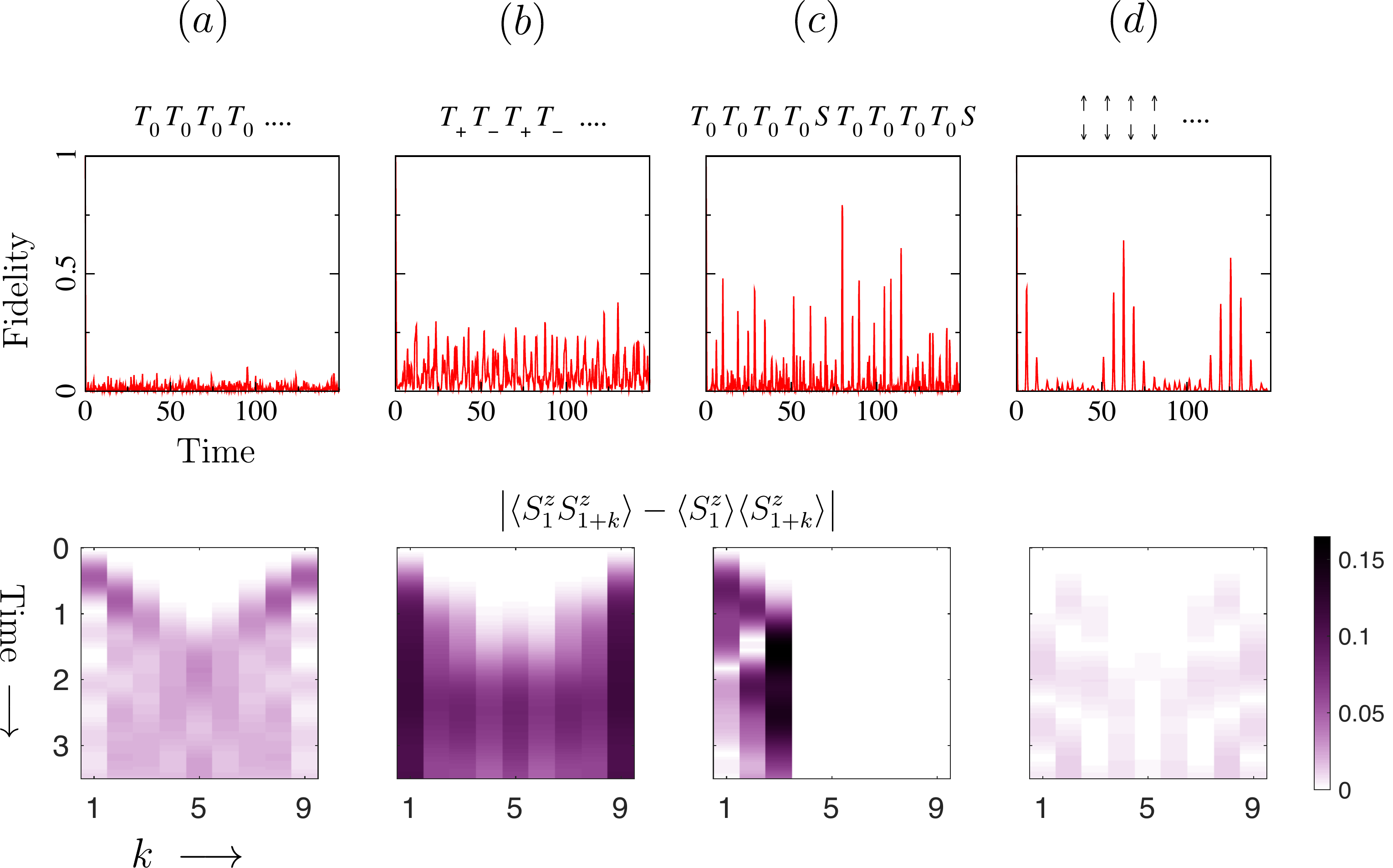}
\caption{Time evolution of  the fully frustrated ladder with $10$ rungs ($N=20$, $J=1$, $J'=1.1$), for four different initial states.  Each column corresponds to a different initial state, as indicated.  Upper row:  fidelity $\vert\langle \psi(0)\vert\psi(t)\rangle\vert^2$.  Lower row: space-time density plots of the absolute value of the connected correlation function, $\langle S_1^z(t) S_{1+k}^z(t)\rangle_c$, between a spin on rung $1$ and a spin on rung $1+k$.  Rung $1$ is connected to rungs $2$ and $10$, corresponding to $k=1$ and $k=9$, the rightmost and leftmost sites appearing in the density plots.  
 \label{fig:ffl_dynamics}
}
\end{figure*}

We now present features of the exact quantum dynamics for a periodic chain of $N=20$, or ten rungs, prepared in different initial states. The results are shown in Fig.~\ref{fig:ffl_dynamics}.  The different columns correspond to different initial states.    The four panels in the top row  show the return probability or fidelity, $F(t)\equiv \vert\langle \psi (0)\vert\psi (t)\rangle\vert^2$, where $\psi(t)$ is the state of the system at time $t$. In order to study the spreading of correlations, we also present the absolute value of the connected correlation function $\langle S_1^z(t) S_{1+k}^z(t)\rangle_c\equiv \vert \langle S_{1}^z(t)S_{1+k}^z(t)\rangle - \langle S^z_1(t)\rangle \langle S^z_{1+k}(t)\rangle \vert$ (bottom row).
The subscript here is the rung index: this quantity  measures correlations between a site of the rung labelled $1$ and a site on the rung $1+k$.  In each case, the site on the same leg of the rung is used.  We use $J=1$ (so that time is measured in units of $J^{-1}$) and $J'=1.1$.  For these parameters, the all-singlet state is not the ground state.

As discussed above, when all the rungs are in the local $S=1$ sector, there is a mapping to the Haldane chain which is non-integrable and should behave like a generic random matrix model in the middle of the spectrum.  Column (a) of Fig.~\ref{fig:ffl_dynamics} shows results for the initial state with all rungs in the total $S^z =0$ triplet state $\vert T_0\rangle = (1/\sqrt{2})(\vert\uparrow\downarrow\rangle + \vert \downarrow\uparrow\rangle)$. In this case, the fidelity drops rapidly from $F(0)=1$ with time on a timescale set by the exchange and fluctuates close to zero as expected for a thermalizing system that should retain little memory of its initial state. The correlator in the bottom row shows rapid spreading of correlations on a well-defined light cone centered on site $1$ and emanating in both directions on the periodic chain so that both paths of the light cone meet at site $k=5$ on a timescale of the order of the exchange. Some oscillations are visible in the correlation function at later times which are presumably a finite size effect. Similar behavior is observed for a second initial state in the sector with all rungs having $S=1$. In panel (b) we take the state with alternating rungs in the $S^z=1$ and $S^z=-1$ states, denoted as $\vert T_+\rangle =\vert\uparrow\uparrow\rangle$ and $\vert T_-\rangle =\vert\downarrow\downarrow\rangle$. Once again, the fidelity falls rapidly and correlations spread on a light cone --- in this case to a largely featureless time-independent state at longer times. We have confirmed that the amplitude of the fidelity fluctuations decays with increasing system size for cases presented in (a) and (b) (as can also be seen by comparing with the fidelity in column (c) as explained below).

Column (c) is for the initial state $T_0T_0T_0T_0ST_0T_0T_0T_0S$ where $S$ denotes a rung in the singlet state. According to our proposed scenario, the presence of the two rungs in singlet states effectively splits the chain into a pair of fragments each composed of four triplet rungs that themselves have nontrivial dynamics. The blocking of the spread of information by the singlet rungs is clearly shown in the lower panel --- correlations with rung $1$ are nonvanishing only with rungs at $k=1$, $2$ and $3$.  The fidelity drop and large amplitude fluctuations are as expected for a chain of an effective length of four rungs. 

As a final example, we consider a state that lives in a linear combination of different singlet and triplet sectors, as one expects for a generic state. Specifically we take the initial state to be a product state, with each rung in the state $\vert\uparrow\downarrow\rangle$.  In other words, the initial state has $S^z=0$ on each rung with equal weight in the singlet and triplet sectors so that many-body eigenstate has contributions in all possible singlet and triplet sectors. Our expectation based on the foregoing is that the small weight in the sector with all rungs in triplet states will be subject to thermalizing dynamics (as in cases (a) and (b)). The rest of the state will undergo some degree of dynamical localization because of the presence of amplitude in mixed triplet-singlet sectors.  The results are shown in column (d). Evidently the system does not straightforwardly thermalize. Instead, the fidelity shows clear periodic recurrences up to about $0.7$ suggesting a lack of complete thermalization. The correlation function does exhibit a feature similar to the light cone of columns (a) and (b). However, in this case, the feature is much less pronounced, the longer time correlations are much weaker than in those other cases. The expectation is that generic states for larger systems will have exponentially small weight in the all-triplet sector. As this is the only truly thermalizing part of the wavefunction, correlations will tend to be trapped within small regions. The numerical results thus confirm our proposed picture, up to usual finite-size limitations.

We now address the question of whether there are two-dimensional analogues of the physics described above. Fig.~\ref{fig:dimer_lattices}(f) is a two-dimensional lattice that has $J'$ bonds with the connectivity of the fully frustrated ladder. The construction, exemplified here by the square lattice, can be generalized to a bilayer of any lattice in two-dimensions. In such cases, there is local total spin conservation on each $J$ bond connecting the two layers. As we did for the chain, we now consider the different sectors on each $J$ bond. Within each sector there is  generally a set of conserved singlet and triplet clusters on the lattice. If we imagine preparing a quantum state within a given sector, one would find information propagation within each connected triplet cluster and that further propagation would be blocked by singlet $J$ bonds at the boundaries of each cluster. The question of whether dynamical localization takes place maps to a percolation problem.  Taking the whole Hilbert space of states, the probability that a $J$ bond is ``occupied" by a triplet is $3/4$ while singlet $J$ bonds are effectively absent. If the site percolation threshold $p^*$ is less than $3/4$, typical clusters percolate and information can propagate to infinity and the system will thermalize. If instead $p^*>3/4$, dynamically connected clusters have a characteristic length scale that is an effective localization length. 

\begin{figure}[ht]
\subfloat[Star Lattice]{\includegraphics[width=.8\linewidth]{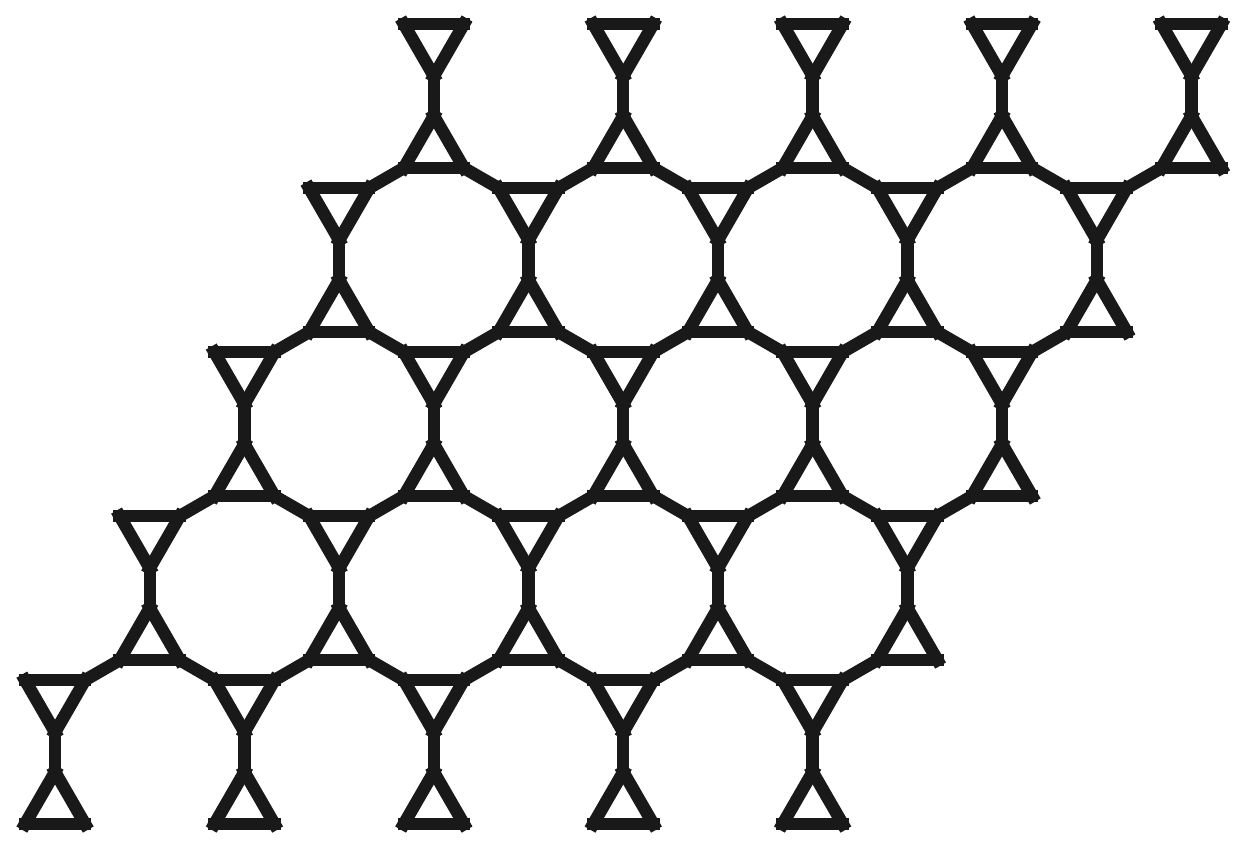}  \label{fig:star}} \\
\subfloat[Martini Lattice]{\includegraphics[width=.8\linewidth]{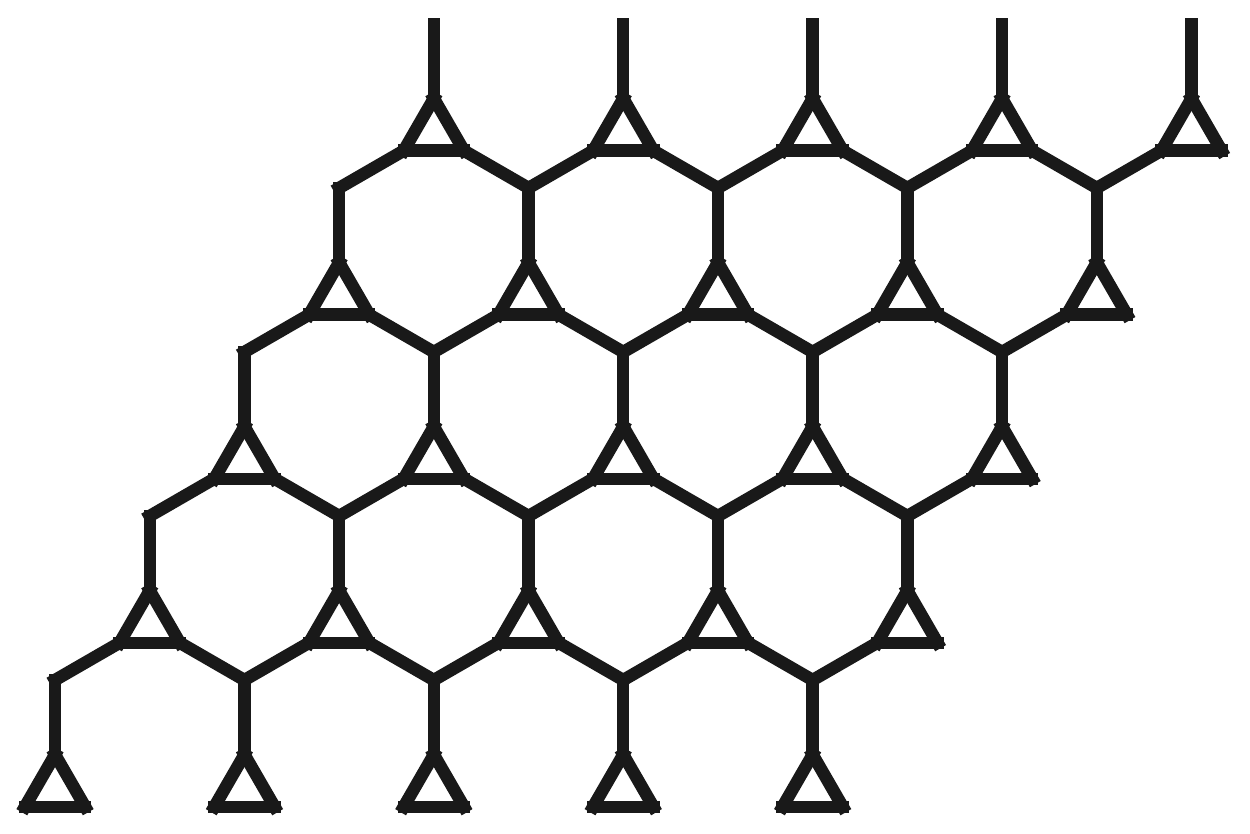}  \label{fig:martini}}
\caption{Panels showing view in projection of lattices argued to belong to class of Heisenberg models with disorder-free localization. The full model is a bilayer of each of the above lattices with ``fully frustrated" couplings between the two layers meaning that $J$ couples the layers site-for-site and $J'$ couples nearest neighbors within each layer and between layers. The ``fully frustrated" square lattice bilayer is shown in Fig.~\ref{fig:dimer_lattices} for reference. }
\label{fig:2D_DFL}
\end{figure}

Having argued that the problem of frustration-induced localization in 2D is reduced to a search for lattices with $p^* >3/4$, we first note that this condition is not satisfied by most lattices. For example, the square lattice bilayer of Fig.~\ref{fig:dimer_lattices}(f) has $p^* =0.5927$ \cite{RevModPhys.64.961}. However, there are lattices with low connectivity and large loops that do satisfy the condition. Examples include the so-called star lattice \cite{richter2004starlattice,PhysRevB.98.155108} with $p^*=0.807904$ \cite{PhysRevB.53.6401} (Fig.~\ref{fig:2D_DFL}(a)) and the martini lattice \cite{PhysRevE.85.062101} with $p^* = 0.764826$ \cite{PhysRevE.73.016107} (Fig.~\ref{fig:2D_DFL}(b)). The dynamics of typical states on such lattices is strictly speaking athermal though the associated length scale may be large and for practical purposes local observables may be close to their thermal values. The unit cell of the fully frustrated bilayer martini lattice has $8$ sites while the fully frustrated bilayer star lattice unit cell has $24$.  So, it would be challenging to numerically access the physics we have argued to exist in these models.  

\section{Examples of Many-Body Quantum Scars: Class (II)}
\label{sec:scars2}

\subsection{Sawtooth Chain}

\begin{figure}[tbp]
\centering
\includegraphics[width=\columnwidth]{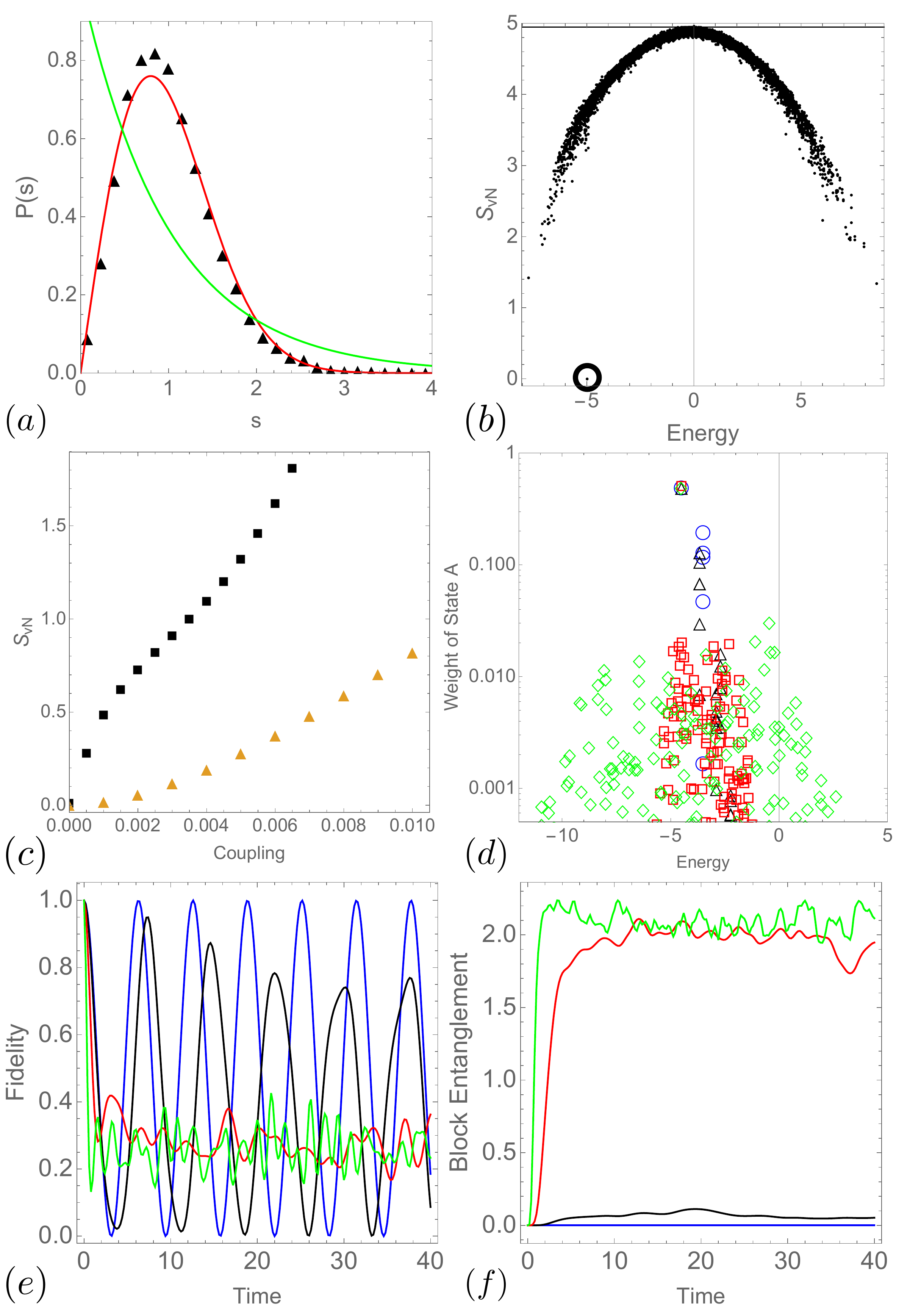}
\caption{Aspects of the sawtooth chain. (a) Level spacing statistics for $N=20$, $J=3$, $J'=1$, $\lambda=-0.5$, compared with predictions for GOE (red) and Poisson (green). (b) Entanglement entropies in all eigenstates, for $J=1$, $J'=1.5$, $\lambda=0.5$, and $N=16$. Scar state is highlighted with a circle. The horizontal line at $4.949$ is the random state entanglement for the $(8,8)$ bipartition.  (c) Effect of perturbations on the entanglement of the scar state: $J=1.5$, $J'=1.0$, $\lambda=0.5$ and $J_p$ (squares), $\eta$ (triangles). Dynamical signatures of scar states for an $N=12$ chain: (d-f)  $J'/J=0.05$ (blue), $J'/J=0.5$ (black), $J'/J=1.5$ (red), $J'/J=4.5$ (green). (d) Square of overlaps between the initial state (``state A'') and different eigenstates, plotted against eigen-energies. (e) The fidelity and (f) the block entanglement as  functions of time.
\label{fig:sawtooth}
}
\end{figure}

Our first example of a model with a many-body scar is the sawtooth chain \cite{PhysRevB.53.6401,PhysRevB.67.054412}, which  we have introduced in detail in Section~\ref{sec:models}.  We demonstrated that the singlet dimer covering on the $J$ bonds is an exact eigenstate of fixed energy that can be tuned so that it is arbitrarily located in the spectrum relative to the ground state, e.g., its neighbouring states can be made to have high effective temperature. Fig.~\ref{fig:sawtooth}(a) shows the distribution of consecutive level spacings normalized to the mean of the distribution, $P(s)$. The approximate prediction for this quantity for random matrices of the gaussian orthogonal ensemble (GOE) is 
\[
P(s) = \frac{\pi}{2}s\exp\left( -\frac{\pi}{4}s^2 \right) .
\]
The figure shows that this prediction is compatible with the $N=16$ total $S^z = 0$ sector for the XXZ model $J=3$, $J'=1$ and $\lambda=-0.5$. The spectra were computed using periodic boundary conditions and were separated into momentum sectors.  Using XXZ couplings  ($\lambda\neq0$) ensures that the total spin is not a good quantum number, so that there are no total spin symmetry sectors needing to be separated.  
The result of Fig.~\ref{fig:sawtooth}(a) is expected in this model which has no local conserved quantities, in contrast to integrable models which show Poissonian level spacing statistics. 

The entanglement entropy between two half-systems ($8$ connected sites each) for the XXZ model as a function of energy is shown in Fig.~\ref{fig:sawtooth}(b).   The couplings used are such that the eigenstate with singlet  coverings lies in the middle of the spectrum.  We have highlighted this  scar state in the Figure by circling the data point.  The scar state has zero entanglement because the entanglement partitioning border cuts through $J'$ bonds. Other than the single scar state, the entanglement is typical of non-integrable systems with no conserved quantities:  the points form an ``arch'' with the entanglements in the middle of the spectrum being close to that of a random state of the same size, and with low  (area law) entanglement at the spectral edges.

We have examined the effect of perturbations away from the sawtooth XXZ model on a periodic $N=16$ chain. Fig.~\ref{fig:sawtooth}(c) shows the scar entanglement for two types of perturbation: one where the two $J'$ bonds become $J'+\eta$ and $J'-\eta$ and the other where a new Heisenberg exchange with coupling $J_p$ is included between corner vertices on neighboring triangles. The effect of the latter type of coupling is more dramatic with the entanglement rising to about $40\%$ of the random-state value for $J_p/J =0.07$. 

The PXP model  has several scar states \cite{2018NatPh..14..745T} and it is possible to observe their presence through dynamical observables by preparing an initial state with significant overlap with the scar states. One finds that the dynamics exhibits Rabi oscillations reflecting unitary evolution within the scar subspace even though this subspace is distributed in energy across the many-body spectrum. In contrast, the sawtooth chain with periodic boundary conditions has a single scar state. Even so, we ask whether this state might have observable dynamical features. To this end, we prepare a state $\vert \Psi_0 \rangle$ in the dimer state except for a single bond that is excited into a product state $\vert 01\rangle$ with weight in the singlet and triplet sectors.  This is referred to ``state A'' in Fig.~\ref{fig:sawtooth}.  The overlap of this state with all eigenstates is shown in Fig.~\ref{fig:sawtooth} (d) for different couplings $J'/J$ (in the Heisenberg model $\lambda=0$). For small coupling, $J'/J=0.05$, close to the decoupled dimer limit, the overlap is concentrated in the dimer state and its low-lying excited states. As the coupling increases, the overlap distribution broadens across the whole spectrum with the most dramatic broadening taking place at the threshold $(J'/J)_c \approx 0.5$ where the exact singlet covering ceases to be the ground state. Turning now to the dynamics, we find that the fidelity $\left|\langle\Psi_0\vert\Psi(t)\rangle\right|^2$ exhibits strong oscillations for $J'/J=0.05$, close to the decoupled dimer limit (Fig.~\ref{fig:sawtooth}(e)) that persist out to times at least of order one thousand times the period of oscillation. This degree of coherence is to be expected as the initial state is predominantly a mixture of the ground state and low-lying excited states. As the coupling increases beyond the critical coupling $(J'/J)_c$ the oscillations are progressively damped reaching a plateau in times of order $1$ when $J'/J\gtrsim 1$. The fidelity in the plateau is the (non-vanishing) weight of the scar state admixed into the initial state. This result is therefore the analogue of the coherent oscillations seen in the PXP model but in the limit where the number of scar states goes to one. These results are mirrored by the time dependence of the entanglement which remains small for long times for $J'/J=0.05$ and $J'/J=0.5$ because  the dimer covering has low energies and the overlap distribution is narrow in energy (Fig.~\ref{fig:sawtooth} (f)). For larger couplings, the entanglement quickly reaches a plateau close to the random-state value as most of the weight of the initial state approaches a random state. In summary, we have selected a simple and natural initial state $\vert \Psi(0)\rangle = \alpha \vert {\rm Singlet \hspace{1mm} covering}\rangle + \ldots$ that has significant weight $\vert \alpha \vert^2$ with the scar state. The dynamics of this state is consistent with the thermalization of the state apart from the residual part coming from the scar $\vert {\rm Singlet \hspace{1mm} covering}\rangle$.

\begin{figure}[tbp]
\centering
\includegraphics[width=0.85\columnwidth]{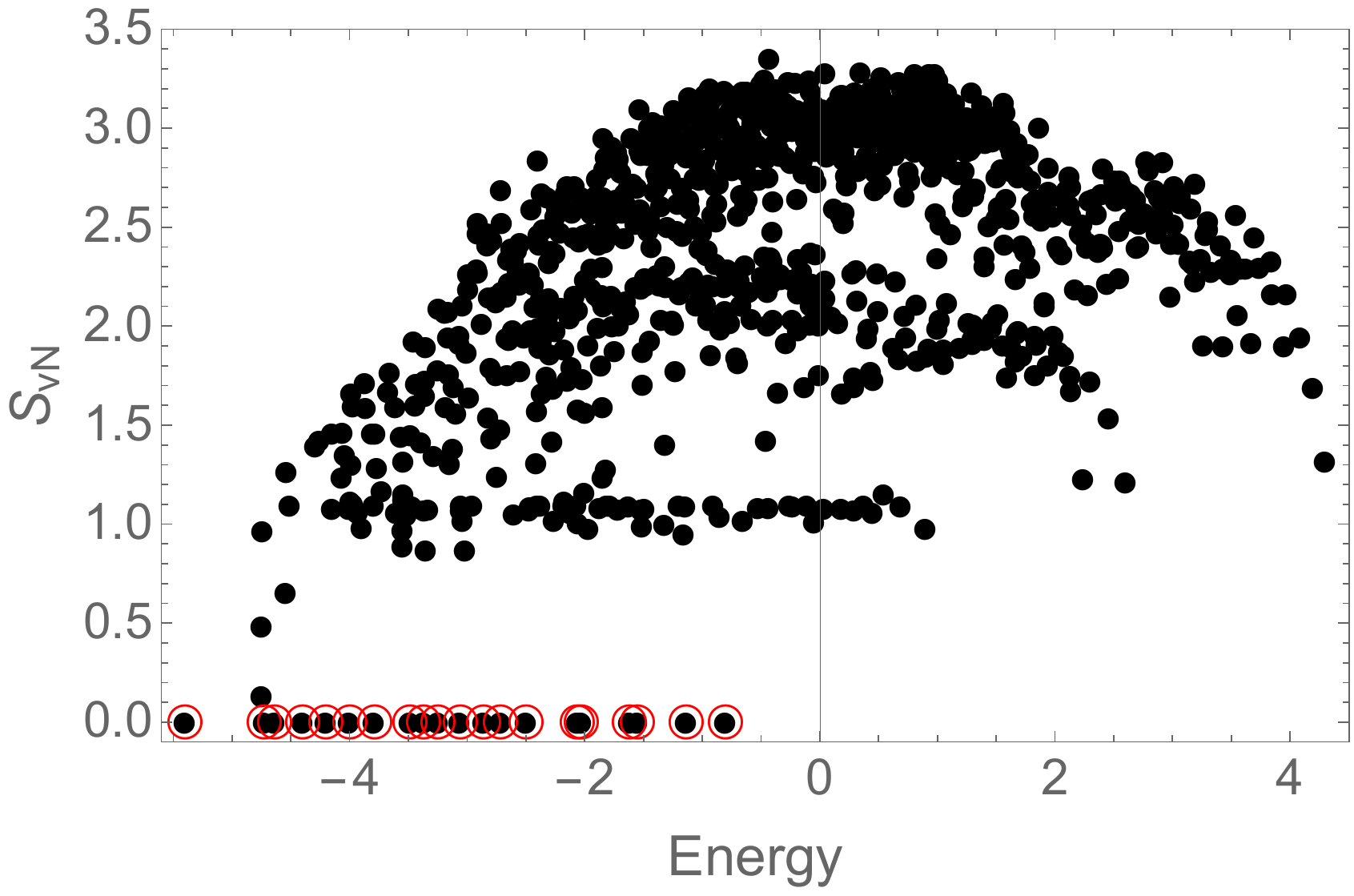}
\caption{Entanglement of the $N=12$ sawtooth chain with open boundary conditions and entanglement cut in the middle of the chain (black points). The couplings are $J=1.0$ and $J'=1.2$. The zero entanglement states for this cut have energies equal to those of the full spectrum on chain of length $N=6$ (red circles) up to a constant shift.
\label{fig:sawtooth_obc}
}
\end{figure}

\paragraph*{Open boundary:} A look at the chain with open boundary conditions is illuminating. As before, we consider a sawtooth chain with $N$ sites with entanglement computed on a bipartition that cuts $J'$ bonds, but now the chain has open boundaries and the bipartition divides the chain into two equal (identical) blocks.  We know that one dimerized state exists in the spectrum of the periodic chain with zero entanglement on cuts through $J'$ bonds. On the open chain, many zero entanglement states are present in the spectrum $-$ their number depending on the location of the cut along the chain. These states can be rationalized as follows. In order for the state to have zero entanglement, the state must be separable at the location of the single cut on the open chain, say dividing the chain into $n$ sites on the left and $N-n$ on the right. Closer investigation reveals that the right-hand-side of the chain in these states has a simple dimer covering imposed by the dangling $J$ bond at the right-hand edge. The left-hand-side is in an eigenstate of the $n$-site open chain. It follows (i) that the number of zero entanglement states on the open chain equals the total number of states on the open $n$ site chain and (ii) that the energy of each zero entanglement states on the open chain is $E^{\rm Scar}_{n\vert N-n}=E^{\rm OBC}_{n}+E^{\rm Dimer}_{N-n}$ where $E^{\rm Scar}_{n\vert N-n}$ is the energy of the zero entanglement state on the $(n,N-n)$ bipartition, $E^{\rm OBC}_{n}$ is the energy of an eigenstate on the $n$ site open chain and $E^{\rm Dimer}_{N-n}$ is the energy of the singlet covering on the $N-n$ open chain which is a constant. This point is illustrated in Fig.~\ref{fig:sawtooth_obc} which shows the entanglement on a subsystem of six sites from an open chain of length $N=12$. The energies of the zero entanglement states on this cut are in one-to-one correspondence to the full spectrum on the $N=6$ open chain (also shown). 

\subsection{Maple Leaf Lattice}

\begin{figure}[tbp]
\centering
\includegraphics[width=1.\columnwidth]{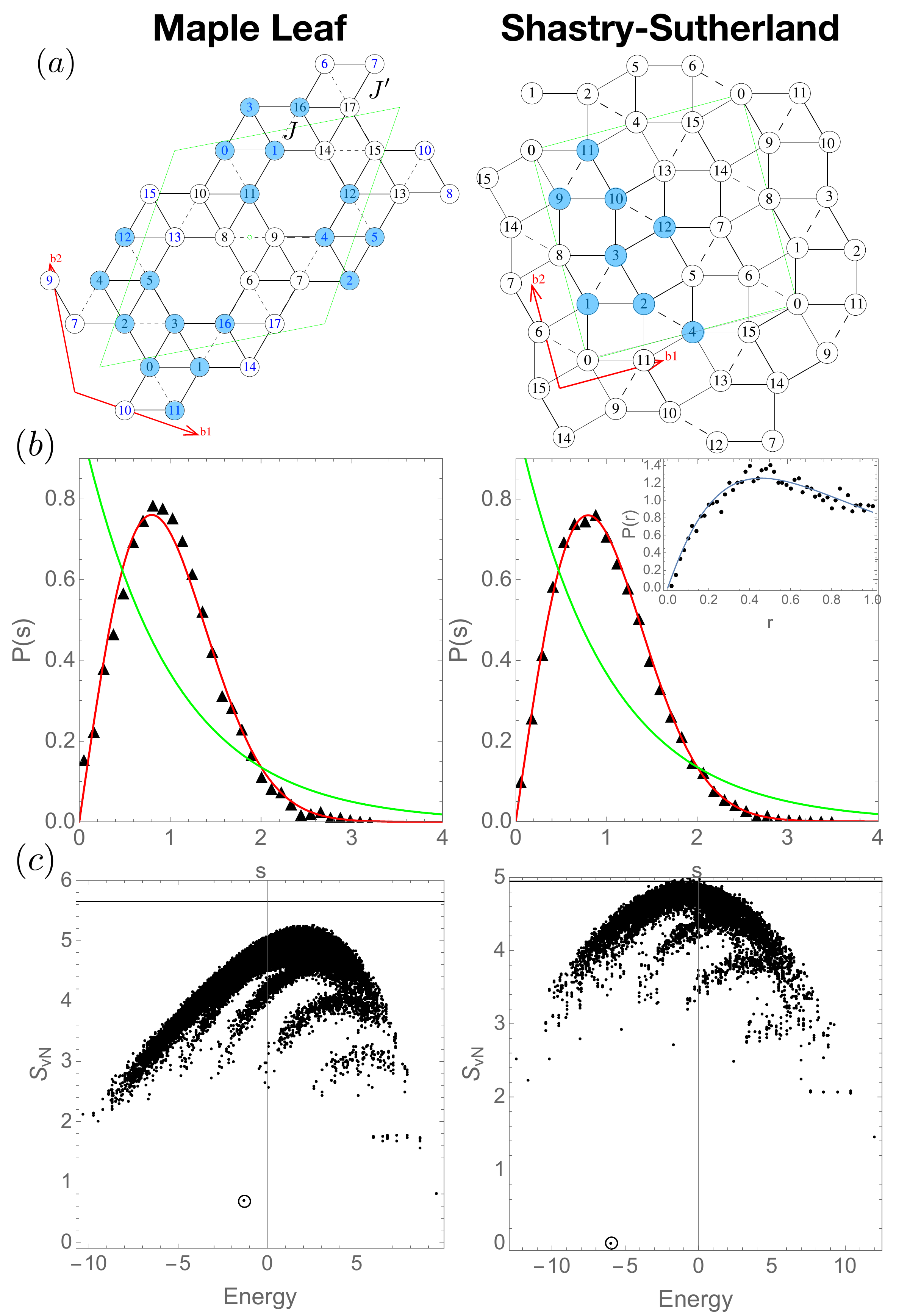}
\caption{Left column: Maple leaf lattice with $18$ sites. Right column: Shastry-Sutherland lattice with either $20$ sites (b) or $16$ sites (a,c). 
(a) $J$ ($J'$) bonds are dashed (full).  One of the entanglement bipartitions is distinguished by blue shaded sites.  The trapezoids demarcate the finite-size systems used for numerical diagonalization, with periodic boundary conditions. 
(b) Level spacing distributions, compared with Poissonian (green) and GOE (red) predictions.  Left: maple leaf XXZ, $J=3$, $J'=1$, $\lambda=-0.5$.  Statistics from the middle $1/6$th of the spectrum in the total $S^z=0$ sector.  Right: Shastry-Sutherland, $N=20$, $J=3$, $J'=1$, $\lambda=0$.  
Right inset shows distribution of ratios of level spacings. 
(c) Entanglement entropy of each eigenstate, against eigen-energies.  Scar states are highlighted. The random state entanglement is indicated as a horizontal line in both panels. Left: maple leaf lattice,   $J=0.2$,  $J'=1$. $\lambda=0$.  Right: Shastry-Sutherland lattice, $N=16$,  $J=1$, $J'=1.25$, $\lambda=0$.  
\label{fig:mllshsu}
}
\end{figure}

The maple leaf lattice is a five-coordinated two dimensional edge-shared triangular lattice obtained by periodically depleting $1/7$th of the sites from the regular triangular lattice (Fig.~\ref{fig:mllshsu}(a,left)) \cite{richter2004quantum,PhysRevB.60.1064,Fennell_2011}. The lattice has six sublattices and three symmetry-distinct nearest neighbor Heisenberg couplings. For our purposes, we set two of these couplings to be equal.  Thus we have a $J$-$J'$ Heisenberg model (Fig.~\ref{fig:mllshsu}(a,left)).  This model is known to have the singlet covering on the $J$ bonds as an exact eigenstate, which is the ground state for $J'/J \lesssim 0.69$ \cite{PhysRevB.84.104406}. 

The level spacing statistics computed from the eigen-energies in the middle of the spectrum for an $18$ site lattice are compatible with random matrix predictions and the non-integrability of the model (Fig. ~\ref{fig:mllshsu}(b,left)).  For the level spacing results, we have broken SU$(2)$ symmetry by using an anisotropic coupling, i.e., using the XXZ model, in order to avoid having multiple sectors corresponding to different total spin values. 

If Fig.~\ref{fig:mllshsu}(c,left) we present the entanglement for the Heisenberg model ($\lambda=0$) in the total $S^z=0$ sector.  For the Heisenberg model, the total spin is a good quantum number.  This results in separate ``arches'' corresponding to different total spin sectors. 
Fig.~\ref{fig:mllshsu}(c,left) features the protected singlet state at intermediate energies in the spectrum.
This scar state is highlighted in the figure.  The partition (separating the two blocks between which the entanglement is calculated) cuts one singlet bond, so the scar state entanglement is $\log 2$.  This is well below the random-matrix value, which is close to $5$.

\subsection{Shastry-Sutherland Model}

The Shastry-Sutherland model \cite{shastry1981exact,PhysRevLett.82.3168,albrecht1996first,PhysRevB.72.104425,PhysRevResearch.1.033038,mcclarty2017topological} is a $J$-$J'$ model defined on the four sublattice 2D lattice shown in Fig.~\ref{fig:dimer_lattices}(f).  This model is realized to a good approximation in SrCu$_2($BO$_3)_2$ with $J'/J \sim 0.6$ \cite{PhysRevLett.82.3168,albrecht1996first,PhysRevB.72.104425,mcclarty2017topological}, and its ground state and thermodynamic properties have been of intense interest in the field of frustrated magnetism. 

When $J'=0$, the lattice decouples into isolated pairs of $J$-coupled spins, and the ground state has singlets on each $J$ bond. The argument outlined in Section~\ref{sec:models} tells us that the singlet covering remains an exact eigenstate for any value of $J'$ and extensive numerical studies have shown that this state is the ground state for $J'/J \lesssim 0.7$ \cite{PhysRevB.87.115144}. For larger values of $J'/J$, this eigenstate is no longer the ground state and is instead a scar state.  

The level spacing distribution for a $20$ site lattice is shown in Fig.~\ref{fig:mllshsu}(b,right).  The spectrum has been split into symmetry sectors and the level spacing computed for each separately for the middle one-sixth of the spectrum and then combined into the full distribution. The distribution is well described by that of the GOE result consistent with the fact that the model is non-integrable. The inset to Fig.~\ref{fig:mllshsu}(b) shows the distribution of $r$ values defined by \cite{Oganesyan_Huse_PRB2007, Atas_Bogomolny_Roux_PRL2013}
\[
 r_n \equiv \frac{{\rm Min}(s_n, s_{n+1})}{{\rm Max}(s_n, s_{n+1})}
\]
where $s_n = E_{n+1}-E_n$ and the eigenvalues $E_n$ are ordered. Again the distribution is compatible with the GOE result with mean $\langle r\rangle = 0.537$. 

Fig.~\ref{fig:mllshsu}(c,right) shows the entanglement computed for the $16$ spin lattice shown in Fig.~\ref{fig:mllshsu}(a,right) on a connected partition encompassing $8$ spins and in the total $S^z=0$ sector.  The entanglement in each eigenstate is plotted against the eigen-energies.  Other than a single scar state, the entanglements are arranged in several arches, corresponding to different sectors of total spin, as in the maple leaf case.  
The boundary between partitions is such that it avoids cutting singlet bonds, so that the entanglement of the scar state is zero. For an arbitrary cut, the entanglement of the scar state would scale as the area of the boundary, in contrast to volume law for neighboring states in the middle of the spectrum.  The isolated  zero entanglement state has fixed ($J'$-independent) energy and its location relative to the middle of the spectrum can be tuned so that it lies among the mid-spectrum states in the lower (upper) half of the spectrum for antiferromagnetic (ferromagnetic) couplings.

\section{Many-Body Quantum Scars from the Square Kagome: Class (III)}
\label{sec:scars3}

In Section~\ref{sec:models} we listed three example of models from class (III). In one dimension we mentioned the sawtooth chain with distinguished bonds on alternate valleys, and the bowtie chain. We now discuss an example of a class (III) model in two dimensions.

\begin{figure}[tbp]
\centering
\includegraphics[width=1.\columnwidth]{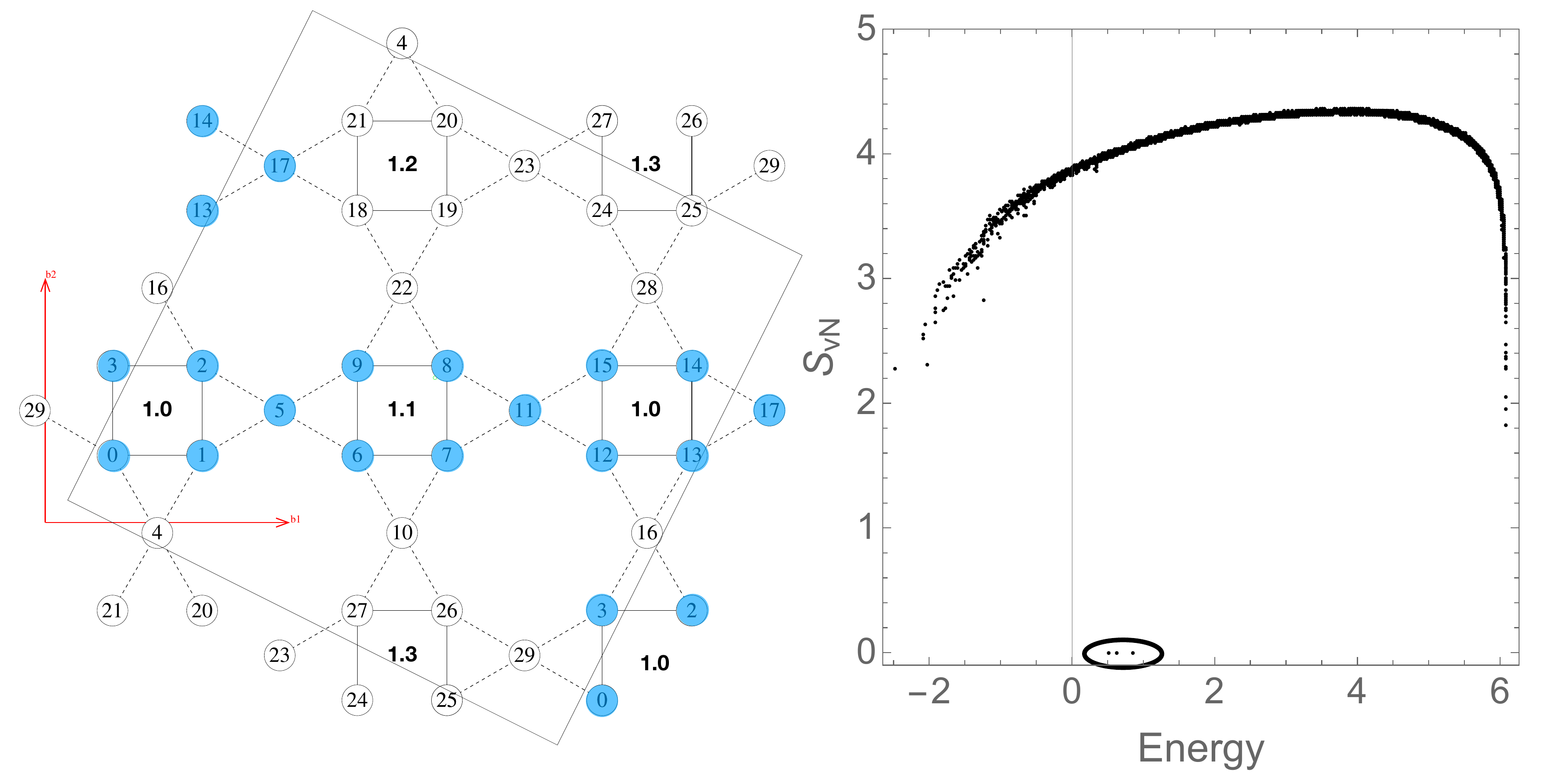}
\caption{(Left) Finite square kagome lattice of $30$ sites used for the exact diagonalization study. The entanglement bipartition is indicated by the shading of the lattice sites.  The exchange coupling $J_n$ on the square plaquettes is shown for each square, while $J'=1.8$, $\lambda=0.5$ throughout. (Right) Entanglement entropy within the total $S^z=22s$ sector on the $30$ site system showing three scar states with zero entanglement entropy.
\label{fig:square_kagome}
}
\end{figure}

The square kagome lattice \cite{schnack2001independent,richter2009heisenberg,PhysRevB.88.195109,PhysRevResearch.1.033147,SquareKagome1} is a two-dimensional lattice of corner-sharing triangles with a six-site primitive cell. Like the kagome lattice, it has coordination number four but, whereas the kagome lattice has an underlying triangular Bravais lattice and triangular and hexagonal polygonal units, the square kagome has a square Bravais cell and triangular and square units. The latter are crucial to the existence of scar states in the $J$-$J'$ XXZ model on this lattice where $J$ is on the square edges and $J'$ on all other bonds \cite{,DerzhkoRichter,schnack2001independent}. Evidently this has the same kind of frustrated triangular unit that we have seen throughout this paper.

It is known that this model has a two-thirds magnetization plateau that can be reached from the fully polarized high-field limit through the condensation of a flat band of magnons. These localized magnon states live on the square plaquettes and exact eigenstates for such states starting from the fully magnetized state $\vert 11\ldots 1\rangle$ are obtained as $\sum_m (-)^m S^-_m \vert 11\ldots 1\rangle$ where $m$ is taken anti-clockwise around a square plaquette. The ground state in the $2/3$rd magnetization sector is obtained by tiling every square plaquette with such localized magnons. 

In order to construct quantum many-body scar states we may simply place localized magnons on a subset of the square plaquettes. For example, if we tile all but one of the plaquettes with localized magnons we shall have a many-body scar state with degeneracy equal to the number of unit cells. In order to obtain multiple scar states as in the PXP model we may enlarge the unit cell by taking the $J$ exchange to be different on different square plaquettes. As a concrete example, we consider the $30$ site system shown in Fig.~\ref{fig:square_kagome} with the crystallographic unit cell enlarged by choosing the exchange on square plaquettes to be $J_1=J_5=1.0$, $J_2 = 1.1$, $J_3= 1.2$ and $J_4 = 1.3$ as shown. For this system size the saturation magnetization is $S^z = 30s$ with $s=1/2$.  We carry out diagonalization in the sector with all but one of the square plaquettes in a localized magnon state --- the $S^z = 22s$ sector. If the $J_n$ were equal there would be a five-fold degeneracy of the localized magnon states corresponding to the five ways of choosing the position of the fully polarized plaquette. By enlarging the unit cell in the way indicated this degeneracy is broken down to $1+1+1+2$ and the two-fold degenerate states can mix leading to a finite entanglement. The remaining three states are manifestly scar states appearing at distinct energies with zero entanglement.  For our choice of $J'=1.8$, these appear roughly in the middle of the spectrum (Fig.~\ref{fig:square_kagome} right panel), thus forming scars.

\section{Summary and Conclusions}
\label{sec:conclusions}

Geometrical frustration has long been one of the central ideas in condensed matter physics with important connections to various low energy exotic classical and quantum states of matter. Here, we have described how geometrical frustration can also lead to unusual high energy states. We have divided the presentation into three classes of phenomena each giving a large class of models exhibiting anomalous thermalization in at least some mid-spectrum states. In the first class, geometrical frustration leads to an extensive number of local conservation laws that is, however, smaller than the number of degrees of freedom. This class therefore consists of non-integrable models with highly structured Hilbert spaces. We have shown that standard ETH scaling is violated for one example from this class --- the fully frustrated ladder --- which, instead, most closely resembles the behavior seen in integrable models. A more detailed examination of the fully frustrated ladder reveals that it is an example of disorder-free localization --- in which correlations and entanglement spreading are dynamically inhibited on the scale of a few lattice spacings. This example generalizes straightforwardly to any one-dimensional model with frustrated units, carrying locally conserved spin protecting localized singlets, separated from one another by any set of arbitrarily coupled spins. We have also argued that aspects of this physics carry over to certain 2D models including the fully frustrated bilayer models on star and martini lattices. We leave a detailed numerical analysis of these 2D models as a problem for the future. The second class of models we have considered has many-body quantum scar states that are product states of singlets and examples include the sawtooth chain, the famous Shastry-Sutherland model and the maple leaf lattice. Each of these examples has a single many-body scar state and we have studied the dynamical signatures of such states. We have shown through several examples that this physics is insensitive to the choice of $J'/J$ and the anisotropic coupling. We have also shown that such models naturally fall within the framework of Shiraishi and Mori that uses local projectors to construct Hamiltonians with scar states. The final class is composed of flat band models exhibiting localized magnon states. As for classes (I) and (II), there are many such models and all rely on magnetic frustration as the mechanism for the existence of athermal states. We gave three examples of such models and investigated one of these in detail 
showing that it can be engineering so that arbitrarily many scar states with distinct energies appear in the middle of the spectrum. One interesting open question regarding class (III) models is whether there is a framework analogous to Shiraishi-Mori that can generalize the physics to other models.

\section{Acknowledgements}

The work of A.S. is partly supported through the Max Planck Partner Group program between the Indian Association for the Cultivation of Science (Kolkata) and the Max Planck Institute for the Physics of Complex Systems (Dresden). P. A. M. thanks Carlo Danieli for a careful reading of the manuscript. J. R. thanks the DFG for financial support (grant RI615/25-1).

\bibliography{references}

\begin{thebibliography}{132}%
\makeatletter
\providecommand \@ifxundefined [1]{%
 \@ifx{#1\undefined}
}%
\providecommand \@ifnum [1]{%
 \ifnum #1\expandafter \@firstoftwo
 \else \expandafter \@secondoftwo
 \fi
}%
\providecommand \@ifx [1]{%
 \ifx #1\expandafter \@firstoftwo
 \else \expandafter \@secondoftwo
 \fi
}%
\providecommand \natexlab [1]{#1}%
\providecommand \enquote  [1]{``#1''}%
\providecommand \bibnamefont  [1]{#1}%
\providecommand \bibfnamefont [1]{#1}%
\providecommand \citenamefont [1]{#1}%
\providecommand \href@noop [0]{\@secondoftwo}%
\providecommand \href [0]{\begingroup \@sanitize@url \@href}%
\providecommand \@href[1]{\@@startlink{#1}\@@href}%
\providecommand \@@href[1]{\endgroup#1\@@endlink}%
\providecommand \@sanitize@url [0]{\catcode `\\12\catcode `\$12\catcode
  `\&12\catcode `\#12\catcode `\^12\catcode `\_12\catcode `\%12\relax}%
\providecommand \@@startlink[1]{}%
\providecommand \@@endlink[0]{}%
\providecommand \url  [0]{\begingroup\@sanitize@url \@url }%
\providecommand \@url [1]{\endgroup\@href {#1}{\urlprefix }}%
\providecommand \urlprefix  [0]{URL }%
\providecommand \Eprint [0]{\href }%
\providecommand \doibase [0]{https://doi.org/}%
\providecommand \selectlanguage [0]{\@gobble}%
\providecommand \bibinfo  [0]{\@secondoftwo}%
\providecommand \bibfield  [0]{\@secondoftwo}%
\providecommand \translation [1]{[#1]}%
\providecommand \BibitemOpen [0]{}%
\providecommand \bibitemStop [0]{}%
\providecommand \bibitemNoStop [0]{.\EOS\space}%
\providecommand \EOS [0]{\spacefactor3000\relax}%
\providecommand \BibitemShut  [1]{\csname bibitem#1\endcsname}%
\let\auto@bib@innerbib\@empty
\bibitem [{\citenamefont {D'Alessio}\ \emph {et~al.}(2016)\citenamefont
  {D'Alessio}, \citenamefont {Kafri}, \citenamefont {Polkovnikov},\ and\
  \citenamefont {Rigol}}]{DAlessio_review2016}%
  \BibitemOpen
  \bibfield  {author} {\bibinfo {author} {\bibfnamefont {L.}~\bibnamefont
  {D'Alessio}}, \bibinfo {author} {\bibfnamefont {Y.}~\bibnamefont {Kafri}},
  \bibinfo {author} {\bibfnamefont {A.}~\bibnamefont {Polkovnikov}},\ and\
  \bibinfo {author} {\bibfnamefont {M.}~\bibnamefont {Rigol}},\ }\bibfield
  {title} {\bibinfo {title} {From quantum chaos and eigenstate thermalization
  to statistical mechanics and thermodynamics},\ }\href
  {https://doi.org/10.1080/00018732.2016.1198134} {\bibfield  {journal}
  {\bibinfo  {journal} {Advances in Physics}\ }\textbf {\bibinfo {volume}
  {65}},\ \bibinfo {pages} {239} (\bibinfo {year} {2016})}\BibitemShut
  {NoStop}%
\bibitem [{\citenamefont {Deutsch}(1991)}]{PhysRevA.43.2046}%
  \BibitemOpen
  \bibfield  {author} {\bibinfo {author} {\bibfnamefont {J.~M.}\ \bibnamefont
  {Deutsch}},\ }\bibfield  {title} {\bibinfo {title} {Quantum statistical
  mechanics in a closed system},\ }\href
  {https://doi.org/10.1103/PhysRevA.43.2046} {\bibfield  {journal} {\bibinfo
  {journal} {Phys. Rev. A}\ }\textbf {\bibinfo {volume} {43}},\ \bibinfo
  {pages} {2046} (\bibinfo {year} {1991})}\BibitemShut {NoStop}%
\bibitem [{\citenamefont {Srednicki}(1994)}]{PhysRevE.50.888}%
  \BibitemOpen
  \bibfield  {author} {\bibinfo {author} {\bibfnamefont {M.}~\bibnamefont
  {Srednicki}},\ }\bibfield  {title} {\bibinfo {title} {Chaos and quantum
  thermalization},\ }\href {https://doi.org/10.1103/PhysRevE.50.888} {\bibfield
   {journal} {\bibinfo  {journal} {Phys. Rev. E}\ }\textbf {\bibinfo {volume}
  {50}},\ \bibinfo {pages} {888} (\bibinfo {year} {1994})}\BibitemShut
  {NoStop}%
\bibitem [{\citenamefont {Rigol}\ \emph {et~al.}(2008)\citenamefont {Rigol},
  \citenamefont {Dunjko},\ and\ \citenamefont {Olshanii}}]{Rigol_Nature2008}%
  \BibitemOpen
  \bibfield  {author} {\bibinfo {author} {\bibfnamefont {M.}~\bibnamefont
  {Rigol}}, \bibinfo {author} {\bibfnamefont {V.}~\bibnamefont {Dunjko}},\ and\
  \bibinfo {author} {\bibfnamefont {M.}~\bibnamefont {Olshanii}},\ }\bibfield
  {title} {\bibinfo {title} {Thermalization and its mechanism for generic
  isolated quantum systems.},\ }\href {https://doi.org/10.1038/nature06838}
  {\bibfield  {journal} {\bibinfo  {journal} {Nature}\ }\textbf {\bibinfo
  {volume} {452}},\ \bibinfo {pages} {854} (\bibinfo {year}
  {2008})}\BibitemShut {NoStop}%
\bibitem [{\citenamefont {Reimann}(2015)}]{Reimann_NJP2015}%
  \BibitemOpen
  \bibfield  {author} {\bibinfo {author} {\bibfnamefont {P.}~\bibnamefont
  {Reimann}},\ }\bibfield  {title} {\bibinfo {title} {Eigenstate
  thermalization: Deutsch's approach and beyond},\ }\href
  {http://stacks.iop.org/1367-2630/17/i=5/a=055025} {\bibfield  {journal}
  {\bibinfo  {journal} {New Journal of Physics}\ }\textbf {\bibinfo {volume}
  {17}},\ \bibinfo {pages} {055025} (\bibinfo {year} {2015})}\BibitemShut
  {NoStop}%
\bibitem [{\citenamefont {{Deutsch}}(2018)}]{Deutsch_RepProgPhys2018}%
  \BibitemOpen
  \bibfield  {author} {\bibinfo {author} {\bibfnamefont {J.~M.}\ \bibnamefont
  {{Deutsch}}},\ }\bibfield  {title} {\bibinfo {title} {{Eigenstate
  thermalization hypothesis}},\ }\href
  {https://doi.org/10.1088/1361-6633/aac9f1} {\bibfield  {journal} {\bibinfo
  {journal} {Reports on Progress in Physics}\ }\textbf {\bibinfo {volume}
  {81}},\ \bibinfo {eid} {082001} (\bibinfo {year} {2018})},\ \Eprint
  {https://arxiv.org/abs/1805.01616} {arXiv:1805.01616 [quant-ph]} \BibitemShut
  {NoStop}%
\bibitem [{\citenamefont {Vidmar}\ and\ \citenamefont
  {Rigol}(2016)}]{vidmar2016generalized}%
  \BibitemOpen
  \bibfield  {author} {\bibinfo {author} {\bibfnamefont {L.}~\bibnamefont
  {Vidmar}}\ and\ \bibinfo {author} {\bibfnamefont {M.}~\bibnamefont {Rigol}},\
  }\bibfield  {title} {\bibinfo {title} {Generalized {G}ibbs ensemble in
  integrable lattice models},\ }\href@noop {} {\bibfield  {journal} {\bibinfo
  {journal} {Journal of Statistical Mechanics: Theory and Experiment}\ }\textbf
  {\bibinfo {volume} {2016}},\ \bibinfo {pages} {064007} (\bibinfo {year}
  {2016})}\BibitemShut {NoStop}%
\bibitem [{\citenamefont {Nandkishore}\ and\ \citenamefont
  {Huse}(2015)}]{nandkishore2015many}%
  \BibitemOpen
  \bibfield  {author} {\bibinfo {author} {\bibfnamefont {R.}~\bibnamefont
  {Nandkishore}}\ and\ \bibinfo {author} {\bibfnamefont {D.~A.}\ \bibnamefont
  {Huse}},\ }\bibfield  {title} {\bibinfo {title} {Many-body localization and
  thermalization in quantum statistical mechanics},\ }\href@noop {} {\bibfield
  {journal} {\bibinfo  {journal} {Annu. Rev. Condens. Matter Phys.}\ }\textbf
  {\bibinfo {volume} {6}},\ \bibinfo {pages} {15} (\bibinfo {year}
  {2015})}\BibitemShut {NoStop}%
\bibitem [{\citenamefont {Smith}\ \emph {et~al.}(2017)\citenamefont {Smith},
  \citenamefont {Knolle}, \citenamefont {Kovrizhin},\ and\ \citenamefont
  {Moessner}}]{PhysRevLett.118.266601}%
  \BibitemOpen
  \bibfield  {author} {\bibinfo {author} {\bibfnamefont {A.}~\bibnamefont
  {Smith}}, \bibinfo {author} {\bibfnamefont {J.}~\bibnamefont {Knolle}},
  \bibinfo {author} {\bibfnamefont {D.~L.}\ \bibnamefont {Kovrizhin}},\ and\
  \bibinfo {author} {\bibfnamefont {R.}~\bibnamefont {Moessner}},\ }\bibfield
  {title} {\bibinfo {title} {Disorder-free localization},\ }\href
  {https://doi.org/10.1103/PhysRevLett.118.266601} {\bibfield  {journal}
  {\bibinfo  {journal} {Phys. Rev. Lett.}\ }\textbf {\bibinfo {volume} {118}},\
  \bibinfo {pages} {266601} (\bibinfo {year} {2017})}\BibitemShut {NoStop}%
\bibitem [{\citenamefont {Yao}\ \emph {et~al.}(2016)\citenamefont {Yao},
  \citenamefont {Laumann}, \citenamefont {Cirac}, \citenamefont {Lukin},\ and\
  \citenamefont {Moore}}]{Yao_Lauman_Cirac_Lukin_Moore_PRL2016}%
  \BibitemOpen
  \bibfield  {author} {\bibinfo {author} {\bibfnamefont {N.~Y.}\ \bibnamefont
  {Yao}}, \bibinfo {author} {\bibfnamefont {C.~R.}\ \bibnamefont {Laumann}},
  \bibinfo {author} {\bibfnamefont {J.~I.}\ \bibnamefont {Cirac}}, \bibinfo
  {author} {\bibfnamefont {M.~D.}\ \bibnamefont {Lukin}},\ and\ \bibinfo
  {author} {\bibfnamefont {J.~E.}\ \bibnamefont {Moore}},\ }\bibfield  {title}
  {\bibinfo {title} {Quasi-many-body localization in translation-invariant
  systems},\ }\href {https://doi.org/10.1103/PhysRevLett.117.240601} {\bibfield
   {journal} {\bibinfo  {journal} {Phys. Rev. Lett.}\ }\textbf {\bibinfo
  {volume} {117}},\ \bibinfo {pages} {240601} (\bibinfo {year}
  {2016})}\BibitemShut {NoStop}%
\bibitem [{\citenamefont {Schiulaz}\ \emph {et~al.}(2015)\citenamefont
  {Schiulaz}, \citenamefont {Silva},\ and\ \citenamefont
  {M\"uller}}]{Silva_Mueller_PRB2015}%
  \BibitemOpen
  \bibfield  {author} {\bibinfo {author} {\bibfnamefont {M.}~\bibnamefont
  {Schiulaz}}, \bibinfo {author} {\bibfnamefont {A.}~\bibnamefont {Silva}},\
  and\ \bibinfo {author} {\bibfnamefont {M.}~\bibnamefont {M\"uller}},\
  }\bibfield  {title} {\bibinfo {title} {Dynamics in many-body localized
  quantum systems without disorder},\ }\href
  {https://doi.org/10.1103/PhysRevB.91.184202} {\bibfield  {journal} {\bibinfo
  {journal} {Phys. Rev. B}\ }\textbf {\bibinfo {volume} {91}},\ \bibinfo
  {pages} {184202} (\bibinfo {year} {2015})}\BibitemShut {NoStop}%
\bibitem [{\citenamefont {Mondaini}\ and\ \citenamefont
  {Cai}(2017)}]{Mondaini_Cai_PRB2017}%
  \BibitemOpen
  \bibfield  {author} {\bibinfo {author} {\bibfnamefont {R.}~\bibnamefont
  {Mondaini}}\ and\ \bibinfo {author} {\bibfnamefont {Z.}~\bibnamefont {Cai}},\
  }\bibfield  {title} {\bibinfo {title} {Many-body self-localization in a
  translation-invariant {H}amiltonian},\ }\href
  {https://doi.org/10.1103/PhysRevB.96.035153} {\bibfield  {journal} {\bibinfo
  {journal} {Phys. Rev. B}\ }\textbf {\bibinfo {volume} {96}},\ \bibinfo
  {pages} {035153} (\bibinfo {year} {2017})}\BibitemShut {NoStop}%
\bibitem [{\citenamefont {Brenes}\ \emph {et~al.}(2018)\citenamefont {Brenes},
  \citenamefont {Dalmonte}, \citenamefont {Heyl},\ and\ \citenamefont
  {Scardicchio}}]{PhysRevLett.120.030601}%
  \BibitemOpen
  \bibfield  {author} {\bibinfo {author} {\bibfnamefont {M.}~\bibnamefont
  {Brenes}}, \bibinfo {author} {\bibfnamefont {M.}~\bibnamefont {Dalmonte}},
  \bibinfo {author} {\bibfnamefont {M.}~\bibnamefont {Heyl}},\ and\ \bibinfo
  {author} {\bibfnamefont {A.}~\bibnamefont {Scardicchio}},\ }\bibfield
  {title} {\bibinfo {title} {Many-body localization dynamics from gauge
  invariance},\ }\href {https://doi.org/10.1103/PhysRevLett.120.030601}
  {\bibfield  {journal} {\bibinfo  {journal} {Phys. Rev. Lett.}\ }\textbf
  {\bibinfo {volume} {120}},\ \bibinfo {pages} {030601} (\bibinfo {year}
  {2018})}\BibitemShut {NoStop}%
\bibitem [{\citenamefont {Lan}\ \emph {et~al.}(2018)\citenamefont {Lan},
  \citenamefont {van Horssen}, \citenamefont {Powell},\ and\ \citenamefont
  {Garrahan}}]{PhysRevLett.121.040603}%
  \BibitemOpen
  \bibfield  {author} {\bibinfo {author} {\bibfnamefont {Z.}~\bibnamefont
  {Lan}}, \bibinfo {author} {\bibfnamefont {M.}~\bibnamefont {van Horssen}},
  \bibinfo {author} {\bibfnamefont {S.}~\bibnamefont {Powell}},\ and\ \bibinfo
  {author} {\bibfnamefont {J.~P.}\ \bibnamefont {Garrahan}},\ }\bibfield
  {title} {\bibinfo {title} {Quantum slow relaxation and metastability due to
  dynamical constraints},\ }\href
  {https://doi.org/10.1103/PhysRevLett.121.040603} {\bibfield  {journal}
  {\bibinfo  {journal} {Phys. Rev. Lett.}\ }\textbf {\bibinfo {volume} {121}},\
  \bibinfo {pages} {040603} (\bibinfo {year} {2018})}\BibitemShut {NoStop}%
\bibitem [{\citenamefont {van Nieuwenburg}\ \emph {et~al.}(2019)\citenamefont
  {van Nieuwenburg}, \citenamefont {Baum},\ and\ \citenamefont
  {Refael}}]{vanNieuwenburg9269}%
  \BibitemOpen
  \bibfield  {author} {\bibinfo {author} {\bibfnamefont {E.}~\bibnamefont {van
  Nieuwenburg}}, \bibinfo {author} {\bibfnamefont {Y.}~\bibnamefont {Baum}},\
  and\ \bibinfo {author} {\bibfnamefont {G.}~\bibnamefont {Refael}},\
  }\bibfield  {title} {\bibinfo {title} {From {B}loch oscillations to many-body
  localization in clean interacting systems},\ }\href
  {https://doi.org/10.1073/pnas.1819316116} {\bibfield  {journal} {\bibinfo
  {journal} {Proceedings of the National Academy of Sciences}\ }\textbf
  {\bibinfo {volume} {116}},\ \bibinfo {pages} {9269} (\bibinfo {year}
  {2019})}\BibitemShut {NoStop}%
\bibitem [{\citenamefont {Sala}\ \emph {et~al.}(2020)\citenamefont {Sala},
  \citenamefont {Rakovszky}, \citenamefont {Verresen}, \citenamefont {Knap},\
  and\ \citenamefont {Pollmann}}]{PhysRevX.10.011047}%
  \BibitemOpen
  \bibfield  {author} {\bibinfo {author} {\bibfnamefont {P.}~\bibnamefont
  {Sala}}, \bibinfo {author} {\bibfnamefont {T.}~\bibnamefont {Rakovszky}},
  \bibinfo {author} {\bibfnamefont {R.}~\bibnamefont {Verresen}}, \bibinfo
  {author} {\bibfnamefont {M.}~\bibnamefont {Knap}},\ and\ \bibinfo {author}
  {\bibfnamefont {F.}~\bibnamefont {Pollmann}},\ }\bibfield  {title} {\bibinfo
  {title} {Ergodicity breaking arising from {H}ilbert space fragmentation in
  dipole-conserving {H}amiltonians},\ }\href
  {https://doi.org/10.1103/PhysRevX.10.011047} {\bibfield  {journal} {\bibinfo
  {journal} {Phys. Rev. X}\ }\textbf {\bibinfo {volume} {10}},\ \bibinfo
  {pages} {011047} (\bibinfo {year} {2020})}\BibitemShut {NoStop}%
\bibitem [{\citenamefont {Moudgalya}\ \emph {et~al.}(2019)\citenamefont
  {Moudgalya}, \citenamefont {Prem}, \citenamefont {Nandkishore}, \citenamefont
  {Regnault},\ and\ \citenamefont {Bernevig}}]{moudgalya2019thermalization}%
  \BibitemOpen
  \bibfield  {author} {\bibinfo {author} {\bibfnamefont {S.}~\bibnamefont
  {Moudgalya}}, \bibinfo {author} {\bibfnamefont {A.}~\bibnamefont {Prem}},
  \bibinfo {author} {\bibfnamefont {R.}~\bibnamefont {Nandkishore}}, \bibinfo
  {author} {\bibfnamefont {N.}~\bibnamefont {Regnault}},\ and\ \bibinfo
  {author} {\bibfnamefont {B.~A.}\ \bibnamefont {Bernevig}},\ }\href@noop {}
  {\bibinfo {title} {Thermalization and its absence within krylov subspaces of
  a constrained {H}amiltonian}} (\bibinfo {year} {2019}),\ \Eprint
  {https://arxiv.org/abs/1910.14048} {arXiv:1910.14048 [cond-mat.str-el]}
  \BibitemShut {NoStop}%
\bibitem [{\citenamefont {Yang}\ \emph {et~al.}(2020)\citenamefont {Yang},
  \citenamefont {Liu}, \citenamefont {Gorshkov},\ and\ \citenamefont
  {Iadecola}}]{PhysRevLett.124.207602}%
  \BibitemOpen
  \bibfield  {author} {\bibinfo {author} {\bibfnamefont {Z.-C.}\ \bibnamefont
  {Yang}}, \bibinfo {author} {\bibfnamefont {F.}~\bibnamefont {Liu}}, \bibinfo
  {author} {\bibfnamefont {A.~V.}\ \bibnamefont {Gorshkov}},\ and\ \bibinfo
  {author} {\bibfnamefont {T.}~\bibnamefont {Iadecola}},\ }\bibfield  {title}
  {\bibinfo {title} {{H}ilbert-space fragmentation from strict confinement},\
  }\href {https://doi.org/10.1103/PhysRevLett.124.207602} {\bibfield  {journal}
  {\bibinfo  {journal} {Phys. Rev. Lett.}\ }\textbf {\bibinfo {volume} {124}},\
  \bibinfo {pages} {207602} (\bibinfo {year} {2020})}\BibitemShut {NoStop}%
\bibitem [{\citenamefont {Mazza}\ \emph {et~al.}(2019)\citenamefont {Mazza},
  \citenamefont {Perfetto}, \citenamefont {Lerose}, \citenamefont {Collura},\
  and\ \citenamefont {Gambassi}}]{PhysRevB.99.180302}%
  \BibitemOpen
  \bibfield  {author} {\bibinfo {author} {\bibfnamefont {P.~P.}\ \bibnamefont
  {Mazza}}, \bibinfo {author} {\bibfnamefont {G.}~\bibnamefont {Perfetto}},
  \bibinfo {author} {\bibfnamefont {A.}~\bibnamefont {Lerose}}, \bibinfo
  {author} {\bibfnamefont {M.}~\bibnamefont {Collura}},\ and\ \bibinfo {author}
  {\bibfnamefont {A.}~\bibnamefont {Gambassi}},\ }\bibfield  {title} {\bibinfo
  {title} {Suppression of transport in nondisordered quantum spin chains due to
  confined excitations},\ }\href {https://doi.org/10.1103/PhysRevB.99.180302}
  {\bibfield  {journal} {\bibinfo  {journal} {Phys. Rev. B}\ }\textbf {\bibinfo
  {volume} {99}},\ \bibinfo {pages} {180302} (\bibinfo {year}
  {2019})}\BibitemShut {NoStop}%
\bibitem [{\citenamefont {Kuno}\ \emph
  {et~al.}(2020{\natexlab{a}})\citenamefont {Kuno}, \citenamefont {Orito},\
  and\ \citenamefont {Ichinose}}]{Kuno_2020}%
  \BibitemOpen
  \bibfield  {author} {\bibinfo {author} {\bibfnamefont {Y.}~\bibnamefont
  {Kuno}}, \bibinfo {author} {\bibfnamefont {T.}~\bibnamefont {Orito}},\ and\
  \bibinfo {author} {\bibfnamefont {I.}~\bibnamefont {Ichinose}},\ }\bibfield
  {title} {\bibinfo {title} {Flat-band many-body localization and ergodicity
  breaking in the {C}reutz ladder},\ }\href
  {https://doi.org/10.1088/1367-2630/ab6352} {\bibfield  {journal} {\bibinfo
  {journal} {New Journal of Physics}\ }\textbf {\bibinfo {volume} {22}},\
  \bibinfo {pages} {013032} (\bibinfo {year} {2020}{\natexlab{a}})}\BibitemShut
  {NoStop}%
\bibitem [{\citenamefont {{Karpov}}\ \emph {et~al.}(2020)\citenamefont
  {{Karpov}}, \citenamefont {{Verdel}}, \citenamefont {{Huang}}, \citenamefont
  {{Schmitt}},\ and\ \citenamefont {{Heyl}}}]{2020arXiv200304901K}%
  \BibitemOpen
  \bibfield  {author} {\bibinfo {author} {\bibfnamefont {P.}~\bibnamefont
  {{Karpov}}}, \bibinfo {author} {\bibfnamefont {R.}~\bibnamefont {{Verdel}}},
  \bibinfo {author} {\bibfnamefont {Y.~P.}\ \bibnamefont {{Huang}}}, \bibinfo
  {author} {\bibfnamefont {M.}~\bibnamefont {{Schmitt}}},\ and\ \bibinfo
  {author} {\bibfnamefont {M.}~\bibnamefont {{Heyl}}},\ }\bibfield  {title}
  {\bibinfo {title} {{Disorder-free localization in an interacting
  two-dimensional lattice gauge theory}},\ }\href@noop {} {\bibfield  {journal}
  {\bibinfo  {journal} {arXiv e-prints}\ } (\bibinfo {year} {2020})},\ \Eprint
  {https://arxiv.org/abs/2003.04901} {arXiv:2003.04901 [cond-mat.str-el]}
  \BibitemShut {NoStop}%
\bibitem [{\citenamefont {Magnifico}\ \emph {et~al.}(2020)\citenamefont
  {Magnifico}, \citenamefont {Dalmonte}, \citenamefont {Facchi}, \citenamefont
  {Pascazio}, \citenamefont {Pepe},\ and\ \citenamefont
  {Ercolessi}}]{Magnifico_Ercolessi_Quantum2020_Schwinger}%
  \BibitemOpen
  \bibfield  {author} {\bibinfo {author} {\bibfnamefont {G.}~\bibnamefont
  {Magnifico}}, \bibinfo {author} {\bibfnamefont {M.}~\bibnamefont {Dalmonte}},
  \bibinfo {author} {\bibfnamefont {P.}~\bibnamefont {Facchi}}, \bibinfo
  {author} {\bibfnamefont {S.}~\bibnamefont {Pascazio}}, \bibinfo {author}
  {\bibfnamefont {F.~V.}\ \bibnamefont {Pepe}},\ and\ \bibinfo {author}
  {\bibfnamefont {E.}~\bibnamefont {Ercolessi}},\ }\bibfield  {title} {\bibinfo
  {title} {Real time dynamics and confinement in the $\mathbb{Z}_n$
  {S}chwinger-{W}eyl lattice model for 1+1 qed},\ }\href@noop {} {\bibfield
  {journal} {\bibinfo  {journal} {Quantum}\ }\textbf {\bibinfo {volume} {4}},\
  \bibinfo {pages} {281} (\bibinfo {year} {2020})}\BibitemShut {NoStop}%
\bibitem [{\citenamefont {Schulz}\ \emph {et~al.}(2019)\citenamefont {Schulz},
  \citenamefont {Hooley}, \citenamefont {Moessner},\ and\ \citenamefont
  {Pollmann}}]{Schulz_Hooley_Moessner_Pollmann_PRL2019}%
  \BibitemOpen
  \bibfield  {author} {\bibinfo {author} {\bibfnamefont {M.}~\bibnamefont
  {Schulz}}, \bibinfo {author} {\bibfnamefont {C.~A.}\ \bibnamefont {Hooley}},
  \bibinfo {author} {\bibfnamefont {R.}~\bibnamefont {Moessner}},\ and\
  \bibinfo {author} {\bibfnamefont {F.}~\bibnamefont {Pollmann}},\ }\bibfield
  {title} {\bibinfo {title} {Stark many-body localization},\ }\href
  {https://doi.org/10.1103/PhysRevLett.122.040606} {\bibfield  {journal}
  {\bibinfo  {journal} {Phys. Rev. Lett.}\ }\textbf {\bibinfo {volume} {122}},\
  \bibinfo {pages} {040606} (\bibinfo {year} {2019})}\BibitemShut {NoStop}%
\bibitem [{\citenamefont {Ribeiro}\ \emph {et~al.}(2020)\citenamefont
  {Ribeiro}, \citenamefont {Lazarides},\ and\ \citenamefont
  {Haque}}]{Ribeiro_Lazarides_Haque_PRL2020}%
  \BibitemOpen
  \bibfield  {author} {\bibinfo {author} {\bibfnamefont {P.}~\bibnamefont
  {Ribeiro}}, \bibinfo {author} {\bibfnamefont {A.}~\bibnamefont {Lazarides}},\
  and\ \bibinfo {author} {\bibfnamefont {M.}~\bibnamefont {Haque}},\ }\bibfield
   {title} {\bibinfo {title} {Many-body quantum dynamics of initially trapped
  systems due to a stark potential: Thermalization versus {B}loch
  oscillations},\ }\href {https://doi.org/10.1103/PhysRevLett.124.110603}
  {\bibfield  {journal} {\bibinfo  {journal} {Phys. Rev. Lett.}\ }\textbf
  {\bibinfo {volume} {124}},\ \bibinfo {pages} {110603} (\bibinfo {year}
  {2020})}\BibitemShut {NoStop}%
\bibitem [{\citenamefont {Kuno}\ \emph
  {et~al.}(2020{\natexlab{b}})\citenamefont {Kuno}, \citenamefont {Orito},\
  and\ \citenamefont {Ichinose}}]{Kuno_Orito_Ichinose_NJP2020_Creuzladder}%
  \BibitemOpen
  \bibfield  {author} {\bibinfo {author} {\bibfnamefont {Y.}~\bibnamefont
  {Kuno}}, \bibinfo {author} {\bibfnamefont {T.}~\bibnamefont {Orito}},\ and\
  \bibinfo {author} {\bibfnamefont {I.}~\bibnamefont {Ichinose}},\ }\bibfield
  {title} {\bibinfo {title} {Flat-band many-body localization and ergodicity
  breaking in the {C}reutz ladder},\ }\href@noop {} {\bibfield  {journal}
  {\bibinfo  {journal} {New Journal of Physics}\ }\textbf {\bibinfo {volume}
  {22}},\ \bibinfo {pages} {013032} (\bibinfo {year}
  {2020}{\natexlab{b}})}\BibitemShut {NoStop}%
\bibitem [{\citenamefont {Danieli}\ \emph {et~al.}(2020)\citenamefont
  {Danieli}, \citenamefont {Andreanov},\ and\ \citenamefont
  {Flach}}]{danieli2020manybody}%
  \BibitemOpen
  \bibfield  {author} {\bibinfo {author} {\bibfnamefont {C.}~\bibnamefont
  {Danieli}}, \bibinfo {author} {\bibfnamefont {A.}~\bibnamefont {Andreanov}},\
  and\ \bibinfo {author} {\bibfnamefont {S.}~\bibnamefont {Flach}},\
  }\href@noop {} {\bibinfo {title} {Many-body flatband localization}} (\bibinfo
  {year} {2020}),\ \Eprint {https://arxiv.org/abs/2004.11928} {arXiv:2004.11928
  [cond-mat.stat-mech]} \BibitemShut {NoStop}%
\bibitem [{\citenamefont {Santos}(2009)}]{LeaSantos_JMP2009_transport}%
  \BibitemOpen
  \bibfield  {author} {\bibinfo {author} {\bibfnamefont {L.~F.}\ \bibnamefont
  {Santos}},\ }\bibfield  {title} {\bibinfo {title} {Transport and control in
  one-dimensional systems},\ }\href@noop {} {\bibfield  {journal} {\bibinfo
  {journal} {Journal of mathematical physics}\ }\textbf {\bibinfo {volume}
  {50}},\ \bibinfo {pages} {095211} (\bibinfo {year} {2009})}\BibitemShut
  {NoStop}%
\bibitem [{\citenamefont {Haque}(2010)}]{Haque_PRA2010_locking}%
  \BibitemOpen
  \bibfield  {author} {\bibinfo {author} {\bibfnamefont {M.}~\bibnamefont
  {Haque}},\ }\bibfield  {title} {\bibinfo {title} {Self-similar spectral
  structures and edge-locking hierarchy in open-boundary spin chains},\ }\href
  {https://doi.org/10.1103/PhysRevA.82.012108} {\bibfield  {journal} {\bibinfo
  {journal} {Phys. Rev. A}\ }\textbf {\bibinfo {volume} {82}},\ \bibinfo
  {pages} {012108} (\bibinfo {year} {2010})}\BibitemShut {NoStop}%
\bibitem [{\citenamefont {Choudhury}\ \emph {et~al.}(2018)\citenamefont
  {Choudhury}, \citenamefont {Kim},\ and\ \citenamefont
  {Zhou}}]{Choudhury_Kim_Zhou_arXiv2018}%
  \BibitemOpen
  \bibfield  {author} {\bibinfo {author} {\bibfnamefont {S.}~\bibnamefont
  {Choudhury}}, \bibinfo {author} {\bibfnamefont {E.-a.}\ \bibnamefont {Kim}},\
  and\ \bibinfo {author} {\bibfnamefont {Q.}~\bibnamefont {Zhou}},\ }\bibfield
  {title} {\bibinfo {title} {Frustration induced quasi-many-body localization
  without disorder},\ }\href@noop {} {\bibfield  {journal} {\bibinfo  {journal}
  {arXiv preprint arXiv:1807.05969}\ } (\bibinfo {year} {2018})}\BibitemShut
  {NoStop}%
\bibitem [{\citenamefont {{Turner}}\ \emph {et~al.}(2018)\citenamefont
  {{Turner}}, \citenamefont {{Michailidis}}, \citenamefont {{Abanin}},
  \citenamefont {{Serbyn}},\ and\ \citenamefont
  {{Papi{\'c}}}}]{2018NatPh..14..745T}%
  \BibitemOpen
  \bibfield  {author} {\bibinfo {author} {\bibfnamefont {C.~J.}\ \bibnamefont
  {{Turner}}}, \bibinfo {author} {\bibfnamefont {A.~A.}\ \bibnamefont
  {{Michailidis}}}, \bibinfo {author} {\bibfnamefont {D.~A.}\ \bibnamefont
  {{Abanin}}}, \bibinfo {author} {\bibfnamefont {M.}~\bibnamefont {{Serbyn}}},\
  and\ \bibinfo {author} {\bibfnamefont {Z.}~\bibnamefont {{Papi{\'c}}}},\
  }\bibfield  {title} {\bibinfo {title} {{Weak ergodicity breaking from quantum
  many-body scars}},\ }\href {https://doi.org/10.1038/s41567-018-0137-5}
  {\bibfield  {journal} {\bibinfo  {journal} {Nature Physics}\ }\textbf
  {\bibinfo {volume} {14}},\ \bibinfo {pages} {745} (\bibinfo {year}
  {2018})}\BibitemShut {NoStop}%
\bibitem [{\citenamefont {Turner}\ \emph {et~al.}(2018)\citenamefont {Turner},
  \citenamefont {Michailidis}, \citenamefont {Abanin}, \citenamefont {Serbyn},\
  and\ \citenamefont {Papi\ifmmode~\acute{c}\else
  \'{c}\fi{}}}]{PhysRevB.98.155134}%
  \BibitemOpen
  \bibfield  {author} {\bibinfo {author} {\bibfnamefont {C.~J.}\ \bibnamefont
  {Turner}}, \bibinfo {author} {\bibfnamefont {A.~A.}\ \bibnamefont
  {Michailidis}}, \bibinfo {author} {\bibfnamefont {D.~A.}\ \bibnamefont
  {Abanin}}, \bibinfo {author} {\bibfnamefont {M.}~\bibnamefont {Serbyn}},\
  and\ \bibinfo {author} {\bibfnamefont {Z.}~\bibnamefont
  {Papi\ifmmode~\acute{c}\else \'{c}\fi{}}},\ }\bibfield  {title} {\bibinfo
  {title} {Quantum scarred eigenstates in a {R}ydberg atom chain: Entanglement,
  breakdown of thermalization, and stability to perturbations},\ }\href
  {https://doi.org/10.1103/PhysRevB.98.155134} {\bibfield  {journal} {\bibinfo
  {journal} {Phys. Rev. B}\ }\textbf {\bibinfo {volume} {98}},\ \bibinfo
  {pages} {155134} (\bibinfo {year} {2018})}\BibitemShut {NoStop}%
\bibitem [{\citenamefont {Fendley}\ \emph {et~al.}(2004)\citenamefont
  {Fendley}, \citenamefont {Sengupta},\ and\ \citenamefont
  {Sachdev}}]{PhysRevB.69.075106}%
  \BibitemOpen
  \bibfield  {author} {\bibinfo {author} {\bibfnamefont {P.}~\bibnamefont
  {Fendley}}, \bibinfo {author} {\bibfnamefont {K.}~\bibnamefont {Sengupta}},\
  and\ \bibinfo {author} {\bibfnamefont {S.}~\bibnamefont {Sachdev}},\
  }\bibfield  {title} {\bibinfo {title} {Competing density-wave orders in a
  one-dimensional hard-boson model},\ }\href
  {https://doi.org/10.1103/PhysRevB.69.075106} {\bibfield  {journal} {\bibinfo
  {journal} {Phys. Rev. B}\ }\textbf {\bibinfo {volume} {69}},\ \bibinfo
  {pages} {075106} (\bibinfo {year} {2004})}\BibitemShut {NoStop}%
\bibitem [{\citenamefont {Lesanovsky}\ and\ \citenamefont
  {Katsura}(2012)}]{Lesanovsky_Katsura_PRA2012}%
  \BibitemOpen
  \bibfield  {author} {\bibinfo {author} {\bibfnamefont {I.}~\bibnamefont
  {Lesanovsky}}\ and\ \bibinfo {author} {\bibfnamefont {H.}~\bibnamefont
  {Katsura}},\ }\bibfield  {title} {\bibinfo {title} {Interacting {F}ibonacci
  anyons in a {R}ydberg gas},\ }\href
  {https://doi.org/10.1103/PhysRevA.86.041601} {\bibfield  {journal} {\bibinfo
  {journal} {Phys. Rev. A}\ }\textbf {\bibinfo {volume} {86}},\ \bibinfo
  {pages} {041601} (\bibinfo {year} {2012})}\BibitemShut {NoStop}%
\bibitem [{\citenamefont {Bernien}\ \emph {et~al.}(2017)\citenamefont
  {Bernien}, \citenamefont {Schwartz}, \citenamefont {Keesling}, \citenamefont
  {Levine}, \citenamefont {Omran}, \citenamefont {Pichler}, \citenamefont
  {Choi}, \citenamefont {Zibrov}, \citenamefont {Endres}, \citenamefont
  {Greiner}, \citenamefont {Vuleti{\'c}},\ and\ \citenamefont
  {Lukin}}]{bernien2017probing}%
  \BibitemOpen
  \bibfield  {author} {\bibinfo {author} {\bibfnamefont {H.}~\bibnamefont
  {Bernien}}, \bibinfo {author} {\bibfnamefont {S.}~\bibnamefont {Schwartz}},
  \bibinfo {author} {\bibfnamefont {A.}~\bibnamefont {Keesling}}, \bibinfo
  {author} {\bibfnamefont {H.}~\bibnamefont {Levine}}, \bibinfo {author}
  {\bibfnamefont {A.}~\bibnamefont {Omran}}, \bibinfo {author} {\bibfnamefont
  {H.}~\bibnamefont {Pichler}}, \bibinfo {author} {\bibfnamefont
  {S.}~\bibnamefont {Choi}}, \bibinfo {author} {\bibfnamefont {A.~S.}\
  \bibnamefont {Zibrov}}, \bibinfo {author} {\bibfnamefont {M.}~\bibnamefont
  {Endres}}, \bibinfo {author} {\bibfnamefont {M.}~\bibnamefont {Greiner}},
  \bibinfo {author} {\bibfnamefont {V.}~\bibnamefont {Vuleti{\'c}}},\ and\
  \bibinfo {author} {\bibfnamefont {M.~D.}\ \bibnamefont {Lukin}},\ }\bibfield
  {title} {\bibinfo {title} {Probing many-body dynamics on a 51-atom quantum
  simulator},\ }\href {https://doi.org/10.1038/nature24622} {\bibfield
  {journal} {\bibinfo  {journal} {Nature}\ }\textbf {\bibinfo {volume} {551}},\
  \bibinfo {pages} {579} (\bibinfo {year} {2017})}\BibitemShut {NoStop}%
\bibitem [{\citenamefont {Heller}(1984)}]{PhysRevLett.53.1515}%
  \BibitemOpen
  \bibfield  {author} {\bibinfo {author} {\bibfnamefont {E.~J.}\ \bibnamefont
  {Heller}},\ }\bibfield  {title} {\bibinfo {title} {Bound-state eigenfunctions
  of classically chaotic {H}amiltonian systems: Scars of periodic orbits},\
  }\href {https://doi.org/10.1103/PhysRevLett.53.1515} {\bibfield  {journal}
  {\bibinfo  {journal} {Phys. Rev. Lett.}\ }\textbf {\bibinfo {volume} {53}},\
  \bibinfo {pages} {1515} (\bibinfo {year} {1984})}\BibitemShut {NoStop}%
\bibitem [{\citenamefont {Choi}\ \emph {et~al.}(2019)\citenamefont {Choi},
  \citenamefont {Turner}, \citenamefont {Pichler}, \citenamefont {Ho},
  \citenamefont {Michailidis}, \citenamefont {Papi{\'c}}, \citenamefont
  {Serbyn}, \citenamefont {Lukin},\ and\ \citenamefont
  {Abanin}}]{PhysRevLett.122.220603}%
  \BibitemOpen
  \bibfield  {author} {\bibinfo {author} {\bibfnamefont {S.}~\bibnamefont
  {Choi}}, \bibinfo {author} {\bibfnamefont {C.~J.}\ \bibnamefont {Turner}},
  \bibinfo {author} {\bibfnamefont {H.}~\bibnamefont {Pichler}}, \bibinfo
  {author} {\bibfnamefont {W.~W.}\ \bibnamefont {Ho}}, \bibinfo {author}
  {\bibfnamefont {A.~A.}\ \bibnamefont {Michailidis}}, \bibinfo {author}
  {\bibfnamefont {Z.}~\bibnamefont {Papi{\'c}}}, \bibinfo {author}
  {\bibfnamefont {M.}~\bibnamefont {Serbyn}}, \bibinfo {author} {\bibfnamefont
  {M.~D.}\ \bibnamefont {Lukin}},\ and\ \bibinfo {author} {\bibfnamefont
  {D.~A.}\ \bibnamefont {Abanin}},\ }\bibfield  {title} {\bibinfo {title}
  {Emergent su(2) dynamics and perfect quantum many-body scars},\ }\href
  {https://doi.org/10.1103/PhysRevLett.122.220603} {\bibfield  {journal}
  {\bibinfo  {journal} {Phys. Rev. Lett.}\ }\textbf {\bibinfo {volume} {122}},\
  \bibinfo {pages} {220603} (\bibinfo {year} {2019})}\BibitemShut {NoStop}%
\bibitem [{\citenamefont {Ho}\ \emph {et~al.}(2019)\citenamefont {Ho},
  \citenamefont {Choi}, \citenamefont {Pichler},\ and\ \citenamefont
  {Lukin}}]{PhysRevLett.122.040603}%
  \BibitemOpen
  \bibfield  {author} {\bibinfo {author} {\bibfnamefont {W.~W.}\ \bibnamefont
  {Ho}}, \bibinfo {author} {\bibfnamefont {S.}~\bibnamefont {Choi}}, \bibinfo
  {author} {\bibfnamefont {H.}~\bibnamefont {Pichler}},\ and\ \bibinfo {author}
  {\bibfnamefont {M.~D.}\ \bibnamefont {Lukin}},\ }\bibfield  {title} {\bibinfo
  {title} {Periodic orbits, entanglement, and quantum many-body scars in
  constrained models: Matrix product state approach},\ }\href
  {https://doi.org/10.1103/PhysRevLett.122.040603} {\bibfield  {journal}
  {\bibinfo  {journal} {Phys. Rev. Lett.}\ }\textbf {\bibinfo {volume} {122}},\
  \bibinfo {pages} {040603} (\bibinfo {year} {2019})}\BibitemShut {NoStop}%
\bibitem [{\citenamefont {Khemani}\ \emph {et~al.}(2019)\citenamefont
  {Khemani}, \citenamefont {Laumann},\ and\ \citenamefont
  {Chandran}}]{Khemani_Lauman_Chandran_PRB2019}%
  \BibitemOpen
  \bibfield  {author} {\bibinfo {author} {\bibfnamefont {V.}~\bibnamefont
  {Khemani}}, \bibinfo {author} {\bibfnamefont {C.~R.}\ \bibnamefont
  {Laumann}},\ and\ \bibinfo {author} {\bibfnamefont {A.}~\bibnamefont
  {Chandran}},\ }\bibfield  {title} {\bibinfo {title} {Signatures of
  integrability in the dynamics of {R}ydberg-blockaded chains},\ }\href
  {https://doi.org/10.1103/PhysRevB.99.161101} {\bibfield  {journal} {\bibinfo
  {journal} {Phys. Rev. B}\ }\textbf {\bibinfo {volume} {99}},\ \bibinfo
  {pages} {161101} (\bibinfo {year} {2019})}\BibitemShut {NoStop}%
\bibitem [{\citenamefont {Lin}\ and\ \citenamefont
  {Motrunich}(2019)}]{Lin_Motrunich_PRL2019_exactscars}%
  \BibitemOpen
  \bibfield  {author} {\bibinfo {author} {\bibfnamefont {C.-J.}\ \bibnamefont
  {Lin}}\ and\ \bibinfo {author} {\bibfnamefont {O.~I.}\ \bibnamefont
  {Motrunich}},\ }\bibfield  {title} {\bibinfo {title} {Exact quantum many-body
  scar states in the {R}ydberg-blockaded atom chain},\ }\href
  {https://doi.org/10.1103/PhysRevLett.122.173401} {\bibfield  {journal}
  {\bibinfo  {journal} {Phys. Rev. Lett.}\ }\textbf {\bibinfo {volume} {122}},\
  \bibinfo {pages} {173401} (\bibinfo {year} {2019})}\BibitemShut {NoStop}%
\bibitem [{\citenamefont {Iadecola}\ \emph {et~al.}(2019)\citenamefont
  {Iadecola}, \citenamefont {Schecter},\ and\ \citenamefont
  {Xu}}]{Iadecola_Schecter_Xu_PRB2019}%
  \BibitemOpen
  \bibfield  {author} {\bibinfo {author} {\bibfnamefont {T.}~\bibnamefont
  {Iadecola}}, \bibinfo {author} {\bibfnamefont {M.}~\bibnamefont {Schecter}},\
  and\ \bibinfo {author} {\bibfnamefont {S.}~\bibnamefont {Xu}},\ }\bibfield
  {title} {\bibinfo {title} {Quantum many-body scars from magnon
  condensation},\ }\href {https://doi.org/10.1103/PhysRevB.100.184312}
  {\bibfield  {journal} {\bibinfo  {journal} {Phys. Rev. B}\ }\textbf {\bibinfo
  {volume} {100}},\ \bibinfo {pages} {184312} (\bibinfo {year}
  {2019})}\BibitemShut {NoStop}%
\bibitem [{\citenamefont {Shiraishi}(2019)}]{shiraishi2019connection}%
  \BibitemOpen
  \bibfield  {author} {\bibinfo {author} {\bibfnamefont {N.}~\bibnamefont
  {Shiraishi}},\ }\bibfield  {title} {\bibinfo {title} {Connection between
  quantum-many-body scars and the {A}ffleck--{K}ennedy--{L}ieb--{T}asaki model
  from the viewpoint of embedded {H}amiltonians},\ }\href@noop {} {\bibfield
  {journal} {\bibinfo  {journal} {Journal of Statistical Mechanics: Theory and
  Experiment}\ }\textbf {\bibinfo {volume} {2019}},\ \bibinfo {pages} {083103}
  (\bibinfo {year} {2019})}\BibitemShut {NoStop}%
\bibitem [{\citenamefont {{Moudgalya}}\ \emph {et~al.}(2019)\citenamefont
  {{Moudgalya}}, \citenamefont {{Bernevig}},\ and\ \citenamefont
  {{Regnault}}}]{Bernevig_Regnault_arXiv1906_thintorus}%
  \BibitemOpen
  \bibfield  {author} {\bibinfo {author} {\bibfnamefont {S.}~\bibnamefont
  {{Moudgalya}}}, \bibinfo {author} {\bibfnamefont {B.~A.}\ \bibnamefont
  {{Bernevig}}},\ and\ \bibinfo {author} {\bibfnamefont {N.}~\bibnamefont
  {{Regnault}}},\ }\bibfield  {title} {\bibinfo {title} {{Quantum Many-body
  Scars in a Landau Level on a Thin Torus}},\ }\href@noop {} {\bibfield
  {journal} {\bibinfo  {journal} {arXiv e-prints}\ } (\bibinfo {year}
  {2019})},\ \Eprint {https://arxiv.org/abs/1906.05292} {arXiv:1906.05292
  [cond-mat.str-el]} \BibitemShut {NoStop}%
\bibitem [{\citenamefont {Lin}\ \emph {et~al.}(2020{\natexlab{a}})\citenamefont
  {Lin}, \citenamefont {Chandran},\ and\ \citenamefont
  {Motrunich}}]{Lin_Chandran_Motrunich_PRResearch2020}%
  \BibitemOpen
  \bibfield  {author} {\bibinfo {author} {\bibfnamefont {C.-J.}\ \bibnamefont
  {Lin}}, \bibinfo {author} {\bibfnamefont {A.}~\bibnamefont {Chandran}},\ and\
  \bibinfo {author} {\bibfnamefont {O.~I.}\ \bibnamefont {Motrunich}},\
  }\bibfield  {title} {\bibinfo {title} {Slow thermalization of exact quantum
  many-body scar states under perturbations},\ }\href
  {https://doi.org/10.1103/PhysRevResearch.2.033044} {\bibfield  {journal}
  {\bibinfo  {journal} {Phys. Rev. Research}\ }\textbf {\bibinfo {volume}
  {2}},\ \bibinfo {pages} {033044} (\bibinfo {year}
  {2020}{\natexlab{a}})}\BibitemShut {NoStop}%
\bibitem [{\citenamefont {Bull}\ \emph {et~al.}(2020)\citenamefont {Bull},
  \citenamefont {Desaules},\ and\ \citenamefont {Papi\ifmmode~\acute{c}\else
  \'{c}\fi{}}}]{Bull_Desaules_Papic_PRB2020}%
  \BibitemOpen
  \bibfield  {author} {\bibinfo {author} {\bibfnamefont {K.}~\bibnamefont
  {Bull}}, \bibinfo {author} {\bibfnamefont {J.-Y.}\ \bibnamefont {Desaules}},\
  and\ \bibinfo {author} {\bibfnamefont {Z.}~\bibnamefont
  {Papi\ifmmode~\acute{c}\else \'{c}\fi{}}},\ }\bibfield  {title} {\bibinfo
  {title} {Quantum scars as embeddings of weakly broken {L}ie algebra
  representations},\ }\href {https://doi.org/10.1103/PhysRevB.101.165139}
  {\bibfield  {journal} {\bibinfo  {journal} {Phys. Rev. B}\ }\textbf {\bibinfo
  {volume} {101}},\ \bibinfo {pages} {165139} (\bibinfo {year}
  {2020})}\BibitemShut {NoStop}%
\bibitem [{\citenamefont {Michailidis}\ \emph
  {et~al.}(2020{\natexlab{a}})\citenamefont {Michailidis}, \citenamefont
  {Turner}, \citenamefont {Papi\ifmmode~\acute{c}\else \'{c}\fi{}},
  \citenamefont {Abanin},\ and\ \citenamefont
  {Serbyn}}]{Papic_Abanin_PRX2020_slow}%
  \BibitemOpen
  \bibfield  {author} {\bibinfo {author} {\bibfnamefont {A.~A.}\ \bibnamefont
  {Michailidis}}, \bibinfo {author} {\bibfnamefont {C.~J.}\ \bibnamefont
  {Turner}}, \bibinfo {author} {\bibfnamefont {Z.}~\bibnamefont
  {Papi\ifmmode~\acute{c}\else \'{c}\fi{}}}, \bibinfo {author} {\bibfnamefont
  {D.~A.}\ \bibnamefont {Abanin}},\ and\ \bibinfo {author} {\bibfnamefont
  {M.}~\bibnamefont {Serbyn}},\ }\bibfield  {title} {\bibinfo {title} {Slow
  quantum thermalization and many-body revivals from mixed phase space},\
  }\href {https://doi.org/10.1103/PhysRevX.10.011055} {\bibfield  {journal}
  {\bibinfo  {journal} {Phys. Rev. X}\ }\textbf {\bibinfo {volume} {10}},\
  \bibinfo {pages} {011055} (\bibinfo {year} {2020}{\natexlab{a}})}\BibitemShut
  {NoStop}%
\bibitem [{\citenamefont {Turner}\ \emph {et~al.}(2020)\citenamefont {Turner},
  \citenamefont {Desaules}, \citenamefont {Bull},\ and\ \citenamefont
  {Papić}}]{turner2020correspondence}%
  \BibitemOpen
  \bibfield  {author} {\bibinfo {author} {\bibfnamefont {C.~J.}\ \bibnamefont
  {Turner}}, \bibinfo {author} {\bibfnamefont {J.-Y.}\ \bibnamefont
  {Desaules}}, \bibinfo {author} {\bibfnamefont {K.}~\bibnamefont {Bull}},\
  and\ \bibinfo {author} {\bibfnamefont {Z.}~\bibnamefont {Papić}},\
  }\href@noop {} {\bibinfo {title} {Correspondence principle for many-body
  scars in ultracold {R}ydberg atoms}} (\bibinfo {year} {2020}),\ \Eprint
  {https://arxiv.org/abs/2006.13207} {arXiv:2006.13207 [quant-ph]} \BibitemShut
  {NoStop}%
\bibitem [{\citenamefont {Moudgalya}\ \emph
  {et~al.}(2018{\natexlab{a}})\citenamefont {Moudgalya}, \citenamefont
  {Regnault},\ and\ \citenamefont {Bernevig}}]{PhysRevB.98.235156}%
  \BibitemOpen
  \bibfield  {author} {\bibinfo {author} {\bibfnamefont {S.}~\bibnamefont
  {Moudgalya}}, \bibinfo {author} {\bibfnamefont {N.}~\bibnamefont
  {Regnault}},\ and\ \bibinfo {author} {\bibfnamefont {B.~A.}\ \bibnamefont
  {Bernevig}},\ }\bibfield  {title} {\bibinfo {title} {Entanglement of exact
  excited states of {A}ffleck-{K}ennedy-{L}ieb-{T}asaki models: Exact results,
  many-body scars, and violation of the strong eigenstate thermalization
  hypothesis},\ }\href {https://doi.org/10.1103/PhysRevB.98.235156} {\bibfield
  {journal} {\bibinfo  {journal} {Phys. Rev. B}\ }\textbf {\bibinfo {volume}
  {98}},\ \bibinfo {pages} {235156} (\bibinfo {year}
  {2018}{\natexlab{a}})}\BibitemShut {NoStop}%
\bibitem [{\citenamefont {Moudgalya}\ \emph
  {et~al.}(2018{\natexlab{b}})\citenamefont {Moudgalya}, \citenamefont
  {Rachel}, \citenamefont {Bernevig},\ and\ \citenamefont
  {Regnault}}]{PhysRevB.98.235155}%
  \BibitemOpen
  \bibfield  {author} {\bibinfo {author} {\bibfnamefont {S.}~\bibnamefont
  {Moudgalya}}, \bibinfo {author} {\bibfnamefont {S.}~\bibnamefont {Rachel}},
  \bibinfo {author} {\bibfnamefont {B.~A.}\ \bibnamefont {Bernevig}},\ and\
  \bibinfo {author} {\bibfnamefont {N.}~\bibnamefont {Regnault}},\ }\bibfield
  {title} {\bibinfo {title} {Exact excited states of nonintegrable models},\
  }\href {https://doi.org/10.1103/PhysRevB.98.235155} {\bibfield  {journal}
  {\bibinfo  {journal} {Phys. Rev. B}\ }\textbf {\bibinfo {volume} {98}},\
  \bibinfo {pages} {235155} (\bibinfo {year} {2018}{\natexlab{b}})}\BibitemShut
  {NoStop}%
\bibitem [{\citenamefont {Mark}\ \emph {et~al.}(2020)\citenamefont {Mark},
  \citenamefont {Lin},\ and\ \citenamefont
  {Motrunich}}]{Mark_Lin_Motrunich_PRB2020}%
  \BibitemOpen
  \bibfield  {author} {\bibinfo {author} {\bibfnamefont {D.~K.}\ \bibnamefont
  {Mark}}, \bibinfo {author} {\bibfnamefont {C.-J.}\ \bibnamefont {Lin}},\ and\
  \bibinfo {author} {\bibfnamefont {O.~I.}\ \bibnamefont {Motrunich}},\
  }\bibfield  {title} {\bibinfo {title} {Unified structure for exact towers of
  scar states in the {A}ffleck-{K}ennedy-{L}ieb-{T}asaki and other models},\
  }\href {https://doi.org/10.1103/PhysRevB.101.195131} {\bibfield  {journal}
  {\bibinfo  {journal} {Phys. Rev. B}\ }\textbf {\bibinfo {volume} {101}},\
  \bibinfo {pages} {195131} (\bibinfo {year} {2020})}\BibitemShut {NoStop}%
\bibitem [{\citenamefont {Moudgalya}\ \emph
  {et~al.}(2020{\natexlab{a}})\citenamefont {Moudgalya}, \citenamefont
  {O'Brien}, \citenamefont {Bernevig}, \citenamefont {Fendley},\ and\
  \citenamefont {Regnault}}]{Moudgalya_Bernevig_Fendley_Regnault_2020}%
  \BibitemOpen
  \bibfield  {author} {\bibinfo {author} {\bibfnamefont {S.}~\bibnamefont
  {Moudgalya}}, \bibinfo {author} {\bibfnamefont {E.}~\bibnamefont {O'Brien}},
  \bibinfo {author} {\bibfnamefont {B.~A.}\ \bibnamefont {Bernevig}}, \bibinfo
  {author} {\bibfnamefont {P.}~\bibnamefont {Fendley}},\ and\ \bibinfo {author}
  {\bibfnamefont {N.}~\bibnamefont {Regnault}},\ }\bibfield  {title} {\bibinfo
  {title} {Large classes of quantum scarred {H}amiltonians from matrix product
  states},\ }\href@noop {} {\bibfield  {journal} {\bibinfo  {journal} {arXiv
  preprint arXiv:2002.11725}\ } (\bibinfo {year}
  {2020}{\natexlab{a}})}\BibitemShut {NoStop}%
\bibitem [{\citenamefont {James}\ \emph {et~al.}(2019)\citenamefont {James},
  \citenamefont {Konik},\ and\ \citenamefont
  {Robinson}}]{PhysRevLett.122.130603}%
  \BibitemOpen
  \bibfield  {author} {\bibinfo {author} {\bibfnamefont {A.~J.~A.}\
  \bibnamefont {James}}, \bibinfo {author} {\bibfnamefont {R.~M.}\ \bibnamefont
  {Konik}},\ and\ \bibinfo {author} {\bibfnamefont {N.~J.}\ \bibnamefont
  {Robinson}},\ }\bibfield  {title} {\bibinfo {title} {Nonthermal states
  arising from confinement in one and two dimensions},\ }\href
  {https://doi.org/10.1103/PhysRevLett.122.130603} {\bibfield  {journal}
  {\bibinfo  {journal} {Phys. Rev. Lett.}\ }\textbf {\bibinfo {volume} {122}},\
  \bibinfo {pages} {130603} (\bibinfo {year} {2019})}\BibitemShut {NoStop}%
\bibitem [{\citenamefont {Robinson}\ \emph {et~al.}(2019)\citenamefont
  {Robinson}, \citenamefont {James},\ and\ \citenamefont
  {Konik}}]{PhysRevB.99.195108}%
  \BibitemOpen
  \bibfield  {author} {\bibinfo {author} {\bibfnamefont {N.~J.}\ \bibnamefont
  {Robinson}}, \bibinfo {author} {\bibfnamefont {A.~J.~A.}\ \bibnamefont
  {James}},\ and\ \bibinfo {author} {\bibfnamefont {R.~M.}\ \bibnamefont
  {Konik}},\ }\bibfield  {title} {\bibinfo {title} {Signatures of rare states
  and thermalization in a theory with confinement},\ }\href
  {https://doi.org/10.1103/PhysRevB.99.195108} {\bibfield  {journal} {\bibinfo
  {journal} {Phys. Rev. B}\ }\textbf {\bibinfo {volume} {99}},\ \bibinfo
  {pages} {195108} (\bibinfo {year} {2019})}\BibitemShut {NoStop}%
\bibitem [{\citenamefont {Nachtergaele}\ \emph {et~al.}(2020)\citenamefont
  {Nachtergaele}, \citenamefont {Warzel},\ and\ \citenamefont
  {Young}}]{Nachtergaele_Warzel_arxiv2020_thintorus}%
  \BibitemOpen
  \bibfield  {author} {\bibinfo {author} {\bibfnamefont {B.}~\bibnamefont
  {Nachtergaele}}, \bibinfo {author} {\bibfnamefont {S.}~\bibnamefont
  {Warzel}},\ and\ \bibinfo {author} {\bibfnamefont {A.}~\bibnamefont
  {Young}},\ }\bibfield  {title} {\bibinfo {title} {Low-complexity eigenstates
  of a $\nu= 1/3$ fractional quantum hall system},\ }\href@noop {} {\bibfield
  {journal} {\bibinfo  {journal} {arXiv preprint arXiv:2006.00300}\ } (\bibinfo
  {year} {2020})}\BibitemShut {NoStop}%
\bibitem [{\citenamefont {{Vafek}}\ \emph {et~al.}(2017)\citenamefont
  {{Vafek}}, \citenamefont {{Regnault}},\ and\ \citenamefont
  {{Bernevig}}}]{2017ScPP....3...43V}%
  \BibitemOpen
  \bibfield  {author} {\bibinfo {author} {\bibfnamefont {O.}~\bibnamefont
  {{Vafek}}}, \bibinfo {author} {\bibfnamefont {N.}~\bibnamefont
  {{Regnault}}},\ and\ \bibinfo {author} {\bibfnamefont {B.~A.}\ \bibnamefont
  {{Bernevig}}},\ }\bibfield  {title} {\bibinfo {title} {{Entanglement of exact
  excited eigenstates of the {H}ubbard model in arbitrary dimension}},\ }\href
  {https://doi.org/10.21468/SciPostPhys.3.6.043} {\bibfield  {journal}
  {\bibinfo  {journal} {SciPost Physics}\ }\textbf {\bibinfo {volume} {3}},\
  \bibinfo {eid} {043} (\bibinfo {year} {2017})}\BibitemShut {NoStop}%
\bibitem [{\citenamefont {Iadecola}\ and\ \citenamefont {\ifmmode
  \check{Z}\else \v{Z}\fi{}nidari\ifmmode~\check{c}\else
  \v{c}\fi{}}(2019)}]{PhysRevLett.123.036403}%
  \BibitemOpen
  \bibfield  {author} {\bibinfo {author} {\bibfnamefont {T.}~\bibnamefont
  {Iadecola}}\ and\ \bibinfo {author} {\bibfnamefont {M.}~\bibnamefont
  {\ifmmode \check{Z}\else \v{Z}\fi{}nidari\ifmmode~\check{c}\else
  \v{c}\fi{}}},\ }\bibfield  {title} {\bibinfo {title} {Exact localized and
  ballistic eigenstates in disordered chaotic spin ladders and the
  {Fermi-Hubbard} model},\ }\href
  {https://doi.org/10.1103/PhysRevLett.123.036403} {\bibfield  {journal}
  {\bibinfo  {journal} {Phys. Rev. Lett.}\ }\textbf {\bibinfo {volume} {123}},\
  \bibinfo {pages} {036403} (\bibinfo {year} {2019})}\BibitemShut {NoStop}%
\bibitem [{\citenamefont {Mark}\ and\ \citenamefont
  {Motrunich}(2020)}]{Mark_Motrunich_2020_etapairing}%
  \BibitemOpen
  \bibfield  {author} {\bibinfo {author} {\bibfnamefont {D.~K.}\ \bibnamefont
  {Mark}}\ and\ \bibinfo {author} {\bibfnamefont {O.~I.}\ \bibnamefont
  {Motrunich}},\ }\bibfield  {title} {\bibinfo {title} {Eta-pairing states as
  true scars in an extended {H}ubbard model},\ }\href@noop {} {\bibfield
  {journal} {\bibinfo  {journal} {arXiv preprint arXiv:2004.13800}\ } (\bibinfo
  {year} {2020})}\BibitemShut {NoStop}%
\bibitem [{\citenamefont {Moudgalya}\ \emph
  {et~al.}(2020{\natexlab{b}})\citenamefont {Moudgalya}, \citenamefont
  {Regnault},\ and\ \citenamefont
  {Bernevig}}]{Moudgalya_Regnault_Bernevig_2020_etapairing}%
  \BibitemOpen
  \bibfield  {author} {\bibinfo {author} {\bibfnamefont {S.}~\bibnamefont
  {Moudgalya}}, \bibinfo {author} {\bibfnamefont {N.}~\bibnamefont
  {Regnault}},\ and\ \bibinfo {author} {\bibfnamefont {B.~A.}\ \bibnamefont
  {Bernevig}},\ }\bibfield  {title} {\bibinfo {title} {Eta-pairing in {H}ubbard
  models: From spectrum generating algebras to quantum many-body scars},\
  }\href@noop {} {\bibfield  {journal} {\bibinfo  {journal} {arXiv preprint
  arXiv:2004.13727}\ } (\bibinfo {year} {2020}{\natexlab{b}})}\BibitemShut
  {NoStop}%
\bibitem [{\citenamefont {{Schecter}}\ and\ \citenamefont
  {{Iadecola}}(2019)}]{Schecter_Iadecola_PRL2019_xy}%
  \BibitemOpen
  \bibfield  {author} {\bibinfo {author} {\bibfnamefont {M.}~\bibnamefont
  {{Schecter}}}\ and\ \bibinfo {author} {\bibfnamefont {T.}~\bibnamefont
  {{Iadecola}}},\ }\bibfield  {title} {\bibinfo {title} {{Weak Ergodicity
  Breaking and Quantum Many-Body Scars in Spin-1 X Y Magnets}},\ }\href
  {https://doi.org/10.1103/PhysRevLett.123.147201} {\bibfield  {journal}
  {\bibinfo  {journal} {\prl}\ }\textbf {\bibinfo {volume} {123}},\ \bibinfo
  {eid} {147201} (\bibinfo {year} {2019})}\BibitemShut {NoStop}%
\bibitem [{\citenamefont {Chattopadhyay}\ \emph {et~al.}(2020)\citenamefont
  {Chattopadhyay}, \citenamefont {Pichler}, \citenamefont {Lukin},\ and\
  \citenamefont {Ho}}]{Pichler_Lukin_Ho_PRB2020_xy}%
  \BibitemOpen
  \bibfield  {author} {\bibinfo {author} {\bibfnamefont {S.}~\bibnamefont
  {Chattopadhyay}}, \bibinfo {author} {\bibfnamefont {H.}~\bibnamefont
  {Pichler}}, \bibinfo {author} {\bibfnamefont {M.~D.}\ \bibnamefont {Lukin}},\
  and\ \bibinfo {author} {\bibfnamefont {W.~W.}\ \bibnamefont {Ho}},\
  }\bibfield  {title} {\bibinfo {title} {Quantum many-body scars from virtual
  entangled pairs},\ }\href {https://doi.org/10.1103/PhysRevB.101.174308}
  {\bibfield  {journal} {\bibinfo  {journal} {Phys. Rev. B}\ }\textbf {\bibinfo
  {volume} {101}},\ \bibinfo {pages} {174308} (\bibinfo {year}
  {2020})}\BibitemShut {NoStop}%
\bibitem [{\citenamefont {Mukherjee}\ \emph
  {et~al.}(2020{\natexlab{a}})\citenamefont {Mukherjee}, \citenamefont {Nandy},
  \citenamefont {Sen}, \citenamefont {Sen},\ and\ \citenamefont
  {Sengupta}}]{PhysRevB.101.245107}%
  \BibitemOpen
  \bibfield  {author} {\bibinfo {author} {\bibfnamefont {B.}~\bibnamefont
  {Mukherjee}}, \bibinfo {author} {\bibfnamefont {S.}~\bibnamefont {Nandy}},
  \bibinfo {author} {\bibfnamefont {A.}~\bibnamefont {Sen}}, \bibinfo {author}
  {\bibfnamefont {D.}~\bibnamefont {Sen}},\ and\ \bibinfo {author}
  {\bibfnamefont {K.}~\bibnamefont {Sengupta}},\ }\bibfield  {title} {\bibinfo
  {title} {Collapse and revival of quantum many-body scars via {F}loquet
  engineering},\ }\href {https://doi.org/10.1103/PhysRevB.101.245107}
  {\bibfield  {journal} {\bibinfo  {journal} {Phys. Rev. B}\ }\textbf {\bibinfo
  {volume} {101}},\ \bibinfo {pages} {245107} (\bibinfo {year}
  {2020}{\natexlab{a}})}\BibitemShut {NoStop}%
\bibitem [{\citenamefont {Sugiura}\ \emph {et~al.}(2019)\citenamefont
  {Sugiura}, \citenamefont {Kuwahara},\ and\ \citenamefont
  {Saito}}]{sugiura2019manybody}%
  \BibitemOpen
  \bibfield  {author} {\bibinfo {author} {\bibfnamefont {S.}~\bibnamefont
  {Sugiura}}, \bibinfo {author} {\bibfnamefont {T.}~\bibnamefont {Kuwahara}},\
  and\ \bibinfo {author} {\bibfnamefont {K.}~\bibnamefont {Saito}},\
  }\href@noop {} {\bibinfo {title} {Many-body scar state intrinsic to
  periodically driven system: Rigorous results}} (\bibinfo {year} {2019}),\
  \Eprint {https://arxiv.org/abs/1911.06092} {arXiv:1911.06092
  [cond-mat.stat-mech]} \BibitemShut {NoStop}%
\bibitem [{\citenamefont {Pai}\ and\ \citenamefont
  {Pretko}(2019)}]{Pai_Pretko_PRL2019}%
  \BibitemOpen
  \bibfield  {author} {\bibinfo {author} {\bibfnamefont {S.}~\bibnamefont
  {Pai}}\ and\ \bibinfo {author} {\bibfnamefont {M.}~\bibnamefont {Pretko}},\
  }\bibfield  {title} {\bibinfo {title} {Dynamical scar states in driven
  fracton systems},\ }\href {https://doi.org/10.1103/PhysRevLett.123.136401}
  {\bibfield  {journal} {\bibinfo  {journal} {Phys. Rev. Lett.}\ }\textbf
  {\bibinfo {volume} {123}},\ \bibinfo {pages} {136401} (\bibinfo {year}
  {2019})}\BibitemShut {NoStop}%
\bibitem [{\citenamefont {Mukherjee}\ \emph
  {et~al.}(2020{\natexlab{b}})\citenamefont {Mukherjee}, \citenamefont {Sen},
  \citenamefont {Sen},\ and\ \citenamefont {Sengupta}}]{mukherjee2020dynamics}%
  \BibitemOpen
  \bibfield  {author} {\bibinfo {author} {\bibfnamefont {B.}~\bibnamefont
  {Mukherjee}}, \bibinfo {author} {\bibfnamefont {A.}~\bibnamefont {Sen}},
  \bibinfo {author} {\bibfnamefont {D.}~\bibnamefont {Sen}},\ and\ \bibinfo
  {author} {\bibfnamefont {K.}~\bibnamefont {Sengupta}},\ }\bibfield  {title}
  {\bibinfo {title} {Dynamics of the vacuum state in a periodically driven
  {R}ydberg chain},\ }\href {https://doi.org/10.1103/PhysRevB.102.075123}
  {\bibfield  {journal} {\bibinfo  {journal} {Phys. Rev. B}\ }\textbf {\bibinfo
  {volume} {102}},\ \bibinfo {pages} {075123} (\bibinfo {year}
  {2020}{\natexlab{b}})}\BibitemShut {NoStop}%
\bibitem [{\citenamefont {Zhao}\ \emph {et~al.}(2020)\citenamefont {Zhao},
  \citenamefont {Vovrosh}, \citenamefont {Mintert},\ and\ \citenamefont
  {Knolle}}]{Mintert_Knolle_PRL2020}%
  \BibitemOpen
  \bibfield  {author} {\bibinfo {author} {\bibfnamefont {H.}~\bibnamefont
  {Zhao}}, \bibinfo {author} {\bibfnamefont {J.}~\bibnamefont {Vovrosh}},
  \bibinfo {author} {\bibfnamefont {F.}~\bibnamefont {Mintert}},\ and\ \bibinfo
  {author} {\bibfnamefont {J.}~\bibnamefont {Knolle}},\ }\bibfield  {title}
  {\bibinfo {title} {Quantum many-body scars in optical lattices},\ }\href
  {https://doi.org/10.1103/PhysRevLett.124.160604} {\bibfield  {journal}
  {\bibinfo  {journal} {Phys. Rev. Lett.}\ }\textbf {\bibinfo {volume} {124}},\
  \bibinfo {pages} {160604} (\bibinfo {year} {2020})}\BibitemShut {NoStop}%
\bibitem [{\citenamefont {Mizuta}\ \emph {et~al.}(2020)\citenamefont {Mizuta},
  \citenamefont {Takasan},\ and\ \citenamefont {Kawakami}}]{mizuta2020exact}%
  \BibitemOpen
  \bibfield  {author} {\bibinfo {author} {\bibfnamefont {K.}~\bibnamefont
  {Mizuta}}, \bibinfo {author} {\bibfnamefont {K.}~\bibnamefont {Takasan}},\
  and\ \bibinfo {author} {\bibfnamefont {N.}~\bibnamefont {Kawakami}},\
  }\href@noop {} {\bibinfo {title} {Exact floquet quantum many-body scars under
  {R}ydberg blockade}} (\bibinfo {year} {2020}),\ \Eprint
  {https://arxiv.org/abs/2004.04431} {arXiv:2004.04431 [cond-mat.stat-mech]}
  \BibitemShut {NoStop}%
\bibitem [{\citenamefont {Ok}\ \emph {et~al.}(2019)\citenamefont {Ok},
  \citenamefont {Choo}, \citenamefont {Mudry}, \citenamefont {Castelnovo},
  \citenamefont {Chamon},\ and\ \citenamefont
  {Neupert}}]{PhysRevResearch.1.033144}%
  \BibitemOpen
  \bibfield  {author} {\bibinfo {author} {\bibfnamefont {S.}~\bibnamefont
  {Ok}}, \bibinfo {author} {\bibfnamefont {K.}~\bibnamefont {Choo}}, \bibinfo
  {author} {\bibfnamefont {C.}~\bibnamefont {Mudry}}, \bibinfo {author}
  {\bibfnamefont {C.}~\bibnamefont {Castelnovo}}, \bibinfo {author}
  {\bibfnamefont {C.}~\bibnamefont {Chamon}},\ and\ \bibinfo {author}
  {\bibfnamefont {T.}~\bibnamefont {Neupert}},\ }\bibfield  {title} {\bibinfo
  {title} {Topological many-body scar states in dimensions one, two, and
  three},\ }\href {https://doi.org/10.1103/PhysRevResearch.1.033144} {\bibfield
   {journal} {\bibinfo  {journal} {Phys. Rev. Research}\ }\textbf {\bibinfo
  {volume} {1}},\ \bibinfo {pages} {033144} (\bibinfo {year}
  {2019})}\BibitemShut {NoStop}%
\bibitem [{\citenamefont {Khemani}\ \emph {et~al.}(2020)\citenamefont
  {Khemani}, \citenamefont {Hermele},\ and\ \citenamefont
  {Nandkishore}}]{PhysRevB.101.174204}%
  \BibitemOpen
  \bibfield  {author} {\bibinfo {author} {\bibfnamefont {V.}~\bibnamefont
  {Khemani}}, \bibinfo {author} {\bibfnamefont {M.}~\bibnamefont {Hermele}},\
  and\ \bibinfo {author} {\bibfnamefont {R.}~\bibnamefont {Nandkishore}},\
  }\bibfield  {title} {\bibinfo {title} {Localization from {H}ilbert space
  shattering: From theory to physical realizations},\ }\href
  {https://doi.org/10.1103/PhysRevB.101.174204} {\bibfield  {journal} {\bibinfo
   {journal} {Phys. Rev. B}\ }\textbf {\bibinfo {volume} {101}},\ \bibinfo
  {pages} {174204} (\bibinfo {year} {2020})}\BibitemShut {NoStop}%
\bibitem [{\citenamefont {Michailidis}\ \emph
  {et~al.}(2020{\natexlab{b}})\citenamefont {Michailidis}, \citenamefont
  {Turner}, \citenamefont {Papi\ifmmode~\acute{c}\else \'{c}\fi{}},
  \citenamefont {Abanin},\ and\ \citenamefont
  {Serbyn}}]{Papic_Abanin_Serbyn_PRResearch2020}%
  \BibitemOpen
  \bibfield  {author} {\bibinfo {author} {\bibfnamefont {A.~A.}\ \bibnamefont
  {Michailidis}}, \bibinfo {author} {\bibfnamefont {C.~J.}\ \bibnamefont
  {Turner}}, \bibinfo {author} {\bibfnamefont {Z.}~\bibnamefont
  {Papi\ifmmode~\acute{c}\else \'{c}\fi{}}}, \bibinfo {author} {\bibfnamefont
  {D.~A.}\ \bibnamefont {Abanin}},\ and\ \bibinfo {author} {\bibfnamefont
  {M.}~\bibnamefont {Serbyn}},\ }\bibfield  {title} {\bibinfo {title}
  {Stabilizing two-dimensional quantum scars by deformation and
  synchronization},\ }\href {https://doi.org/10.1103/PhysRevResearch.2.022065}
  {\bibfield  {journal} {\bibinfo  {journal} {Phys. Rev. Research}\ }\textbf
  {\bibinfo {volume} {2}},\ \bibinfo {pages} {022065} (\bibinfo {year}
  {2020}{\natexlab{b}})}\BibitemShut {NoStop}%
\bibitem [{\citenamefont {Lin}\ \emph {et~al.}(2020{\natexlab{b}})\citenamefont
  {Lin}, \citenamefont {Calvera},\ and\ \citenamefont
  {Hsieh}}]{Lin_Calvera_Hsieh_PRB2020}%
  \BibitemOpen
  \bibfield  {author} {\bibinfo {author} {\bibfnamefont {C.-J.}\ \bibnamefont
  {Lin}}, \bibinfo {author} {\bibfnamefont {V.}~\bibnamefont {Calvera}},\ and\
  \bibinfo {author} {\bibfnamefont {T.~H.}\ \bibnamefont {Hsieh}},\ }\bibfield
  {title} {\bibinfo {title} {Quantum many-body scar states in two-dimensional
  {R}ydberg atom arrays},\ }\href {https://doi.org/10.1103/PhysRevB.101.220304}
  {\bibfield  {journal} {\bibinfo  {journal} {Phys. Rev. B}\ }\textbf {\bibinfo
  {volume} {101}},\ \bibinfo {pages} {220304} (\bibinfo {year}
  {2020}{\natexlab{b}})}\BibitemShut {NoStop}%
\bibitem [{\citenamefont {Hudomal}\ \emph {et~al.}(2020)\citenamefont
  {Hudomal}, \citenamefont {Vasić}, \citenamefont {Regnault},\ and\
  \citenamefont {Papić}}]{Hudomal_2020}%
  \BibitemOpen
  \bibfield  {author} {\bibinfo {author} {\bibfnamefont {A.}~\bibnamefont
  {Hudomal}}, \bibinfo {author} {\bibfnamefont {I.}~\bibnamefont {Vasić}},
  \bibinfo {author} {\bibfnamefont {N.}~\bibnamefont {Regnault}},\ and\
  \bibinfo {author} {\bibfnamefont {Z.}~\bibnamefont {Papić}},\ }\bibfield
  {title} {\bibinfo {title} {Quantum scars of bosons with correlated hopping},\
  }\bibfield  {journal} {\bibinfo  {journal} {Communications Physics}\ }\textbf
  {\bibinfo {volume} {3}},\ \href {https://doi.org/10.1038/s42005-020-0364-9}
  {10.1038/s42005-020-0364-9} (\bibinfo {year} {2020})\BibitemShut {NoStop}%
\bibitem [{\citenamefont {Lee}\ \emph {et~al.}(2020)\citenamefont {Lee},
  \citenamefont {Melendrez}, \citenamefont {Pal},\ and\ \citenamefont
  {Changlani}}]{PhysRevB.101.241111}%
  \BibitemOpen
  \bibfield  {author} {\bibinfo {author} {\bibfnamefont {K.}~\bibnamefont
  {Lee}}, \bibinfo {author} {\bibfnamefont {R.}~\bibnamefont {Melendrez}},
  \bibinfo {author} {\bibfnamefont {A.}~\bibnamefont {Pal}},\ and\ \bibinfo
  {author} {\bibfnamefont {H.~J.}\ \bibnamefont {Changlani}},\ }\bibfield
  {title} {\bibinfo {title} {Exact three-colored quantum scars from geometric
  frustration},\ }\href {https://doi.org/10.1103/PhysRevB.101.241111}
  {\bibfield  {journal} {\bibinfo  {journal} {Phys. Rev. B}\ }\textbf {\bibinfo
  {volume} {101}},\ \bibinfo {pages} {241111} (\bibinfo {year}
  {2020})}\BibitemShut {NoStop}%
\bibitem [{\citenamefont {Bull}\ \emph {et~al.}(2019)\citenamefont {Bull},
  \citenamefont {Martin},\ and\ \citenamefont {Papi\ifmmode~\acute{c}\else
  \'{c}\fi{}}}]{PhysRevLett.123.030601}%
  \BibitemOpen
  \bibfield  {author} {\bibinfo {author} {\bibfnamefont {K.}~\bibnamefont
  {Bull}}, \bibinfo {author} {\bibfnamefont {I.}~\bibnamefont {Martin}},\ and\
  \bibinfo {author} {\bibfnamefont {Z.}~\bibnamefont
  {Papi\ifmmode~\acute{c}\else \'{c}\fi{}}},\ }\bibfield  {title} {\bibinfo
  {title} {Systematic construction of scarred many-body dynamics in 1d lattice
  models},\ }\href {https://doi.org/10.1103/PhysRevLett.123.030601} {\bibfield
  {journal} {\bibinfo  {journal} {Phys. Rev. Lett.}\ }\textbf {\bibinfo
  {volume} {123}},\ \bibinfo {pages} {030601} (\bibinfo {year}
  {2019})}\BibitemShut {NoStop}%
\bibitem [{\citenamefont {Shiraishi}\ and\ \citenamefont
  {Mori}(2017)}]{PhysRevLett.119.030601}%
  \BibitemOpen
  \bibfield  {author} {\bibinfo {author} {\bibfnamefont {N.}~\bibnamefont
  {Shiraishi}}\ and\ \bibinfo {author} {\bibfnamefont {T.}~\bibnamefont
  {Mori}},\ }\bibfield  {title} {\bibinfo {title} {Systematic construction of
  counterexamples to the eigenstate thermalization hypothesis},\ }\href
  {https://doi.org/10.1103/PhysRevLett.119.030601} {\bibfield  {journal}
  {\bibinfo  {journal} {Phys. Rev. Lett.}\ }\textbf {\bibinfo {volume} {119}},\
  \bibinfo {pages} {030601} (\bibinfo {year} {2017})}\BibitemShut {NoStop}%
\bibitem [{\citenamefont {Shibata}\ \emph {et~al.}(2020)\citenamefont
  {Shibata}, \citenamefont {Yoshioka},\ and\ \citenamefont
  {Katsura}}]{PhysRevLett.124.180604}%
  \BibitemOpen
  \bibfield  {author} {\bibinfo {author} {\bibfnamefont {N.}~\bibnamefont
  {Shibata}}, \bibinfo {author} {\bibfnamefont {N.}~\bibnamefont {Yoshioka}},\
  and\ \bibinfo {author} {\bibfnamefont {H.}~\bibnamefont {Katsura}},\
  }\bibfield  {title} {\bibinfo {title} {Onsager's scars in disordered spin
  chains},\ }\href {https://doi.org/10.1103/PhysRevLett.124.180604} {\bibfield
  {journal} {\bibinfo  {journal} {Phys. Rev. Lett.}\ }\textbf {\bibinfo
  {volume} {124}},\ \bibinfo {pages} {180604} (\bibinfo {year}
  {2020})}\BibitemShut {NoStop}%
\bibitem [{\citenamefont {Hart}\ \emph {et~al.}(2020)\citenamefont {Hart},
  \citenamefont {Tomasi},\ and\ \citenamefont {Castelnovo}}]{hart2020random}%
  \BibitemOpen
  \bibfield  {author} {\bibinfo {author} {\bibfnamefont {O.}~\bibnamefont
  {Hart}}, \bibinfo {author} {\bibfnamefont {G.~D.}\ \bibnamefont {Tomasi}},\
  and\ \bibinfo {author} {\bibfnamefont {C.}~\bibnamefont {Castelnovo}},\
  }\href@noop {} {\bibinfo {title} {The random quantum comb: from compact
  localized states to many-body scars}} (\bibinfo {year} {2020}),\ \Eprint
  {https://arxiv.org/abs/2005.03036} {arXiv:2005.03036 [cond-mat.str-el]}
  \BibitemShut {NoStop}%
\bibitem [{\citenamefont {Dooley}\ and\ \citenamefont
  {Kells}(2020)}]{dooley2020enhancing}%
  \BibitemOpen
  \bibfield  {author} {\bibinfo {author} {\bibfnamefont {S.}~\bibnamefont
  {Dooley}}\ and\ \bibinfo {author} {\bibfnamefont {G.}~\bibnamefont {Kells}},\
  }\href@noop {} {\bibinfo {title} {Enhancing the effect of quantum many-body
  scars on dynamics by minimising the effective dimension}} (\bibinfo {year}
  {2020}),\ \Eprint {https://arxiv.org/abs/2006.03099} {arXiv:2006.03099
  [quant-ph]} \BibitemShut {NoStop}%
\bibitem [{\citenamefont {Surace}\ \emph {et~al.}(2020)\citenamefont {Surace},
  \citenamefont {Mazza}, \citenamefont {Giudici}, \citenamefont {Lerose},
  \citenamefont {Gambassi},\ and\ \citenamefont
  {Dalmonte}}]{Gambassi_Dalmonte_PRX2020}%
  \BibitemOpen
  \bibfield  {author} {\bibinfo {author} {\bibfnamefont {F.~M.}\ \bibnamefont
  {Surace}}, \bibinfo {author} {\bibfnamefont {P.~P.}\ \bibnamefont {Mazza}},
  \bibinfo {author} {\bibfnamefont {G.}~\bibnamefont {Giudici}}, \bibinfo
  {author} {\bibfnamefont {A.}~\bibnamefont {Lerose}}, \bibinfo {author}
  {\bibfnamefont {A.}~\bibnamefont {Gambassi}},\ and\ \bibinfo {author}
  {\bibfnamefont {M.}~\bibnamefont {Dalmonte}},\ }\bibfield  {title} {\bibinfo
  {title} {Lattice gauge theories and string dynamics in {R}ydberg atom quantum
  simulators},\ }\href {https://doi.org/10.1103/PhysRevX.10.021041} {\bibfield
  {journal} {\bibinfo  {journal} {Phys. Rev. X}\ }\textbf {\bibinfo {volume}
  {10}},\ \bibinfo {pages} {021041} (\bibinfo {year} {2020})}\BibitemShut
  {NoStop}%
\bibitem [{\citenamefont {van Voorden}\ \emph {et~al.}(2020)\citenamefont {van
  Voorden}, \citenamefont {Min\'a\ifmmode~\check{r}\else \v{r}\fi{}},\ and\
  \citenamefont {Schoutens}}]{Schoutens_PRB2020_scars}%
  \BibitemOpen
  \bibfield  {author} {\bibinfo {author} {\bibfnamefont {B.}~\bibnamefont {van
  Voorden}}, \bibinfo {author} {\bibfnamefont {J.~c.~v.}\ \bibnamefont
  {Min\'a\ifmmode~\check{r}\else \v{r}\fi{}}},\ and\ \bibinfo {author}
  {\bibfnamefont {K.}~\bibnamefont {Schoutens}},\ }\bibfield  {title} {\bibinfo
  {title} {Quantum many-body scars in transverse field {I}sing ladders and
  beyond},\ }\href {https://doi.org/10.1103/PhysRevB.101.220305} {\bibfield
  {journal} {\bibinfo  {journal} {Phys. Rev. B}\ }\textbf {\bibinfo {volume}
  {101}},\ \bibinfo {pages} {220305} (\bibinfo {year} {2020})}\BibitemShut
  {NoStop}%
\bibitem [{\citenamefont {Lacroix}\ \emph {et~al.}(2011)\citenamefont
  {Lacroix}, \citenamefont {Mendels},\ and\ \citenamefont
  {Mila}}]{lacroix2011introduction}%
  \BibitemOpen
  \bibfield  {author} {\bibinfo {author} {\bibfnamefont {C.}~\bibnamefont
  {Lacroix}}, \bibinfo {author} {\bibfnamefont {P.}~\bibnamefont {Mendels}},\
  and\ \bibinfo {author} {\bibfnamefont {F.}~\bibnamefont {Mila}},\ }\href@noop
  {} {\emph {\bibinfo {title} {Introduction to frustrated magnetism: materials,
  experiments, theory}}},\ Vol.\ \bibinfo {volume} {164}\ (\bibinfo
  {publisher} {Springer Science \& Business Media},\ \bibinfo {year}
  {2011})\BibitemShut {NoStop}%
\bibitem [{\citenamefont {Starykh}(2015)}]{Starykh_2015}%
  \BibitemOpen
  \bibfield  {author} {\bibinfo {author} {\bibfnamefont {O.~A.}\ \bibnamefont
  {Starykh}},\ }\bibfield  {title} {\bibinfo {title} {Unusual ordered phases of
  highly frustrated magnets: a review},\ }\href
  {https://doi.org/10.1088/0034-4885/78/5/052502} {\bibfield  {journal}
  {\bibinfo  {journal} {Reports on Progress in Physics}\ }\textbf {\bibinfo
  {volume} {78}},\ \bibinfo {pages} {052502} (\bibinfo {year}
  {2015})}\BibitemShut {NoStop}%
\bibitem [{\citenamefont {{Savary}}\ and\ \citenamefont
  {{Balents}}(2017)}]{SavaryBalents}%
  \BibitemOpen
  \bibfield  {author} {\bibinfo {author} {\bibfnamefont {L.}~\bibnamefont
  {{Savary}}}\ and\ \bibinfo {author} {\bibfnamefont {L.}~\bibnamefont
  {{Balents}}},\ }\bibfield  {title} {\bibinfo {title} {{Quantum spin liquids:
  a review}},\ }\href {https://doi.org/10.1088/0034-4885/80/1/016502}
  {\bibfield  {journal} {\bibinfo  {journal} {Reports on Progress in Physics}\
  }\textbf {\bibinfo {volume} {80}},\ \bibinfo {eid} {016502} (\bibinfo {year}
  {2017})}\BibitemShut {NoStop}%
\bibitem [{\citenamefont {Derzhko}\ \emph {et~al.}(2015)\citenamefont
  {Derzhko}, \citenamefont {Richter},\ and\ \citenamefont
  {Maksymenko}}]{Derzhko_2015}%
  \BibitemOpen
  \bibfield  {author} {\bibinfo {author} {\bibfnamefont {O.}~\bibnamefont
  {Derzhko}}, \bibinfo {author} {\bibfnamefont {J.}~\bibnamefont {Richter}},\
  and\ \bibinfo {author} {\bibfnamefont {M.}~\bibnamefont {Maksymenko}},\
  }\bibfield  {title} {\bibinfo {title} {Strongly correlated flat-band systems:
  The route from {H}eisenberg spins to {H}ubbard electrons},\ }\href
  {https://doi.org/10.1142/s0217979215300078} {\bibfield  {journal} {\bibinfo
  {journal} {International Journal of Modern Physics B}\ }\textbf {\bibinfo
  {volume} {29}},\ \bibinfo {pages} {1530007} (\bibinfo {year}
  {2015})}\BibitemShut {NoStop}%
\bibitem [{\citenamefont {Shastry}\ and\ \citenamefont
  {Sutherland}(1981)}]{shastry1981exact}%
  \BibitemOpen
  \bibfield  {author} {\bibinfo {author} {\bibfnamefont {B.~S.}\ \bibnamefont
  {Shastry}}\ and\ \bibinfo {author} {\bibfnamefont {B.}~\bibnamefont
  {Sutherland}},\ }\bibfield  {title} {\bibinfo {title} {Exact ground state of
  a quantum mechanical antiferromagnet},\ }\href@noop {} {\bibfield  {journal}
  {\bibinfo  {journal} {Physica B+C}\ }\textbf {\bibinfo {volume} {108}},\
  \bibinfo {pages} {1069} (\bibinfo {year} {1981})}\BibitemShut {NoStop}%
\bibitem [{\citenamefont {Kageyama}\ \emph {et~al.}(1999)\citenamefont
  {Kageyama}, \citenamefont {Yoshimura}, \citenamefont {Stern}, \citenamefont
  {Mushnikov}, \citenamefont {Onizuka}, \citenamefont {Kato}, \citenamefont
  {Kosuge}, \citenamefont {Slichter}, \citenamefont {Goto},\ and\ \citenamefont
  {Ueda}}]{PhysRevLett.82.3168}%
  \BibitemOpen
  \bibfield  {author} {\bibinfo {author} {\bibfnamefont {H.}~\bibnamefont
  {Kageyama}}, \bibinfo {author} {\bibfnamefont {K.}~\bibnamefont {Yoshimura}},
  \bibinfo {author} {\bibfnamefont {R.}~\bibnamefont {Stern}}, \bibinfo
  {author} {\bibfnamefont {N.~V.}\ \bibnamefont {Mushnikov}}, \bibinfo {author}
  {\bibfnamefont {K.}~\bibnamefont {Onizuka}}, \bibinfo {author} {\bibfnamefont
  {M.}~\bibnamefont {Kato}}, \bibinfo {author} {\bibfnamefont {K.}~\bibnamefont
  {Kosuge}}, \bibinfo {author} {\bibfnamefont {C.~P.}\ \bibnamefont
  {Slichter}}, \bibinfo {author} {\bibfnamefont {T.}~\bibnamefont {Goto}},\
  and\ \bibinfo {author} {\bibfnamefont {Y.}~\bibnamefont {Ueda}},\ }\bibfield
  {title} {\bibinfo {title} {Exact dimer ground state and quantized
  magnetization plateaus in the two-dimensional spin system
  ${\mathrm{srcu}}_{2}({\mathrm{bo}}_{3}){}_{2}$},\ }\href
  {https://doi.org/10.1103/PhysRevLett.82.3168} {\bibfield  {journal} {\bibinfo
   {journal} {Phys. Rev. Lett.}\ }\textbf {\bibinfo {volume} {82}},\ \bibinfo
  {pages} {3168} (\bibinfo {year} {1999})}\BibitemShut {NoStop}%
\bibitem [{\citenamefont {Albrecht}\ and\ \citenamefont
  {Mila}(1996)}]{albrecht1996first}%
  \BibitemOpen
  \bibfield  {author} {\bibinfo {author} {\bibfnamefont {M.}~\bibnamefont
  {Albrecht}}\ and\ \bibinfo {author} {\bibfnamefont {F.}~\bibnamefont
  {Mila}},\ }\bibfield  {title} {\bibinfo {title} {First-order transition
  between magnetic order and valence bond order in a 2d frustrated {H}eisenberg
  model},\ }\href@noop {} {\bibfield  {journal} {\bibinfo  {journal} {EPL
  (Europhysics Letters)}\ }\textbf {\bibinfo {volume} {34}},\ \bibinfo {pages}
  {145} (\bibinfo {year} {1996})}\BibitemShut {NoStop}%
\bibitem [{\citenamefont {Darradi}\ \emph {et~al.}(2005)\citenamefont
  {Darradi}, \citenamefont {Richter},\ and\ \citenamefont
  {Farnell}}]{PhysRevB.72.104425}%
  \BibitemOpen
  \bibfield  {author} {\bibinfo {author} {\bibfnamefont {R.}~\bibnamefont
  {Darradi}}, \bibinfo {author} {\bibfnamefont {J.}~\bibnamefont {Richter}},\
  and\ \bibinfo {author} {\bibfnamefont {D.~J.~J.}\ \bibnamefont {Farnell}},\
  }\bibfield  {title} {\bibinfo {title} {Coupled cluster treatment of the
  shastry-sutherland antiferromagnet},\ }\href
  {https://doi.org/10.1103/PhysRevB.72.104425} {\bibfield  {journal} {\bibinfo
  {journal} {Phys. Rev. B}\ }\textbf {\bibinfo {volume} {72}},\ \bibinfo
  {pages} {104425} (\bibinfo {year} {2005})}\BibitemShut {NoStop}%
\bibitem [{\citenamefont {Wietek}\ \emph {et~al.}(2019)\citenamefont {Wietek},
  \citenamefont {Corboz}, \citenamefont {Wessel}, \citenamefont {Normand},
  \citenamefont {Mila},\ and\ \citenamefont
  {Honecker}}]{PhysRevResearch.1.033038}%
  \BibitemOpen
  \bibfield  {author} {\bibinfo {author} {\bibfnamefont {A.}~\bibnamefont
  {Wietek}}, \bibinfo {author} {\bibfnamefont {P.}~\bibnamefont {Corboz}},
  \bibinfo {author} {\bibfnamefont {S.}~\bibnamefont {Wessel}}, \bibinfo
  {author} {\bibfnamefont {B.}~\bibnamefont {Normand}}, \bibinfo {author}
  {\bibfnamefont {F.}~\bibnamefont {Mila}},\ and\ \bibinfo {author}
  {\bibfnamefont {A.}~\bibnamefont {Honecker}},\ }\bibfield  {title} {\bibinfo
  {title} {Thermodynamic properties of the shastry-sutherland model throughout
  the dimer-product phase},\ }\href
  {https://doi.org/10.1103/PhysRevResearch.1.033038} {\bibfield  {journal}
  {\bibinfo  {journal} {Phys. Rev. Research}\ }\textbf {\bibinfo {volume}
  {1}},\ \bibinfo {pages} {033038} (\bibinfo {year} {2019})}\BibitemShut
  {NoStop}%
\bibitem [{\citenamefont {McClarty}\ \emph {et~al.}(2017)\citenamefont
  {McClarty}, \citenamefont {Kr{\"u}ger}, \citenamefont {Guidi}, \citenamefont
  {Parker}, \citenamefont {Refson}, \citenamefont {Parker}, \citenamefont
  {Prabhakaran},\ and\ \citenamefont {Coldea}}]{mcclarty2017topological}%
  \BibitemOpen
  \bibfield  {author} {\bibinfo {author} {\bibfnamefont {P.~A.}\ \bibnamefont
  {McClarty}}, \bibinfo {author} {\bibfnamefont {F.}~\bibnamefont
  {Kr{\"u}ger}}, \bibinfo {author} {\bibfnamefont {T.}~\bibnamefont {Guidi}},
  \bibinfo {author} {\bibfnamefont {S.}~\bibnamefont {Parker}}, \bibinfo
  {author} {\bibfnamefont {K.}~\bibnamefont {Refson}}, \bibinfo {author}
  {\bibfnamefont {A.}~\bibnamefont {Parker}}, \bibinfo {author} {\bibfnamefont
  {D.}~\bibnamefont {Prabhakaran}},\ and\ \bibinfo {author} {\bibfnamefont
  {R.}~\bibnamefont {Coldea}},\ }\bibfield  {title} {\bibinfo {title}
  {Topological triplon modes and bound states in a shastry--sutherland
  magnet},\ }\href@noop {} {\bibfield  {journal} {\bibinfo  {journal} {Nature
  Physics}\ }\textbf {\bibinfo {volume} {13}},\ \bibinfo {pages} {736}
  (\bibinfo {year} {2017})}\BibitemShut {NoStop}%
\bibitem [{foo()}]{footnote_symmetrysector_etapairing}%
  \BibitemOpen
  \href@noop {} {}\bibinfo {note} {A low-entanglement eigenstate surrounded by
  mid-spectrum ``thermal'' states can only be considered a nontrivial scar if
  it belongs to the same symmetry sector as the surrounding eigenstates. See
  the discussion in Ref.~\cite{Mark_Motrunich_2020_etapairing} regarding eta
  pairing --- the Hubbard model needs to be perturbed in order for the eta
  pairing eigenstates to be regarded as true eigenstates.}\BibitemShut {Stop}%
\bibitem [{\citenamefont {Richter}\ \emph {et~al.}(1998)\citenamefont
  {Richter}, \citenamefont {Ivanov},\ and\ \citenamefont
  {Schulenburg}}]{richter1998antiferromagnetic}%
  \BibitemOpen
  \bibfield  {author} {\bibinfo {author} {\bibfnamefont {J.}~\bibnamefont
  {Richter}}, \bibinfo {author} {\bibfnamefont {N.}~\bibnamefont {Ivanov}},\
  and\ \bibinfo {author} {\bibfnamefont {J.}~\bibnamefont {Schulenburg}},\
  }\bibfield  {title} {\bibinfo {title} {The antiferromagnetic spin-chain with
  competing dimers and plaquettes: numerical versus exact results},\
  }\href@noop {} {\bibfield  {journal} {\bibinfo  {journal} {Journal of
  Physics: Condensed Matter}\ }\textbf {\bibinfo {volume} {10}},\ \bibinfo
  {pages} {3635} (\bibinfo {year} {1998})}\BibitemShut {NoStop}%
\bibitem [{\citenamefont {Koga}\ \emph {et~al.}(2000)\citenamefont {Koga},
  \citenamefont {Okunishi},\ and\ \citenamefont {Kawakami}}]{PhysRevB.62.5558}%
  \BibitemOpen
  \bibfield  {author} {\bibinfo {author} {\bibfnamefont {A.}~\bibnamefont
  {Koga}}, \bibinfo {author} {\bibfnamefont {K.}~\bibnamefont {Okunishi}},\
  and\ \bibinfo {author} {\bibfnamefont {N.}~\bibnamefont {Kawakami}},\
  }\bibfield  {title} {\bibinfo {title} {First-order quantum phase transition
  in the orthogonal-dimer spin chain},\ }\href
  {https://doi.org/10.1103/PhysRevB.62.5558} {\bibfield  {journal} {\bibinfo
  {journal} {Phys. Rev. B}\ }\textbf {\bibinfo {volume} {62}},\ \bibinfo
  {pages} {5558} (\bibinfo {year} {2000})}\BibitemShut {NoStop}%
\bibitem [{\citenamefont {Schulenburg}\ and\ \citenamefont
  {Richter}(2002)}]{PhysRevB.65.054420}%
  \BibitemOpen
  \bibfield  {author} {\bibinfo {author} {\bibfnamefont {J.}~\bibnamefont
  {Schulenburg}}\ and\ \bibinfo {author} {\bibfnamefont {J.}~\bibnamefont
  {Richter}},\ }\bibfield  {title} {\bibinfo {title} {Infinite series of
  magnetization plateaus in the frustrated dimer-plaquette chain},\ }\href
  {https://doi.org/10.1103/PhysRevB.65.054420} {\bibfield  {journal} {\bibinfo
  {journal} {Phys. Rev. B}\ }\textbf {\bibinfo {volume} {65}},\ \bibinfo
  {pages} {054420} (\bibinfo {year} {2002})}\BibitemShut {NoStop}%
\bibitem [{\citenamefont {Tanaka}\ \emph {et~al.}(2014)\citenamefont {Tanaka},
  \citenamefont {Kurita}, \citenamefont {Okada}, \citenamefont {Kunihiro},
  \citenamefont {Shirata}, \citenamefont {Fujii}, \citenamefont {Uekusa},
  \citenamefont {Matsuo}, \citenamefont {Kindo},\ and\ \citenamefont
  {Nojiri}}]{Tanaka_2014}%
  \BibitemOpen
  \bibfield  {author} {\bibinfo {author} {\bibfnamefont {H.}~\bibnamefont
  {Tanaka}}, \bibinfo {author} {\bibfnamefont {N.}~\bibnamefont {Kurita}},
  \bibinfo {author} {\bibfnamefont {M.}~\bibnamefont {Okada}}, \bibinfo
  {author} {\bibfnamefont {E.}~\bibnamefont {Kunihiro}}, \bibinfo {author}
  {\bibfnamefont {Y.}~\bibnamefont {Shirata}}, \bibinfo {author} {\bibfnamefont
  {K.}~\bibnamefont {Fujii}}, \bibinfo {author} {\bibfnamefont
  {H.}~\bibnamefont {Uekusa}}, \bibinfo {author} {\bibfnamefont
  {A.}~\bibnamefont {Matsuo}}, \bibinfo {author} {\bibfnamefont
  {K.}~\bibnamefont {Kindo}},\ and\ \bibinfo {author} {\bibfnamefont
  {H.}~\bibnamefont {Nojiri}},\ }\bibfield  {title} {\bibinfo {title} {Almost
  perfect frustration in the dimer magnet ba2cosi2o6cl2},\ }\href
  {https://doi.org/10.7566/jpsj.83.103701} {\bibfield  {journal} {\bibinfo
  {journal} {Journal of the Physical Society of Japan}\ }\textbf {\bibinfo
  {volume} {83}},\ \bibinfo {pages} {103701} (\bibinfo {year}
  {2014})}\BibitemShut {NoStop}%
\bibitem [{\citenamefont {Stre\ifmmode~\check{c}\else \v{c}\fi{}ka}\ \emph
  {et~al.}(2017)\citenamefont {Stre\ifmmode~\check{c}\else \v{c}\fi{}ka},
  \citenamefont {Richter}, \citenamefont {Derzhko}, \citenamefont
  {Verkholyak},\ and\ \citenamefont {Kar\ifmmode~\check{l}\else
  \v{l}\fi{}ov\'a}}]{PhysRevB.95.224415}%
  \BibitemOpen
  \bibfield  {author} {\bibinfo {author} {\bibfnamefont {J.}~\bibnamefont
  {Stre\ifmmode~\check{c}\else \v{c}\fi{}ka}}, \bibinfo {author} {\bibfnamefont
  {J.}~\bibnamefont {Richter}}, \bibinfo {author} {\bibfnamefont
  {O.}~\bibnamefont {Derzhko}}, \bibinfo {author} {\bibfnamefont
  {T.}~\bibnamefont {Verkholyak}},\ and\ \bibinfo {author} {\bibfnamefont
  {K.}~\bibnamefont {Kar\ifmmode~\check{l}\else \v{l}\fi{}ov\'a}},\ }\bibfield
  {title} {\bibinfo {title} {Diversity of quantum ground states and quantum
  phase transitions of a spin-$\frac{1}{2}$ {H}eisenberg octahedral chain},\
  }\href {https://doi.org/10.1103/PhysRevB.95.224415} {\bibfield  {journal}
  {\bibinfo  {journal} {Phys. Rev. B}\ }\textbf {\bibinfo {volume} {95}},\
  \bibinfo {pages} {224415} (\bibinfo {year} {2017})}\BibitemShut {NoStop}%
\bibitem [{\citenamefont {Sen}\ \emph {et~al.}(1996)\citenamefont {Sen},
  \citenamefont {Shastry}, \citenamefont {Walstedt},\ and\ \citenamefont
  {Cava}}]{PhysRevB.53.6401}%
  \BibitemOpen
  \bibfield  {author} {\bibinfo {author} {\bibfnamefont {D.}~\bibnamefont
  {Sen}}, \bibinfo {author} {\bibfnamefont {B.~S.}\ \bibnamefont {Shastry}},
  \bibinfo {author} {\bibfnamefont {R.~E.}\ \bibnamefont {Walstedt}},\ and\
  \bibinfo {author} {\bibfnamefont {R.}~\bibnamefont {Cava}},\ }\bibfield
  {title} {\bibinfo {title} {Quantum solitons in the sawtooth lattice},\ }\href
  {https://doi.org/10.1103/PhysRevB.53.6401} {\bibfield  {journal} {\bibinfo
  {journal} {Phys. Rev. B}\ }\textbf {\bibinfo {volume} {53}},\ \bibinfo
  {pages} {6401} (\bibinfo {year} {1996})}\BibitemShut {NoStop}%
\bibitem [{\citenamefont {Chen}\ \emph {et~al.}(2003)\citenamefont {Chen},
  \citenamefont {B\"uttner},\ and\ \citenamefont {Voit}}]{PhysRevB.67.054412}%
  \BibitemOpen
  \bibfield  {author} {\bibinfo {author} {\bibfnamefont {S.}~\bibnamefont
  {Chen}}, \bibinfo {author} {\bibfnamefont {H.}~\bibnamefont {B\"uttner}},\
  and\ \bibinfo {author} {\bibfnamefont {J.}~\bibnamefont {Voit}},\ }\bibfield
  {title} {\bibinfo {title} {Ground state and excitation of an asymmetric spin
  ladder model},\ }\href {https://doi.org/10.1103/PhysRevB.67.054412}
  {\bibfield  {journal} {\bibinfo  {journal} {Phys. Rev. B}\ }\textbf {\bibinfo
  {volume} {67}},\ \bibinfo {pages} {054412} (\bibinfo {year}
  {2003})}\BibitemShut {NoStop}%
\bibitem [{\citenamefont {Schulenburg}\ \emph {et~al.}(2002)\citenamefont
  {Schulenburg}, \citenamefont {Honecker}, \citenamefont {Schnack},
  \citenamefont {Richter},\ and\ \citenamefont
  {Schmidt}}]{PhysRevLett.88.167207}%
  \BibitemOpen
  \bibfield  {author} {\bibinfo {author} {\bibfnamefont {J.}~\bibnamefont
  {Schulenburg}}, \bibinfo {author} {\bibfnamefont {A.}~\bibnamefont
  {Honecker}}, \bibinfo {author} {\bibfnamefont {J.}~\bibnamefont {Schnack}},
  \bibinfo {author} {\bibfnamefont {J.}~\bibnamefont {Richter}},\ and\ \bibinfo
  {author} {\bibfnamefont {H.-J.}\ \bibnamefont {Schmidt}},\ }\bibfield
  {title} {\bibinfo {title} {Macroscopic magnetization jumps due to independent
  magnons in frustrated quantum spin lattices},\ }\href
  {https://doi.org/10.1103/PhysRevLett.88.167207} {\bibfield  {journal}
  {\bibinfo  {journal} {Phys. Rev. Lett.}\ }\textbf {\bibinfo {volume} {88}},\
  \bibinfo {pages} {167207} (\bibinfo {year} {2002})}\BibitemShut {NoStop}%
\bibitem [{\citenamefont {Zhitomirsky}\ and\ \citenamefont
  {Tsunetsugu}(2004)}]{PhysRevB.70.100403}%
  \BibitemOpen
  \bibfield  {author} {\bibinfo {author} {\bibfnamefont {M.~E.}\ \bibnamefont
  {Zhitomirsky}}\ and\ \bibinfo {author} {\bibfnamefont {H.}~\bibnamefont
  {Tsunetsugu}},\ }\bibfield  {title} {\bibinfo {title} {Exact low-temperature
  behavior of a kagom\'e antiferromagnet at high fields},\ }\href
  {https://doi.org/10.1103/PhysRevB.70.100403} {\bibfield  {journal} {\bibinfo
  {journal} {Phys. Rev. B}\ }\textbf {\bibinfo {volume} {70}},\ \bibinfo
  {pages} {100403} (\bibinfo {year} {2004})}\BibitemShut {NoStop}%
\bibitem [{\citenamefont {Gelfand}(1991)}]{PhysRevB.43.8644}%
  \BibitemOpen
  \bibfield  {author} {\bibinfo {author} {\bibfnamefont {M.~P.}\ \bibnamefont
  {Gelfand}},\ }\bibfield  {title} {\bibinfo {title} {Linked-tetrahedra spin
  chain: Exact ground state and excitations},\ }\href
  {https://doi.org/10.1103/PhysRevB.43.8644} {\bibfield  {journal} {\bibinfo
  {journal} {Phys. Rev. B}\ }\textbf {\bibinfo {volume} {43}},\ \bibinfo
  {pages} {8644} (\bibinfo {year} {1991})}\BibitemShut {NoStop}%
\bibitem [{\citenamefont {Honecker}\ \emph {et~al.}(2000)\citenamefont
  {Honecker}, \citenamefont {Mila},\ and\ \citenamefont
  {Troyer}}]{honecker2000magnetization}%
  \BibitemOpen
  \bibfield  {author} {\bibinfo {author} {\bibfnamefont {A.}~\bibnamefont
  {Honecker}}, \bibinfo {author} {\bibfnamefont {F.}~\bibnamefont {Mila}},\
  and\ \bibinfo {author} {\bibfnamefont {M.}~\bibnamefont {Troyer}},\
  }\bibfield  {title} {\bibinfo {title} {Magnetization plateaux and jumps in a
  class of frustrated ladders: A simple route to a complex behaviour},\
  }\href@noop {} {\bibfield  {journal} {\bibinfo  {journal} {The European
  Physical Journal B-Condensed Matter and Complex Systems}\ }\textbf {\bibinfo
  {volume} {15}},\ \bibinfo {pages} {227} (\bibinfo {year} {2000})}\BibitemShut
  {NoStop}%
\bibitem [{\citenamefont {Derzhko}\ \emph {et~al.}(2010)\citenamefont
  {Derzhko}, \citenamefont {Krokhmalskii},\ and\ \citenamefont
  {Richter}}]{PhysRevB.82.214412}%
  \BibitemOpen
  \bibfield  {author} {\bibinfo {author} {\bibfnamefont {O.}~\bibnamefont
  {Derzhko}}, \bibinfo {author} {\bibfnamefont {T.}~\bibnamefont
  {Krokhmalskii}},\ and\ \bibinfo {author} {\bibfnamefont {J.}~\bibnamefont
  {Richter}},\ }\bibfield  {title} {\bibinfo {title} {Emergent {I}sing degrees
  of freedom in frustrated two-leg ladder and bilayer $s=\frac{1}{2}$
  {H}eisenberg antiferromagnets},\ }\href
  {https://doi.org/10.1103/PhysRevB.82.214412} {\bibfield  {journal} {\bibinfo
  {journal} {Phys. Rev. B}\ }\textbf {\bibinfo {volume} {82}},\ \bibinfo
  {pages} {214412} (\bibinfo {year} {2010})}\BibitemShut {NoStop}%
\bibitem [{\citenamefont {Alba}\ \emph {et~al.}(2009)\citenamefont {Alba},
  \citenamefont {Fagotti},\ and\ \citenamefont {Calabrese}}]{Alba_2009}%
  \BibitemOpen
  \bibfield  {author} {\bibinfo {author} {\bibfnamefont {V.}~\bibnamefont
  {Alba}}, \bibinfo {author} {\bibfnamefont {M.}~\bibnamefont {Fagotti}},\ and\
  \bibinfo {author} {\bibfnamefont {P.}~\bibnamefont {Calabrese}},\ }\bibfield
  {title} {\bibinfo {title} {Entanglement entropy of excited states},\ }\href
  {https://doi.org/10.1088/1742-5468/2009/10/p10020} {\bibfield  {journal}
  {\bibinfo  {journal} {Journal of Statistical Mechanics: Theory and
  Experiment}\ }\textbf {\bibinfo {volume} {2009}},\ \bibinfo {pages} {P10020}
  (\bibinfo {year} {2009})}\BibitemShut {NoStop}%
\bibitem [{\citenamefont {Beugeling}\ \emph
  {et~al.}(2015{\natexlab{a}})\citenamefont {Beugeling}, \citenamefont
  {Andreanov},\ and\ \citenamefont {Haque}}]{Beugeling_2015}%
  \BibitemOpen
  \bibfield  {author} {\bibinfo {author} {\bibfnamefont {W.}~\bibnamefont
  {Beugeling}}, \bibinfo {author} {\bibfnamefont {A.}~\bibnamefont
  {Andreanov}},\ and\ \bibinfo {author} {\bibfnamefont {M.}~\bibnamefont
  {Haque}},\ }\bibfield  {title} {\bibinfo {title} {Global characteristics of
  all eigenstates of local many-body {H}amiltonians: participation ratio and
  entanglement entropy},\ }\href
  {https://doi.org/10.1088/1742-5468/2015/02/p02002} {\bibfield  {journal}
  {\bibinfo  {journal} {Journal of Statistical Mechanics: Theory and
  Experiment}\ }\textbf {\bibinfo {volume} {2015}},\ \bibinfo {pages} {P02002}
  (\bibinfo {year} {2015}{\natexlab{a}})}\BibitemShut {NoStop}%
\bibitem [{\citenamefont {Vidmar}\ \emph {et~al.}(2017)\citenamefont {Vidmar},
  \citenamefont {Hackl}, \citenamefont {Bianchi},\ and\ \citenamefont
  {Rigol}}]{PhysRevLett.119.020601}%
  \BibitemOpen
  \bibfield  {author} {\bibinfo {author} {\bibfnamefont {L.}~\bibnamefont
  {Vidmar}}, \bibinfo {author} {\bibfnamefont {L.}~\bibnamefont {Hackl}},
  \bibinfo {author} {\bibfnamefont {E.}~\bibnamefont {Bianchi}},\ and\ \bibinfo
  {author} {\bibfnamefont {M.}~\bibnamefont {Rigol}},\ }\bibfield  {title}
  {\bibinfo {title} {Entanglement entropy of eigenstates of quadratic fermionic
  {H}amiltonians},\ }\href {https://doi.org/10.1103/PhysRevLett.119.020601}
  {\bibfield  {journal} {\bibinfo  {journal} {Phys. Rev. Lett.}\ }\textbf
  {\bibinfo {volume} {119}},\ \bibinfo {pages} {020601} (\bibinfo {year}
  {2017})}\BibitemShut {NoStop}%
\bibitem [{\citenamefont {LeBlond}\ \emph {et~al.}(2019)\citenamefont
  {LeBlond}, \citenamefont {Mallayya}, \citenamefont {Vidmar},\ and\
  \citenamefont {Rigol}}]{LeBlond_Mallaya_Vidmar_Rigol_PRE2019}%
  \BibitemOpen
  \bibfield  {author} {\bibinfo {author} {\bibfnamefont {T.}~\bibnamefont
  {LeBlond}}, \bibinfo {author} {\bibfnamefont {K.}~\bibnamefont {Mallayya}},
  \bibinfo {author} {\bibfnamefont {L.}~\bibnamefont {Vidmar}},\ and\ \bibinfo
  {author} {\bibfnamefont {M.}~\bibnamefont {Rigol}},\ }\bibfield  {title}
  {\bibinfo {title} {Entanglement and matrix elements of observables in
  interacting integrable systems},\ }\href
  {https://doi.org/10.1103/PhysRevE.100.062134} {\bibfield  {journal} {\bibinfo
   {journal} {Phys. Rev. E}\ }\textbf {\bibinfo {volume} {100}},\ \bibinfo
  {pages} {062134} (\bibinfo {year} {2019})}\BibitemShut {NoStop}%
\bibitem [{\citenamefont {Srednicki}(1996)}]{srednicki1996thermal}%
  \BibitemOpen
  \bibfield  {author} {\bibinfo {author} {\bibfnamefont {M.}~\bibnamefont
  {Srednicki}},\ }\bibfield  {title} {\bibinfo {title} {Thermal fluctuations in
  quantized chaotic systems},\ }\href@noop {} {\bibfield  {journal} {\bibinfo
  {journal} {Journal of Physics A: Mathematical and General}\ }\textbf
  {\bibinfo {volume} {29}},\ \bibinfo {pages} {L75} (\bibinfo {year}
  {1996})}\BibitemShut {NoStop}%
\bibitem [{\citenamefont {Srednicki}(1999)}]{srednicki1999approach}%
  \BibitemOpen
  \bibfield  {author} {\bibinfo {author} {\bibfnamefont {M.}~\bibnamefont
  {Srednicki}},\ }\bibfield  {title} {\bibinfo {title} {The approach to thermal
  equilibrium in quantized chaotic systems},\ }\href@noop {} {\bibfield
  {journal} {\bibinfo  {journal} {Journal of Physics A: Mathematical and
  General}\ }\textbf {\bibinfo {volume} {32}},\ \bibinfo {pages} {1163}
  (\bibinfo {year} {1999})}\BibitemShut {NoStop}%
\bibitem [{\citenamefont {Neuenhahn}\ and\ \citenamefont
  {Marquardt}(2012)}]{Marquardt_PRE12}%
  \BibitemOpen
  \bibfield  {author} {\bibinfo {author} {\bibfnamefont {C.}~\bibnamefont
  {Neuenhahn}}\ and\ \bibinfo {author} {\bibfnamefont {F.}~\bibnamefont
  {Marquardt}},\ }\bibfield  {title} {\bibinfo {title} {Thermalization of
  interacting fermions and delocalization in fock space},\ }\href
  {https://doi.org/10.1103/PhysRevE.85.060101} {\bibfield  {journal} {\bibinfo
  {journal} {Phys. Rev. E}\ }\textbf {\bibinfo {volume} {85}},\ \bibinfo
  {pages} {060101} (\bibinfo {year} {2012})}\BibitemShut {NoStop}%
\bibitem [{\citenamefont {Beugeling}\ \emph {et~al.}(2014)\citenamefont
  {Beugeling}, \citenamefont {Moessner},\ and\ \citenamefont
  {Haque}}]{Beugeling_scaling_PRE14}%
  \BibitemOpen
  \bibfield  {author} {\bibinfo {author} {\bibfnamefont {W.}~\bibnamefont
  {Beugeling}}, \bibinfo {author} {\bibfnamefont {R.}~\bibnamefont
  {Moessner}},\ and\ \bibinfo {author} {\bibfnamefont {M.}~\bibnamefont
  {Haque}},\ }\bibfield  {title} {\bibinfo {title} {Finite-size scaling of
  eigenstate thermalization},\ }\href
  {https://doi.org/10.1103/PhysRevE.89.042112} {\bibfield  {journal} {\bibinfo
  {journal} {Phys. Rev. E}\ }\textbf {\bibinfo {volume} {89}},\ \bibinfo
  {pages} {042112} (\bibinfo {year} {2014})}\BibitemShut {NoStop}%
\bibitem [{\citenamefont {Beugeling}\ \emph
  {et~al.}(2015{\natexlab{b}})\citenamefont {Beugeling}, \citenamefont
  {Moessner},\ and\ \citenamefont {Haque}}]{Beugeling_offdiag_PRE2015}%
  \BibitemOpen
  \bibfield  {author} {\bibinfo {author} {\bibfnamefont {W.}~\bibnamefont
  {Beugeling}}, \bibinfo {author} {\bibfnamefont {R.}~\bibnamefont
  {Moessner}},\ and\ \bibinfo {author} {\bibfnamefont {M.}~\bibnamefont
  {Haque}},\ }\bibfield  {title} {\bibinfo {title} {Off-diagonal matrix
  elements of local operators in many-body quantum systems},\ }\href
  {https://doi.org/10.1103/PhysRevE.91.012144} {\bibfield  {journal} {\bibinfo
  {journal} {Phys. Rev. E}\ }\textbf {\bibinfo {volume} {91}},\ \bibinfo
  {pages} {012144} (\bibinfo {year} {2015}{\natexlab{b}})}\BibitemShut
  {NoStop}%
\bibitem [{\citenamefont {Ziraldo}\ and\ \citenamefont
  {Santoro}(2013)}]{ziraldo2013relaxation}%
  \BibitemOpen
  \bibfield  {author} {\bibinfo {author} {\bibfnamefont {S.}~\bibnamefont
  {Ziraldo}}\ and\ \bibinfo {author} {\bibfnamefont {G.~E.}\ \bibnamefont
  {Santoro}},\ }\bibfield  {title} {\bibinfo {title} {Relaxation and
  thermalization after a quantum quench: Why localization is important},\
  }\href@noop {} {\bibfield  {journal} {\bibinfo  {journal} {Physical Review
  B}\ }\textbf {\bibinfo {volume} {87}},\ \bibinfo {pages} {064201} (\bibinfo
  {year} {2013})}\BibitemShut {NoStop}%
\bibitem [{\citenamefont {Alba}(2015)}]{Alba_PRB15}%
  \BibitemOpen
  \bibfield  {author} {\bibinfo {author} {\bibfnamefont {V.}~\bibnamefont
  {Alba}},\ }\bibfield  {title} {\bibinfo {title} {Eigenstate thermalization
  hypothesis and integrability in quantum spin chains},\ }\href
  {https://doi.org/10.1103/PhysRevB.91.155123} {\bibfield  {journal} {\bibinfo
  {journal} {Phys. Rev. B}\ }\textbf {\bibinfo {volume} {91}},\ \bibinfo
  {pages} {155123} (\bibinfo {year} {2015})}\BibitemShut {NoStop}%
\bibitem [{\citenamefont {Nandy}\ \emph {et~al.}(2016)\citenamefont {Nandy},
  \citenamefont {Sen}, \citenamefont {Das},\ and\ \citenamefont
  {Dhar}}]{ArnabSenArnabDas_PRB16}%
  \BibitemOpen
  \bibfield  {author} {\bibinfo {author} {\bibfnamefont {S.}~\bibnamefont
  {Nandy}}, \bibinfo {author} {\bibfnamefont {A.}~\bibnamefont {Sen}}, \bibinfo
  {author} {\bibfnamefont {A.}~\bibnamefont {Das}},\ and\ \bibinfo {author}
  {\bibfnamefont {A.}~\bibnamefont {Dhar}},\ }\bibfield  {title} {\bibinfo
  {title} {Eigenstate {G}ibbs ensemble in integrable quantum systems},\ }\href
  {https://doi.org/10.1103/PhysRevB.94.245131} {\bibfield  {journal} {\bibinfo
  {journal} {Phys. Rev. B}\ }\textbf {\bibinfo {volume} {94}},\ \bibinfo
  {pages} {245131} (\bibinfo {year} {2016})}\BibitemShut {NoStop}%
\bibitem [{\citenamefont {Haque}\ and\ \citenamefont
  {McClarty}(2019)}]{HaqueMcClarty_SYKETH}%
  \BibitemOpen
  \bibfield  {author} {\bibinfo {author} {\bibfnamefont {M.}~\bibnamefont
  {Haque}}\ and\ \bibinfo {author} {\bibfnamefont {P.~A.}\ \bibnamefont
  {McClarty}},\ }\bibfield  {title} {\bibinfo {title} {Eigenstate
  thermalization scaling in majorana clusters: From chaotic to integrable
  {Sachdev-Ye-Kitaev} models},\ }\href
  {https://doi.org/10.1103/PhysRevB.100.115122} {\bibfield  {journal} {\bibinfo
   {journal} {Phys. Rev. B}\ }\textbf {\bibinfo {volume} {100}},\ \bibinfo
  {pages} {115122} (\bibinfo {year} {2019})}\BibitemShut {NoStop}%
\bibitem [{\citenamefont {Isichenko}(1992)}]{RevModPhys.64.961}%
  \BibitemOpen
  \bibfield  {author} {\bibinfo {author} {\bibfnamefont {M.~B.}\ \bibnamefont
  {Isichenko}},\ }\bibfield  {title} {\bibinfo {title} {Percolation,
  statistical topography, and transport in random media},\ }\href
  {https://doi.org/10.1103/RevModPhys.64.961} {\bibfield  {journal} {\bibinfo
  {journal} {Rev. Mod. Phys.}\ }\textbf {\bibinfo {volume} {64}},\ \bibinfo
  {pages} {961} (\bibinfo {year} {1992})}\BibitemShut {NoStop}%
\bibitem [{\citenamefont {Richter}\ \emph
  {et~al.}(2004{\natexlab{a}})\citenamefont {Richter}, \citenamefont
  {Schulenburg}, \citenamefont {Honecker},\ and\ \citenamefont
  {Schmalfu{\ss}}}]{richter2004starlattice}%
  \BibitemOpen
  \bibfield  {author} {\bibinfo {author} {\bibfnamefont {J.}~\bibnamefont
  {Richter}}, \bibinfo {author} {\bibfnamefont {J.}~\bibnamefont
  {Schulenburg}}, \bibinfo {author} {\bibfnamefont {A.}~\bibnamefont
  {Honecker}},\ and\ \bibinfo {author} {\bibfnamefont {D.}~\bibnamefont
  {Schmalfu{\ss}}},\ }\bibfield  {title} {\bibinfo {title} {Absence of magnetic
  order for the spin-half {H}eisenberg antiferromagnet on the star lattice},\
  }\href@noop {} {\bibfield  {journal} {\bibinfo  {journal} {Physical Review
  B}\ }\textbf {\bibinfo {volume} {70}},\ \bibinfo {pages} {174454} (\bibinfo
  {year} {2004}{\natexlab{a}})}\BibitemShut {NoStop}%
\bibitem [{\citenamefont {Jahromi}\ and\ \citenamefont
  {Or\'us}(2018)}]{PhysRevB.98.155108}%
  \BibitemOpen
  \bibfield  {author} {\bibinfo {author} {\bibfnamefont {S.~S.}\ \bibnamefont
  {Jahromi}}\ and\ \bibinfo {author} {\bibfnamefont {R.}~\bibnamefont
  {Or\'us}},\ }\bibfield  {title} {\bibinfo {title} {Spin-$\frac{1}{2}$
  {H}eisenberg antiferromagnet on the star lattice: Competing
  valence-bond-solid phases studied by means of tensor networks},\ }\href
  {https://doi.org/10.1103/PhysRevB.98.155108} {\bibfield  {journal} {\bibinfo
  {journal} {Phys. Rev. B}\ }\textbf {\bibinfo {volume} {98}},\ \bibinfo
  {pages} {155108} (\bibinfo {year} {2018})}\BibitemShut {NoStop}%
\bibitem [{\citenamefont {Ding}\ \emph {et~al.}(2012)\citenamefont {Ding},
  \citenamefont {Fu},\ and\ \citenamefont {Guo}}]{PhysRevE.85.062101}%
  \BibitemOpen
  \bibfield  {author} {\bibinfo {author} {\bibfnamefont {C.}~\bibnamefont
  {Ding}}, \bibinfo {author} {\bibfnamefont {Z.}~\bibnamefont {Fu}},\ and\
  \bibinfo {author} {\bibfnamefont {W.}~\bibnamefont {Guo}},\ }\bibfield
  {title} {\bibinfo {title} {Critical points of the o($n$) loop model on the
  martini and the 3-12 lattices},\ }\href
  {https://doi.org/10.1103/PhysRevE.85.062101} {\bibfield  {journal} {\bibinfo
  {journal} {Phys. Rev. E}\ }\textbf {\bibinfo {volume} {85}},\ \bibinfo
  {pages} {062101} (\bibinfo {year} {2012})}\BibitemShut {NoStop}%
\bibitem [{\citenamefont {Scullard}(2006)}]{PhysRevE.73.016107}%
  \BibitemOpen
  \bibfield  {author} {\bibinfo {author} {\bibfnamefont {C.~R.}\ \bibnamefont
  {Scullard}},\ }\bibfield  {title} {\bibinfo {title} {Exact site percolation
  thresholds using a site-to-bond transformation and the star-triangle
  transformation},\ }\href {https://doi.org/10.1103/PhysRevE.73.016107}
  {\bibfield  {journal} {\bibinfo  {journal} {Phys. Rev. E}\ }\textbf {\bibinfo
  {volume} {73}},\ \bibinfo {pages} {016107} (\bibinfo {year}
  {2006})}\BibitemShut {NoStop}%
\bibitem [{\citenamefont {Richter}\ \emph
  {et~al.}(2004{\natexlab{b}})\citenamefont {Richter}, \citenamefont
  {Schulenburg},\ and\ \citenamefont {Honecker}}]{richter2004quantum}%
  \BibitemOpen
  \bibfield  {author} {\bibinfo {author} {\bibfnamefont {J.}~\bibnamefont
  {Richter}}, \bibinfo {author} {\bibfnamefont {J.}~\bibnamefont
  {Schulenburg}},\ and\ \bibinfo {author} {\bibfnamefont {A.}~\bibnamefont
  {Honecker}},\ }\bibfield  {title} {\bibinfo {title} {Quantum magnetism in two
  dimensions: From semi-classical neel order to magnetic disorder},\ }in\
  \href@noop {} {\emph {\bibinfo {booktitle} {Two Dimensions: From
  Semi-classical Neel Order to Magnetic Disorder {\rm in} Quantum Magnetism}}}\
  (\bibinfo  {publisher} {Springer},\ \bibinfo {year} {2004})\ p.~\bibinfo
  {pages} {85}\BibitemShut {NoStop}%
\bibitem [{\citenamefont {Misguich}\ \emph {et~al.}(1999)\citenamefont
  {Misguich}, \citenamefont {Lhuillier}, \citenamefont {Bernu},\ and\
  \citenamefont {Waldtmann}}]{PhysRevB.60.1064}%
  \BibitemOpen
  \bibfield  {author} {\bibinfo {author} {\bibfnamefont {G.}~\bibnamefont
  {Misguich}}, \bibinfo {author} {\bibfnamefont {C.}~\bibnamefont {Lhuillier}},
  \bibinfo {author} {\bibfnamefont {B.}~\bibnamefont {Bernu}},\ and\ \bibinfo
  {author} {\bibfnamefont {C.}~\bibnamefont {Waldtmann}},\ }\bibfield  {title}
  {\bibinfo {title} {Spin-liquid phase of the multiple-spin exchange
  {H}amiltonian on the triangular lattice},\ }\href
  {https://doi.org/10.1103/PhysRevB.60.1064} {\bibfield  {journal} {\bibinfo
  {journal} {Phys. Rev. B}\ }\textbf {\bibinfo {volume} {60}},\ \bibinfo
  {pages} {1064} (\bibinfo {year} {1999})}\BibitemShut {NoStop}%
\bibitem [{\citenamefont {Fennell}\ \emph {et~al.}(2011)\citenamefont
  {Fennell}, \citenamefont {Piatek}, \citenamefont {Stephenson}, \citenamefont
  {Nilsen},\ and\ \citenamefont {R{\o}nnow}}]{Fennell_2011}%
  \BibitemOpen
  \bibfield  {author} {\bibinfo {author} {\bibfnamefont {T.}~\bibnamefont
  {Fennell}}, \bibinfo {author} {\bibfnamefont {J.~O.}\ \bibnamefont {Piatek}},
  \bibinfo {author} {\bibfnamefont {R.~A.}\ \bibnamefont {Stephenson}},
  \bibinfo {author} {\bibfnamefont {G.~J.}\ \bibnamefont {Nilsen}},\ and\
  \bibinfo {author} {\bibfnamefont {H.~M.}\ \bibnamefont {R{\o}nnow}},\
  }\bibfield  {title} {\bibinfo {title} {Spangolite: an s= 1/2 maple leaf
  lattice antiferromagnet?},\ }\href
  {https://doi.org/10.1088/0953-8984/23/16/164201} {\bibfield  {journal}
  {\bibinfo  {journal} {Journal of Physics: Condensed Matter}\ }\textbf
  {\bibinfo {volume} {23}},\ \bibinfo {pages} {164201} (\bibinfo {year}
  {2011})}\BibitemShut {NoStop}%
\bibitem [{\citenamefont {Farnell}\ \emph {et~al.}(2011)\citenamefont
  {Farnell}, \citenamefont {Darradi}, \citenamefont {Schmidt},\ and\
  \citenamefont {Richter}}]{PhysRevB.84.104406}%
  \BibitemOpen
  \bibfield  {author} {\bibinfo {author} {\bibfnamefont {D.~J.~J.}\
  \bibnamefont {Farnell}}, \bibinfo {author} {\bibfnamefont {R.}~\bibnamefont
  {Darradi}}, \bibinfo {author} {\bibfnamefont {R.}~\bibnamefont {Schmidt}},\
  and\ \bibinfo {author} {\bibfnamefont {J.}~\bibnamefont {Richter}},\
  }\bibfield  {title} {\bibinfo {title} {Spin-half {H}eisenberg antiferromagnet
  on two archimedian lattices: From the bounce lattice to the maple-leaf
  lattice and beyond},\ }\href {https://doi.org/10.1103/PhysRevB.84.104406}
  {\bibfield  {journal} {\bibinfo  {journal} {Phys. Rev. B}\ }\textbf {\bibinfo
  {volume} {84}},\ \bibinfo {pages} {104406} (\bibinfo {year}
  {2011})}\BibitemShut {NoStop}%
\bibitem [{\citenamefont {Corboz}\ and\ \citenamefont
  {Mila}(2013)}]{PhysRevB.87.115144}%
  \BibitemOpen
  \bibfield  {author} {\bibinfo {author} {\bibfnamefont {P.}~\bibnamefont
  {Corboz}}\ and\ \bibinfo {author} {\bibfnamefont {F.}~\bibnamefont {Mila}},\
  }\bibfield  {title} {\bibinfo {title} {Tensor network study of the
  shastry-sutherland model in zero magnetic field},\ }\href
  {https://doi.org/10.1103/PhysRevB.87.115144} {\bibfield  {journal} {\bibinfo
  {journal} {Phys. Rev. B}\ }\textbf {\bibinfo {volume} {87}},\ \bibinfo
  {pages} {115144} (\bibinfo {year} {2013})}\BibitemShut {NoStop}%
\bibitem [{\citenamefont {Oganesyan}\ and\ \citenamefont
  {Huse}(2007)}]{Oganesyan_Huse_PRB2007}%
  \BibitemOpen
  \bibfield  {author} {\bibinfo {author} {\bibfnamefont {V.}~\bibnamefont
  {Oganesyan}}\ and\ \bibinfo {author} {\bibfnamefont {D.~A.}\ \bibnamefont
  {Huse}},\ }\bibfield  {title} {\bibinfo {title} {Localization of interacting
  fermions at high temperature},\ }\href
  {https://doi.org/10.1103/PhysRevB.75.155111} {\bibfield  {journal} {\bibinfo
  {journal} {Phys. Rev. B}\ }\textbf {\bibinfo {volume} {75}},\ \bibinfo
  {pages} {155111} (\bibinfo {year} {2007})}\BibitemShut {NoStop}%
\bibitem [{\citenamefont {Atas}\ \emph {et~al.}(2013)\citenamefont {Atas},
  \citenamefont {Bogomolny}, \citenamefont {Giraud},\ and\ \citenamefont
  {Roux}}]{Atas_Bogomolny_Roux_PRL2013}%
  \BibitemOpen
  \bibfield  {author} {\bibinfo {author} {\bibfnamefont {Y.~Y.}\ \bibnamefont
  {Atas}}, \bibinfo {author} {\bibfnamefont {E.}~\bibnamefont {Bogomolny}},
  \bibinfo {author} {\bibfnamefont {O.}~\bibnamefont {Giraud}},\ and\ \bibinfo
  {author} {\bibfnamefont {G.}~\bibnamefont {Roux}},\ }\bibfield  {title}
  {\bibinfo {title} {Distribution of the ratio of consecutive level spacings in
  random matrix ensembles},\ }\href
  {https://doi.org/10.1103/PhysRevLett.110.084101} {\bibfield  {journal}
  {\bibinfo  {journal} {Phys. Rev. Lett.}\ }\textbf {\bibinfo {volume} {110}},\
  \bibinfo {pages} {084101} (\bibinfo {year} {2013})}\BibitemShut {NoStop}%
\bibitem [{\citenamefont {Schnack}\ \emph {et~al.}(2001)\citenamefont
  {Schnack}, \citenamefont {Schmidt}, \citenamefont {Richter},\ and\
  \citenamefont {Schulenburg}}]{schnack2001independent}%
  \BibitemOpen
  \bibfield  {author} {\bibinfo {author} {\bibfnamefont {J.}~\bibnamefont
  {Schnack}}, \bibinfo {author} {\bibfnamefont {H.-J.}\ \bibnamefont
  {Schmidt}}, \bibinfo {author} {\bibfnamefont {J.}~\bibnamefont {Richter}},\
  and\ \bibinfo {author} {\bibfnamefont {J.}~\bibnamefont {Schulenburg}},\
  }\bibfield  {title} {\bibinfo {title} {Independent magnon states on magnetic
  polytopes},\ }\href@noop {} {\bibfield  {journal} {\bibinfo  {journal} {The
  European Physical Journal B-Condensed Matter and Complex Systems}\ }\textbf
  {\bibinfo {volume} {24}},\ \bibinfo {pages} {475} (\bibinfo {year}
  {2001})}\BibitemShut {NoStop}%
\bibitem [{\citenamefont {Richter}\ \emph {et~al.}(2009)\citenamefont
  {Richter}, \citenamefont {Schulenburg}, \citenamefont {Tomczak},\ and\
  \citenamefont {Schmalfu{\ss}}}]{richter2009heisenberg}%
  \BibitemOpen
  \bibfield  {author} {\bibinfo {author} {\bibfnamefont {J.}~\bibnamefont
  {Richter}}, \bibinfo {author} {\bibfnamefont {J.}~\bibnamefont
  {Schulenburg}}, \bibinfo {author} {\bibfnamefont {P.}~\bibnamefont
  {Tomczak}},\ and\ \bibinfo {author} {\bibfnamefont {D.}~\bibnamefont
  {Schmalfu{\ss}}},\ }\bibfield  {title} {\bibinfo {title} {The {H}eisenberg
  antiferromagnet on the square-kagome lattice},\ }\href@noop {} {\bibfield
  {journal} {\bibinfo  {journal} {Condensed Matter Physics}\ }\textbf {\bibinfo
  {volume} {12}},\ \bibinfo {pages} {507} (\bibinfo {year} {2009})}\BibitemShut
  {NoStop}%
\bibitem [{\citenamefont {Rousochatzakis}\ \emph {et~al.}(2013)\citenamefont
  {Rousochatzakis}, \citenamefont {Moessner},\ and\ \citenamefont {van~den
  Brink}}]{PhysRevB.88.195109}%
  \BibitemOpen
  \bibfield  {author} {\bibinfo {author} {\bibfnamefont {I.}~\bibnamefont
  {Rousochatzakis}}, \bibinfo {author} {\bibfnamefont {R.}~\bibnamefont
  {Moessner}},\ and\ \bibinfo {author} {\bibfnamefont {J.}~\bibnamefont
  {van~den Brink}},\ }\bibfield  {title} {\bibinfo {title} {Frustrated
  magnetism and resonating valence bond physics in two-dimensional kagome-like
  magnets},\ }\href {https://doi.org/10.1103/PhysRevB.88.195109} {\bibfield
  {journal} {\bibinfo  {journal} {Phys. Rev. B}\ }\textbf {\bibinfo {volume}
  {88}},\ \bibinfo {pages} {195109} (\bibinfo {year} {2013})}\BibitemShut
  {NoStop}%
\bibitem [{\citenamefont {Lugan}\ \emph {et~al.}(2019)\citenamefont {Lugan},
  \citenamefont {Jaubert},\ and\ \citenamefont
  {Ralko}}]{PhysRevResearch.1.033147}%
  \BibitemOpen
  \bibfield  {author} {\bibinfo {author} {\bibfnamefont {T.}~\bibnamefont
  {Lugan}}, \bibinfo {author} {\bibfnamefont {L.~D.~C.}\ \bibnamefont
  {Jaubert}},\ and\ \bibinfo {author} {\bibfnamefont {A.}~\bibnamefont
  {Ralko}},\ }\bibfield  {title} {\bibinfo {title} {Topological nematic spin
  liquid on the square kagome lattice},\ }\href
  {https://doi.org/10.1103/PhysRevResearch.1.033147} {\bibfield  {journal}
  {\bibinfo  {journal} {Phys. Rev. Research}\ }\textbf {\bibinfo {volume}
  {1}},\ \bibinfo {pages} {033147} (\bibinfo {year} {2019})}\BibitemShut
  {NoStop}%
\bibitem [{\citenamefont {Nakano}\ and\ \citenamefont
  {Sakai}(2013)}]{SquareKagome1}%
  \BibitemOpen
  \bibfield  {author} {\bibinfo {author} {\bibfnamefont {H.}~\bibnamefont
  {Nakano}}\ and\ \bibinfo {author} {\bibfnamefont {T.}~\bibnamefont {Sakai}},\
  }\bibfield  {title} {\bibinfo {title} {The two-dimensional s=1/2 {H}eisenberg
  antiferromagnet on the shuriken lattice –a lattice composed of
  vertex-sharing triangles–},\ }\href
  {https://doi.org/10.7566/JPSJ.82.083709} {\bibfield  {journal} {\bibinfo
  {journal} {Journal of the Physical Society of Japan}\ }\textbf {\bibinfo
  {volume} {82}},\ \bibinfo {pages} {083709} (\bibinfo {year}
  {2013})}\BibitemShut {NoStop}%
\bibitem [{\citenamefont {{Derzhko, O.}}\ and\ \citenamefont {{Richter,
  J.}}(2006)}]{DerzhkoRichter}%
  \BibitemOpen
  \bibfield  {author} {\bibinfo {author} {\bibnamefont {{Derzhko, O.}}}\ and\
  \bibinfo {author} {\bibnamefont {{Richter, J.}}},\ }\bibfield  {title}
  {\bibinfo {title} {Universal low-temperature behavior of frustrated quantum
  antiferromagnets in the vicinity of the saturation field},\ }\href
  {https://doi.org/10.1140/epjb/e2006-00273-y} {\bibfield  {journal} {\bibinfo
  {journal} {Eur. Phys. J. B}\ }\textbf {\bibinfo {volume} {52}},\ \bibinfo
  {pages} {23} (\bibinfo {year} {2006})}\BibitemShut {NoStop}%
\end{thebibliography}%

\end{document}